# Nonlinear Electrodynamics and QED

D.H. Delphenich[†]
Physics Department
University of Wisconsin – River Falls
River Falls, WI 54022

*Abstract. The limits of linear electrodynamics are reviewed, and possible directions of nonlinear extension are explored. The central theme is that the qualitative character of the empirical successes of quantum electrodynamics must be used as a guide for understanding the nature of the nonlinearity of electrodynamics at the subatomic level. Some established theories of nonlinear electrodynamics, namely, those of Mie, Born and Infeld are presented in the language of the modern geometrical and topological methods of mathematical physics. The manner by which spacetime curvature and topology can affect electromagnetism is also reviewed. Finally, the phenomena of nonlinear optics are reviewed as a possible guide to building one's intuition regarding the process of extending electrodynamics into nonlinearity in a manner that is consistent with the qualitative and empirical results of quantum electrodynamics.*

[†] david.delphenich@uwrf.edu

## Contents

### Part I: Quantum Electrodynamics

### Part II: Nonlinear electrodynamics

# Part I: Quantum electrodynamics.

**0. Introduction.** Probably the earliest mathematical statement about the nature of electricity and magnetism was Coulomb's law, which stated that the electrostatic force **E** on a unit "test" charge at a distance $r$ from a source charge $q$ is:

$$(0.1) \qquad \mathbf{E}(x) = k\frac{q}{r^2}\hat{\mathbf{r}}(x) = E(r)\hat{\mathbf{r}}, \qquad x \in \mathbb{R}^3 - \{0\}$$

in which $k$ is a coupling constant and $\hat{\mathbf{r}}$ is a unit vector field that points radially outward from the origin {0}. Notice that this means **E** is also undefined at the origin, which represents a pole singularity.

  The quotation marks around the word "test" are to remind one that the concept of a charged particle that can passively respond to the field of another charge without changing the state of that field assumes that there always exists a particle whose charge is so many orders of magnitude smaller than that of the field source that the effect of the test charge on the field can be neglected. This sort of thinking has a distinctly pre-quantum sort of character, since the real breakdown of classical mechanics seems to relate to precisely the fact that there is a non-zero lower bound to the available test particles and when one has to test one particle with a comparable one, it is meaningless to speak of one-particle mechanics anymore. In effect, the smallest physical system that has any validity is an interacting two-particle system, since one cannot measure the state of a non-interacting particle; to measure is to interact.

  *a. Quantum physics as the limit of Maxwellian electrodynamics.* To jump ahead a few centuries, let us say that around the end of the Nineteenth Century the otherwise complete form of the theory of electromagnetism was, of course, that of Maxwell. In the **E-B** vector form, the system of field equations (without inductions) is:

$$(0.2) \qquad \begin{cases} \nabla \cdot \mathbf{E} = \dfrac{4\pi}{c}\rho & \nabla \cdot \mathbf{B} = 0 \\ \nabla \times \mathbf{E} + \dfrac{1}{c}\dfrac{\partial \mathbf{B}}{\partial t} = 0 & \nabla \times \mathbf{B} - \dfrac{1}{c}\dfrac{\partial \mathbf{E}}{\partial t} = \dfrac{4\pi}{c}\mathbf{j}, \end{cases}$$

in which **B** is the magnetic field strength, $\rho$ is the charge density, and **j** is the electrical current density, which usually takes the form $\mathbf{j} = \rho\mathbf{v}$ when the charge distribution moves with a velocity **v**. One can also put these equations into the more elegant modern form ([1]):

---

[1] For the formulation of electrodynamics in the formalism of exterior differential forms, see Thirring [**1**].

(0.3) $$dF = 0, \qquad \delta F = \frac{4\pi}{c} J,$$

when one assembles the **E** and **B** vectors − or rather their corresponding 1-forms $E = i_\mathbf{E} g$ and $B = i_\mathbf{B} g$ − into the Minkowski 2-form $F$ and the charge density and current into $J$:

(0.4) $$F = \theta \wedge E - {*}(\theta \wedge B), \qquad J = \rho \theta + j.$$

The timelike unit 1-form $\theta = i_\mathbf{t} g$ is metric to the timelike unit vector field **t** that one has chosen to define a time orientation on $T(M)$, the tangent bundle to spacetime; such a choice is equivalent to a choice of proper time axis in each tangent space, or a choice of *rest frame* (which is really an $SO(3)$ equivalence class of frames that share **t** as their common timelike member). Physically, a rest frame can be defined only by the motion of something, and even then, there is no rest frame for a light wave.

The key features of these equations that we shall address are:

i) The equation $\nabla \cdot \mathbf{E} = 4\pi/c \rho$ is really a thinly veiled generalization of Coulomb's law.
ii) The equations are linear in the **E** and **B** fields.
iii) Classically, one assumes that the spacetime vacuum does not support inductions.

One way of looking at the significance of the early experiments that ushered in the age of quantum physics is to say that generally the experiments suggested that Maxwell's equations seemed to break down in the realm of atomic to subatomic processes. In particular, the Maxwellian picture of radiation from accelerating charges was totally inconsistent with the experimental facts of atomic spectroscopy, such as the existence of a ground state of non-zero kinetic energy, the discreteness in the allowed transitions, and the spin of the electron, to name a few.

A common argument to support the use of Maxwell's equations nonetheless is that when one combines Coulomb's law for the atomic nucleus with the Dirac equation for the wavefunction of the spinning atomic electron, one obtains theoretical results that match the experimental facts impeccably. However, a counter-argument could then be that the success of the Dirac equation grew out of the success of the Schrödinger equation, whose precise origin was more fortuitous than profound, as even Schrödinger himself admitted.

One then wonders if some extension of Maxwell's theory might resolve the issue more conclusively. Hence, one considers the best ways to weaken the aforementioned hypotheses, especially linearity.

One might also point out quantum electrodynamics is the most definitive form of the theory of charged particles. However, one should remember that quantum electrodynamics is a phenomenological theory that accounts for the scattering data that result from the interactions of charged particles. In effect, that theory does not explain everything there is to say about charged particles; it merely explains all of the known scattering results. In particular, even Feynman himself admitted that little is known about the actual process that occurs *during* the transition from one atomic electron state to another, only the final outcome. The central thesis of the following discussion is that the phenomenological successes of quantum electrodynamics should be treated as strongly worded hints at how to extend Maxwell's equations into the quantum – i.e., atomic to subatomic – domain.

In what follows, we shall test the limits on all of the above three assumptions, as well as give due credit to the successes of quantum electrodynamics. A key theme will be the way that the polarization of the electromagnetic vacuum state that is consistent with the Dirac theory of charged particles is also a good reason to consider the breakdown of linearity in the field equations in the realm of large field strengths, such as one finds close to the atomic nucleus.

*b. Theoretical extensions of Coulomb's law.* Since Coulomb's result was of a purely empirical nature, let us ask about the current limits of the experimental verification of this law. According to Jackson [**2**], the inverse-square law of electrostatics has been experimentally verified to within unimpeachable accuracy over the range of $10^{-15}$ cm. $< r < 10^9$ cm.

Some of the extensions that were proposed as a way of quantifying the accuracy of the law (compared to others) were:

  *i*) An inverse-something-almost-squared law:

(0.5) $$E(r) \propto \frac{1}{r^{2+\varepsilon}}$$

  *ii*) A non-zero photon mass $m_\gamma$. This would suggest the *Proca Lagrangian:*

(0.6) $$+_{EM} = \tfrac{1}{2} F {\wedge} {*} F + \frac{\mu^2}{4\pi} A {\wedge} {*} A,$$

where:

(0.7) $$\mu = \frac{m_\gamma c}{h}$$

is the Compton wave number of the photon. This gives the *Proca equation* for the electromagnetic field equation:

(0.8) $$\mathbf{u} A + \mu^2 A = 0.$$

In the electrostatic case, one can replace $A$ with the electrostatic potential function $\phi(r)$, which then takes the *Yukawa* form:

(0.9) $$\phi(r) = \frac{e^{-\mu r}}{r},$$

Note that the departure of the Yukawa potential from the Coulomb $1/r$ potential is most pronounced at *large* distances from the source charge, not at small ones. This is also expressed by the statement that when an interaction is mediated by the exchange of a massive gauge particle the interaction must have a finite range that gets smaller with increasing mass. Consequently, since photons are expected to cross billions of light years of deep space, one can get an upper bound on the photon mass from cosmological considerations.

As we shall see in other contexts, the introduction of a mass term usually breaks the $U(1)$ gauge symmetry of the field theory, unless it appears as a result of the Higgs mechanism.

Note also that when the photon has mass, one would expect a vacuum dispersion relation for electromagnetic waves of:

$$(0.10) \qquad \omega = \sqrt{c^2 k^2 + c^2 \mu^2} = ck\sqrt{1 + \frac{\mu^2}{k^2}} \approx ck\left(1 + \frac{1}{2}\left(\frac{\mu}{k}\right)^2 + \ldots\right)$$

as opposed to the usual case of $\omega = ck$. Of course, such effects should not become noticeable until $k \sim \mu$.

*c. Experimental tests of Coulomb's law.* The first avenue of extension was tested experimentally by Williams, Faller, and Hill, who found that $\varepsilon < (2.7 \pm 3.1) \times 10^{-16}$ for values of $r$ in the range of centimeters. The second possibility was tested experimentally in the context of the Earth's magnetic field by Kobzarev and Okun, as well as Goldhaber and Nieto, who established that $m_\gamma < 4 \times 10^{-48}$ gm (compare to the electron, for which $m_e = 9.109 \times 10^{-28}$ gm), which also makes the Compton wavelength of a photon greater than $10^{10}$ cm. Hence, the range of validity for $r$ in this experiment was on the order of several hundred Earth radii.

The experimental verification of Coulomb's law in the sub-centimeter range was inferred from the elastic scattering of electrons and positrons with center-of-mass energies up to 5 GeV. This extended Coulomb down to $r \approx 10^{-15}$ cm.

Although the foregoing results may seem disappointing to anyone who is skeptical of Coulomb's law, one must notice that they would only satisfy the macroscopic physicist or engineer since they still suggest that one must look into either the subatomic or cosmological realms for the departure of electrostatics from the Coulomb law.

*d. Linearity in Maxwell's equations.* There is a third avenue of extension beyond Coulomb's law that is more oriented towards the modern line of thinking about the limits of mathematical models in physics: one can consider that the linear field equations that grow out of Coulomb's law, namely, the Maxwell field equations, are only approximations to nonlinear field equations, and the approximation breaks down somewhere inside of the $10^{-15}$ cm bound.

In physics, one must approach nonlinearity with caution, since most nonlinear differential equations – especially nonlinear partial differential equations – are unlikely to admit concise closed-form solutions, and, in some cases, they may admit no solutions at all, at least ones with a physically interesting character. Consequently, one must consider the source of nonlinearity in a field theory by starting with physically reasonable assumptions and weakening them by analogy with other established physical phenomena, rather than merely generalizing the purely mathematical form of the Lagrangian or field equations in an axiomatic way.

To experimentally establish the limits of linearity in the field equations of electromagnetism, one generally looks to the wavelike solutions. For such solutions, one can observe the manner by which distinct fields combine by observing interference and diffraction effects. As Jackson points out, when one assumes the Maxwell form for the potential in Schrödinger's nonrelativistic wave equation for spinless wave-particles or even Dirac's relativistic wave equation for spin-1/2 particles, one predicts interference and diffraction effects that have been experimentally established quite accurately at the atomic level. Then again, the true origin of the Schrödinger equation is sufficiently debatable that it may simply mask the limits of linearity in the Maxwell equations when dealing with the nuclear electrostatic potential.

The most physically suspicious aspect of Coulomb's law is its pole singularity. This leads to not only unbounded field strengths as one approaches a point source arbitrarily closely, but also an infinite self-energy for the field of any point-like charge. This latter property is still present in quantum electrodynamics, where it provokes one to consider the largely non-rigorous process of renormalization in order to arrive at a finite self-energy. It is in the realm of large field strengths that one might reasonably expect to find the limits of Maxwell's equations.

Field strengths in the laboratory are usually limited by the fact that either a gaseous atmosphere will ionize past about 10 kV/cm or the free charges on a conductor in vacuo, which are bound to the conductor by a fairly weak attraction, will break free of the conductor past a certain field strength, thus partially reducing the source charge distribution, and, with it, the resulting field strength. Hence, the only possible experimental realm in which to find enormous electrostatic field strengths is in the small neighborhoods of elementary charges. For instance, the nuclear electrostatic potential gives a field strength of about $10^{15}$ V/cm at the nuclear surface, for a heavy nucleus, and $10^9$–$10^{15}$ V/cm for the various electron shells.

Since the realm in which we expect to find a departure of Coulomb's law from the "true" law of electrostatics is also the realm in which one expects quantum electrodynamics (QED) to be more appropriate, once we have discussed some of the early attempts at extending Maxwell's theory into the atomic domain, we then summarize some of the qualitative results of QED as a motivation for the most productive extensions of the Maxwell model. We then discuss some of the more noteworthy attempts at nonlinear electrodynamics, namely, those of Mie, Born and Infeld. We then take a cursory look at some of the possible ways that the geometry and topology of spacetime might affect Maxwell's equations. Finally, we survey some of the relevant theoretical notions of nonlinear optics that might have analogues in the processes of QED.

**1. Classical electron models.** Nearly every significant figure of theoretical physics, even well into the age of quantum electrodynamics, made some attempt at constructing a model for the electron ([2]). Of course, at the point in history when these researches were begun, namely, the early Twentieth Century, the only elementary charges that were known were the electron and proton. Moreover, there was, of course, no recognition that one of those particles was a lepton and the other was a hadron that could be decomposed into a bound state of more elementary charges in the form of quarks. Hence, one has to keep in mind the limitations of the term "electron" in such works. Notable among the early attempts were the contributions of Abraham, Lorentz, and Poincaré. However, even as quantum physics became the mainstream approach to elementary matter, there continued to be attempts to find classical models, including Dirac himself, as well as Rohrlich and others. This work also related in an intimate way to the attempts at finding a causal interpretation for the wave mechanics of quantum physics, such as the hydrodynamical interpretation of Madelung, Takabayasi, Bohm, Vigier, Mathisson, Weyssenhoff, and Halbwachs.

---

[2] In all of the classical attempts at modeling the most elementary form of electrically charged matter, the word "electron" was used in a more general sense to mean elementary charge distributions, regardless of sign.

In the earliest models, only the mass, charge, spherical symmetry, and stability of the electron field were considered. Later, Uhlenbeck and Goudsmit introduced the notion of a spinning electron into quantum theory, which was then further developed by Pauli and Dirac, and with it, a magnetic moment, which played a key role in establishing quantum electrodynamics by way of the Lamb shift in atomic spectra.

To summarize the various models we classify them on the following basis:

   *i*) Spatial extension: point-like or extended (spherical shell, ball, "dumb-bell")
   *ii*) Rigidity: non-relativistically rigid, relativistically rigid, elastic, fluid
   *iii*) Spinning or non-spinning.

*a. Abraham-Lorentz-Poincaré model.* The earliest model, by Abraham [3], which was then refined by Lorentz [4] and Poincaré gave the charge distribution the form of an extended non-spinning spherical shell of radius $R_e$. In the eyes of experiment, as long as one remains outside of this radius, the field of this distribution is indistinguishable from that of the point-like or ball-like distribution, since it is only the total charge *e* that dictates the field, not the internal structure (beyond symmetry and stability, of course). A common objective of the classical electron models was to see if it was possible to attribute the mechanical properties of the source particle – i.e., rest mass $m_e$ and momentum *p* – exclusively from the information that one found in the field itself. This carried with the problem of ensuring that the components of the resulting energy-momentum 1-form $p = p_\mu dx^\mu$ actually transformed properly under the action of the Lorentz group for them to collectively define a Lorentz invariant object.

Both the extended and point-like electron models had advantages and limitations in the eyes of the linear Maxwell theory. If the charge distribution was distributed over, say, a spherical shell of radius $R_e$ then it only took a finite amount of work:

(1.1) $$U = \frac{e^2}{R_e}$$

to assemble this distribution. This figure should, moreover, equal the energy that is stored in the **E** field itself:

(1.2) $$U = \frac{1}{8\pi} \int_{R \geq R_e} \|\mathbf{E}\|^2 V_s = \frac{1}{8\pi} \int_{R \geq R_e} E \wedge *_s E,$$

in which $V_s = dx \wedge dy \wedge dz$ is the spatial volume element, $*_s$ is its associated Hodge dual operator, and $E = i_\mathbf{E} g$ is the one-form that is metric dual to the **E** field. (Of course, the metric *g* is only assumed to be the Euclidian one, at this point.)

If one assumed that the rest mass of the particle in question was purely due to this field energy or work done then Einstein's celebrated law gives:

$$(1.3) \qquad m_e = \frac{e^2}{2R_e c^2}.$$

On the basis of measured values for $m_e$ and $e$, one generally identifies:

$$(1.4) \qquad R_e = \frac{e^2}{m_e c^2} = 2.82 \times 10^{-13} \text{ cm}.$$

as the *classical electron radius*. Note that this number is actually larger than the experimental limits on the point-like nature of the electron, so if one wishes to maintain the extended character, one must account for the fact that the "true" $R_e$ must be much smaller than the classical estimate. Indeed, some more recent models are inclined to go as far down as the Planck length ($\sim 10^{-33}$ cm.).

In order to get the momentum of the elementary charge, one can either obtain it as $m_e v^i$, $i = 1, 2, 3$, where $v^i$ is the spatial velocity of the center of charge, or, as Abraham assumed, from integrating the components of the Poynting vector:

$$(1.5) \qquad p_i = \frac{1}{c^2} \int_{R > R_e} (\mathbf{E} \times \mathbf{B})_i V_s,$$

which gave a result of the form:

$$(1.6) \qquad p_i = \tfrac{4}{3} m_e f\!\left(\frac{v}{c}\right) v_i$$

in Abraham's model; the function $f(v/c)$ is a form factor that breaks the Lorentz invariance of the resulting energy-momentum 1-form $p = m_e c^2 dx^0 + p_i dx^i$, and has a non-relativistic limit of 4/3, which contradicts one's expectation that it should be 1.

In order to address the discrepancy, one must address the other drawback to the extended electron model, besides the problem of defining a Lorentz invariant energy-momentum 1-form from purely electromagnetic considerations, and that is the fact that the linearity in Maxwell's equations implies that any extended charge distribution will be inherently unstable, due to the mutual repulsion of any pair of disjoint regions in the charge distribution. Hence, the natural tendency of an extended spherical charge distribution will be to expand without limit, which contradicts the apparent stability of electrons. Note that an analogous question arises in the modeling of atomic nuclei, in which the solution was the postulate of a "strong" interaction that was distinct from electromagnetic interaction and counteracted that force, but only over a range of distances of nuclear magnitude. We emphasize that it is the linearity in the field equations for $\mathbf{E}$ that is mostly responsible in order to draw attention to the possibility that a nonlinear self-interaction term might stabilize the distribution; this also opens the conjectural possibility that the strong force that binds nucleons might be the nonlinear part of the electrostatic repulsion force.

Abraham's conception of an electron amounted to a non-relativistic rigid shell in which the charge distribution was bound to the surface by an unspecified mechanical force of constraint. Poincaré modified this model by adding a mechanical force and momentum to the electromagnetic self-force and momentum specifically. He also loosened the rigid sphere to a flexible one that is held together by a uniform surface tension, which he also showed contributed precisely the desired $-1/3(e^2/2R_e c^2)$ term to the total momentum. However, the limitations of his model were that it was unstable under deformations of its shape and one has to account for the physical nature of the binding force, which would lie beyond the scope of Maxwell's theory.

In light of the aforementioned limitations of the extended electron model, an obvious alternative would be to consider a point-like charge distribution. However, although the instability and Lorentz varying form factor go away, this advantage is immediately counterbalanced by the fact that as $R_e \to 0$, $U \to \infty$. Despite this infinite self-energy, which persists in quantum electrodynamics, and the fact that experiments have yet to get close enough to $R_e$ to distinguish an extended from a point-like distribution, most theoretical physicists implicitly envision a point-like electron, although they also envision that quantum considerations will still tend to "smear" the distribution, nonetheless.

Dirac proposed a point-like model in which the source point existed like a "bubble" in the electromagnetic field, a concept with intriguing topological ramifications. Moreover, the bubble gave a boundary surface to the space on which the field was defined and this boundary also had an associated surface tension. However, Dirac mostly concentrated on trying to exhibit the muon as an excited state of the electron that represented spherical oscillations of this boundary, so he did not pursue this model more completely.

One of the unavoidable aspects of the electron, and the one that relates most definitively to the success of the Dirac equation in modeling the fields of some charged particles is the fact that electrons possess an intrinsic angular momentum or spin. It was established experimentally by Stern and Gerlach, theoretically suggested by Uhlenbeck and Goudsmit, and developed further by Pauli, Dirac, and others. Hence, anyone who still believed in the possibility of a classical model for the electron had to account for that fact somehow. Bohm and Vigier [**5**] attempted to account for the *SU*(2) spin theory of the electron that Pauli had posed by envisioning an electron as a rigid body. Weyssenhoff and Raabe [**6**] derived a set of equations of motion for a relativistic spinning fluid that actually could be shown to define the real form of the Dirac equation in the same way the Madelung equations [**7**, **8**] gave the real form of the Schrödinger or Klein-Gordon equations in terms of the motion of an irrotational, but compressible, fluid that also included a force contribution from a mysterious potential of quantum origin.

*b. Classical theory of radiation.* Besides the failure of Maxwell's theory of electromagnetism to provide an acceptable model of elementary charges as a consequence, another, perhaps even more conspicuous, failing of the theory was the manner by which the Maxwellian theory of radiation contradicted the experiments that defined the historical foundations of quantum theory. In particular, Maxwell did not properly account for the spectrum of blackbody radiation or the spectra of atoms. Furthermore, there were serious flaws in the equations of motion that one obtained for an accelerated charge when one included a term to account for the radiation damping.

At the root of all of this evil is the manner by which electromagnetic radiation is produced (or absorbed); one can also think of this problem as that of discerning the true nature of the sources of the wavelike solutions to the electromagnetic field equations in a more general sense. In Maxwell's theory, just as a magnetic field ultimately originates in the relative velocity **v** of a charge, similarly, an electromagnetic wave originates in its relative acceleration **a**. Classically, the time rate at which a moving charge *e* emits energy is equal to:

$$(1.7) \qquad \frac{dU}{dt} = \frac{2}{3}\frac{e^2}{c^3}\|\mathbf{a}\|^2,$$

which is the relativistic form of the Larmor formula [**2**, **9**, 10]. In particular, when the acceleration is due uniform circular motion with a radius *r*, as was assumed in the Bohr model of atomic electrons, one obtains the classical expression for *synchrotron radiation:*

$$(1.8) \qquad \frac{dU}{dt} = \frac{2}{3}c\frac{e^2}{r^2}(\gamma^2 - 1)^2$$

in which $\gamma = (1 - (v/c)^2)^{-1/2}$ is the Fitzgerald-Lorentz factor.

Of course, the problem with this expression is that there will be a continual loss of energy for an electron in uniform circular motion about a nucleus, which will produce three effects that are completely at odds with atomic spectrum experiments:

*i*) The emitted radiation will have a continuously declining energy, hence, a continuously declining frequency, but the spectra of atomic electrons while they are still bound to the nucleus are discrete in character.

*ii*) The ground state of the atomic electron must have a kinetic energy of zero to be consistent with its ultimate collision with the atomic nucleus. Atomic spectra suggest a non-zero ground state, just as one finds in the quantum harmonic oscillator.

*iii*) Individual atoms should generally have rather short lifetimes that correspond to the small amount of energy that the electrons start out with. However, experiments suggest no such instability for most atoms.

One further source of controversy in the classical theory of radiation was in the nature of the solutions to the equations of motion for a charged particle in an external electromagnetic field when one included the effects of energy loss due to the radiation that would be emitted as a result of its acceleration; since we are talking about a loss of energy, this effect is referred to as *radiation damping*.

The equation of motion amounts to the usual Lorentz force equation with an additional term that accounts for the radiation reaction to the acceleration, which is due to Dirac, so one calls the resulting equation the *Lorentz-Dirac* equation:

$$(1.9) \qquad \frac{dp}{dt} = i_\mathbf{v} F + \frac{2}{3}\frac{e^2}{c^3}\left(\frac{da}{dt} - \frac{1}{c^2}a^2 u\right),$$

where we are representing this equation in terms of the energy-momentum 1-form *p*, Faraday 2-form *F*, co-velocity 1-form $u = i_\mathbf{v} g$, co-acceleration 1-form $a = du/dt \rightleftharpoons L_\mathbf{v} u$, and the time derivative is given by the Lie derivative $L_\mathbf{v}$ along the velocity vector field **v**.

The first suspicious aspect of this equation is that it amounts to a third order ODE, instead of the usual second order one. This means that in order to completely pose the initial value problem one must specify the initial acceleration along with the initial position and velocity. This not only contradicts physical intuition, but actually leads to physically pathological solutions, such as the "runaway electron," which can accelerate to infinite velocity in the absence of external forces, acausal solutions in which electrons appear to "anticipate" their interaction with another charge, and electrons that accelerate in the opposite direction to the applied force that would be consistent with Newton's third law of motion. Although there are physically reasonable ways of restricting the solution space to eliminate these solutions, nevertheless, one still maintains a certain degree of skepticism.

One of the intriguing possibilities that the Lorentz-Dirac equation suggests is the existence of non-radiating modes of accelerated charges. That is, when the mass of the electron is purely electromagnetic it is possible for a slow isolated electron to exhibit oscillations of period $2R_e/c$ indefinitely, with no radiation damping. This could be a hint of a possible way to account for the discrete nature of the bound state electrons of atoms.

A key aspect that both the problems of the form of elementary charges and the form of the radiation reaction in accelerated charges had in common is the fact that in both cases, one must confront the problem of replacing an extended object that is governed by a PDE with a point-like object that is governed by an ODE; this also characterizes the problem of characterizing the classical limit of quantum mechanics – à la Eherenfest – as well as the problem of obtaining the geodesic equations of motion from the Einstein field equations for gravitation. A crucial problem one always has to address at such times is how one can realistically expect to approximate an infinite-dimensional dynamical system, such as a field or wave, by a finite-dimensional one, viz., a point moving in spacetime. Although representing the time evolution of solutions to PDE's by the motion of a point in an infinite-dimensional space (usually, a space of Cauchy data for the solution) as governed by a system of ODE's is commonplace nowadays ([3]), nevertheless, the reduction of infinite-dimensional motions to finite-dimensional ones seems to be fraught with paradoxes, except in the simplest cases, such as rigid motions.

Since it is only natural to feel that there is something atavistic about any discussion of electrodynamics at the atomic level when the accepted modern approach to such matters is quantum electrodynamics, we now pause to selectively summarize the qualitative aspects of that theory that might shed light on the most promising directions in which to go beyond the linearity in Maxwell's equations.

**2. Qualitative aspects of quantum electrodynamics.** Naturally, most sober physicists would regard any attempt to summarize the important aspects of quantum electrodynamics as naïve folly bordering upon sheer psychosis. This is because quantum electrodynamics – much less, quantum field theory, in general – is so vast in scope, and involves so many idiosyncratic techniques for dealing with specialized problems that most people who enter its portals must abandon all hope of mastering more than a few such sets of techniques that pertain to a small number of popular research topics.

However, it is the opinion, if not the polemic, of the author that this state of affairs is testimony to the demoralizing influence that quantum physics has had on theoretical physics, in general. In particular, the accepted rationalization for simply agreeing upon the statistical interpretation of wave mechanics as a basis for further research is the pseudo-philosophy of phenomenology, which simply says that, in the absence of future inspirations, one must be content to deal with what little information can be obtained from experiment alone. This is analogous to the way that behavioral psychology chose to pass over any discussion of the states of the human mind and their dynamics in favor of tediously establishing a large enough catalog of experimentally obtained stimulus-response relationships in which the intermediate block in the system diagram − the state space of the mind − was treated as an enigmatic "black box." It is also the methodology of analyzing the structure of the Earth's interior: one emits phonon pulses into the crust and measures the way that the pulse propagates with other sensors placed in the crust. Ultimately, one tries to reconstruct the "scattering potential" that could have produced the observed response, which defines the *inverse scattering problem,* i.e., reconstructing unknown force potentials from the measured scattering data.

---

[3] See Chernoff and Marsden [**11**] or Kijowski and Tulczjew [**12**].

In the case of quantum field theory, the black box is the detailed structure of elementary matter and its interactions. Because such phenomena, like the initial singularity of the universe itself, lie so far beyond the realm of direct laboratory measurement, the phenomenological approach was chosen since it was closest to the primary tools for reaching the subatomic domain, namely, the measurement of atomic spectra for the bound state problems and high-energy particle scattering experiments for the others.

As a result of the reliance of quantum field theory on the scattering process, the term "quantum field theory" rapidly became a misnomer, since, in effect, the statement of an actual "field theory" was more on the order of the ultimate goal of the theory, not one of its starting assumptions. Here, we are loosely defining a "field theory" to consist of a set of sections of some appropriate fiber bundle (usually a vector bundle for linear field theories) that relate to various linear or nonlinear representations of an internal symmetry group *G*, together with a system of *G*-invariant differential equations for the sections of the bundle, and an increasing inventory of solved boundary-value problems whose physical interpretations are consistent with experiment. Indeed, the only fields, field equations, and boundary-value problems that one encounters in quantum field theories are usually just the free, i.e., non-interacting, fields; the scattering matrix that one obtains in this case is, of course, quite lacking in deeper insight.

Ultimately, one must "close the loop" of theory and phenomenology by a sort of inverse scattering approach to the results of quantum electrodynamics, i.e., to use the qualitative and quantitative successes of quantum electrodynamics as the basis for deeper theoretical conjectures about the nature of the first principles of physics, just as the inverse scattering problem seeks to reconstruct a scattering potential from scattering data. Indeed, from the QED side of the problem what one frequently defines are "effective" potentials (cf., [**13**] in particular) that would account for the scattering data. However, unless these effective potentials can be deduced from deeper first principles, they still have much of the semi-empirical quality of the QED scattering amplitudes that one derives from the Feynman diagrams. A recurring theme of this article is a parallel effort must be made from the non-phenomenological side of the problem to find deeper first principles that will account for the effective potentials. The example of this situation that we shall be primarily concerned with is the example of the phenomenological Lagrangian that Euler and Heisenberg derived for electrodynamics using the methods of QED by including the effects of vacuum polarization versus the Lagrangian that Born and Infeld defined by starting from first principles; the two models are seen to be in close agreement up to a point.

Undoubtedly, one such first principle must be the unavoidable appearance of nonlinearity in any natural system, at least in some parameter regime. In the case of electrodynamics, one expects this regime to be that of large field strengths, such as in the small neighborhoods of elementary charged matter. Hence, a basic assumption that one might make for nonlinear electrodynamics is that the effective potentials of QED are implicitly pointing the way to the most experimentally productive forms of nonlinearity to examine.

One aspect of concentrating on the scattering process instead of the field equations is that since the nonlinearity in those equations is presumably significant only for a short period of time over a small neighborhood of the collision process, if one considers only time-asymptotic states in the distant past and future then one can avoid the nonlinearity by dealing with a linear scattering operator $S$: $H_{in} \to H_{out}$ that takes incoming scattering states to outgoing ones. For a given choice of bases $\{|i>, i = 1, 2, \ldots\}$, $\{|f>, f = 1, 2, \ldots\}$ on the Hilbert spaces $H_{in}$ and $H_{out}$, respectively, each of the matrix elements $<f|S|i>$ of $S$ will represent the complex scattering amplitude whose modulus squared gives the transition probability from the state $|i>$ to the state $|f>$. This is consistent with the statistical interpretation of wave mechanics, which interprets $|<f|i>|^2$, when both states are in the same Hilbert space of wave states, as the probability of finding the physical system in question in the state $|f>$ when it was originally given the state $|i>$.

This brings us back to the perturbation expansions that occupy so much of quantum field theory, because, not only does a different choice of interaction generally produce a different set of Feynman rules, but, moreover, one cannot go past general expressions without choosing a particular scattering process, such as Compton scattering, and evaluating the particular terms in the expansion that it implies. It does not take much imagination to recognize the extent of this program of research; one is, after all, trying to span an infinite-dimensional space of possibilities. Perhaps the nadir of all hope in quantum field theory was the era in the 1960's when some strong interaction physicists, notably Heisenberg, decided that perhaps the theory of strong interactions should be *identical* with the theory of this scattering, or *S*-matrix.

One should, however, point out that modern QED [**14**] has taken up some of the problems that one encounters when one recognizes that the coupling constant for electromagnetism, which one is using as the independent variable in the aforementioned power series, is not always the fine structure constant $\alpha = e^2/\hbar c = 1/137$. When looking at the field of a nucleus of atomic number $Z$, one must expand in terms of $Z\alpha$. Clearly, the convergence of such an expansion should be expected to break around $Z = 137$, and it is intriguing that the nuclei of the periodic table grow increasingly unstable before one reaches that number. However, one can also envision *quasi-molecules*, i.e., bound states of nuclei, whose combined atomic number exceeds the critical value. As discussed in [**13**], for the electric field strengths that are associated with such supercritical $Z$ the process of vacuum polarization gives way to the separation of the electron-positron pairs into an electron cloud that surrounds the nucleus and a positron cloud that is repelled and expands to infinity. In effect, if the net charge is treated as an order parameter, the appearance of a non-zero value for the vacuum expectation value of the charge operator represents a phase transition that one associates with symmetry breaking, whether spontaneous or dynamical. For such supercritical scenarios, perturbative methods must give way to non-perturbative ones, but in this work, we shall be concerned only with the more traditional sub-critical scenario in which the perturbative methods are valid.

The question to the intransigent field theorist who chooses to regard the perturbative evaluation of scattering amplitudes as motivational, not definitive, of a final field theory is then: how does one use the experience that physics has amassed by the force of phenomenology to weaken the assumptions of classical field theory in a direction that will produce results that are consistent with the known experimental facts?

The answer that will be given here for the sake of exploration is that one must concentrate more on the qualitative nature of the results as they relate to natural phenomena than on the mathematical formalism itself. Hence, we summarize some of the qualitative notions that seem to have come out of QED through the years; the details of the evaluation of the scattering amplitudes from the Feynman diagrams will be referred to any good reference [**15-18**]:

*a. Quantum harmonic oscillators.* One of way of understanding the difference between the electromagnetic field of classical theory and that of quantum theory, at least when the field operator formalism is used to represent the "quantized" electromagnetic field, is that the classical electromagnetic field is a continuously distributed system of coupled *classical* harmonic oscillators and the quantum electromagnetic field is a continuously distributed system of coupled *quantum* harmonic oscillators. Hence, before continuing we should summarize some of the key aspects of the latter model, as they relate to the present discussion.

One obtains the quantum harmonic oscillator by canonically quantizing the classical Hamiltonian:

(2.1) $$H = \frac{1}{2m}p^2 + \frac{1}{2}m\omega^2 x^2$$

and thus obtaining the stationary Schrödinger equation $H\psi = E\psi$:

(2.2) $$\frac{-\hbar^2}{2m}\nabla^2\psi + \frac{1}{2}m\omega^2 x^2\psi = E\psi,$$

or:

(2.3) $$\nabla^2\psi + (\frac{2mE}{\hbar^2} - m^2\omega^2 x^2)\psi = 0.$$

Rather than go the route of directly solving this equation by Hermite polynomials we pass on to the introduction of *creation* and *annihilation* operators. One defines them by the expressions:

(2.4) $$a = \sqrt{\frac{m\omega}{2}}x + \frac{i}{\sqrt{2m\omega}}p, \qquad a^\dagger = \sqrt{\frac{m\omega}{2}}x - \frac{i}{\sqrt{2m\omega}}p,$$

respectively.

These are both Hermitian operators, and their commutation relation is:

(2.5) $$[a, a^\dagger] = \hbar.$$

This allows us to rewrite the Hamiltonian in the form:

(2.6) $$H = \tfrac{1}{2}\hbar\omega + \omega a a^\dagger,$$

whose eigenvalues are:

(2.7) $$E_n = (n + \tfrac{1}{2})\hbar\omega, \qquad n = 0, 1, 2, \ldots$$

and if $\psi_0$ is the ground state ($n = 0$) wavefunction:

(2.8) $$\psi_0(x) = \left(\frac{m\omega}{\hbar\pi}\right)^{1/4} e^{-m\omega x^2/2\hbar},$$

then all of the other excited states ($n > 0$) can be obtained by repeated application of the creation operator to the ground state:

(2.9) $$\psi_n(x) = \frac{\hbar^{-n/2}}{\sqrt{n!}} (a^\dagger)^n \psi_0(x).$$

Whereas the classical harmonic oscillator that is governed by the Hamiltonian (1.1) has precisely two modes of oscillation, with frequencies 0 and $\omega$, depending on whether the initial energy in the system is zero or non-zero, respectively, the quantum harmonic oscillator has an infinite sequence of closely spaced oscillatory modes. What is especially important is the fact that the quantum ground state has non-zero energy. In order to get some intuition for the classical limit of the quantum expression, note that if $A$ is the amplitude of the classical oscillation then the value of $n$ for which $E_n = m\omega^2 A^2/2$ is:

(2.10) $$n = \frac{m\omega A^2}{2\hbar} - \frac{1}{2} \sim \frac{1}{\hbar} \sim 10^{27}$$

if we use CGS units. Hence, what the quantization has accomplished is to add a lot of "quantum fluctuations" about the ground state. In the statistical interpretation of wave mechanics, this is attributed to the fact that the non-commutation of $p$ and $x$ implies that the $E = 0$ ground state ($p = x = 0$) violates Heisenberg's uncertainty principle, so it must be physically unattainable.

*b. Quantization of the electromagnetic (wave) field.* By analogy with the way that the classical electromagnetic field can be regarded as a continuous distribution of coupled harmonic oscillators, either by by way of conjugate pairs of phase variables ($\mathbf{E}(x)$, $\mathbf{B}(x)$) at every point (the *configuration space* picture) or conjugate pairs ($a(k)$, $a^*(k)$) of functions at every wave number (the *frequency domain* or *momentum space* picture), in the field operator approach to quantum electrodynamics one regards the aforementioned conjugate pairs as Hermitian operator-valued ([4]).

In practice, the phase space of the electromagnetic fields is defined by choosing a potential 1-form $A$, i.e., a 1-form such that $dA = F$, whose conjugate momentum is the 3-form:

---

[4] More analytically rigorous treatments of quantum field (e.g., [19]) would demand that one does not simply replace real numbers with Hermitian operators as field values, that one must also replace the field components with operator-valued tempered distributions on test functions (smooth, compact support), but few of the physical results – i.e., scattering amplitudes for specific processes – depend on these refinements, at least on the surface of the argument.

(2.12) $$*\pi = \frac{\delta_+}{\delta(\partial_\tau A)} = \frac{\partial_+}{\partial(\partial_\tau A)} = *i_t F \qquad (+ = -\tfrac{1}{2} F \wedge *F, \; \partial_\tau A = L_t A)$$

or its dual 1-form $i_t F = E$, the electric field strength 1-form. However, one immediately sees that the timelike component of $\pi$, namely $\pi_0 = i_t E = 0$, but this is an artifact of our choice of proper-time orientation **t**, so in order to maintain Lorentz invariance of the canonical commutation relations, we use the *Gupta-Bleuler* method of correcting for the aforementioned limitation by adding a term $-\tfrac{1}{2}\lambda \|\delta A\|^2$ to the electromagnetic field Lagrangian ($\lambda$ is essentially a Lagrange multiplier for the constraint on $A$ defined by the Lorentz gauge $\delta A = 0$). The new conjugate momentum becomes:

(2.13) $$\pi = i_t F - \lambda \delta A \theta, \qquad\qquad (\theta = i_t g).$$

In the process of adding the Gupta-Bleuler term, we have introduced a number of unphysical features, which we eliminate by choosing the Feynman "gauge" $\lambda = 1$; in fact, the use of term "gauge" is not precise in this context. Hence:

(2.14) $$\pi = E - \delta A \theta,$$

which differs from $\dot{A} \coloneqq L_t A = i_t dA + d i_t A = E - d\phi$ by a closed 1-form when $\lambda = 1$. Hence, we use $\dot{A}$ as our conjugate variable instead.

One then replaces the canonical variables $A$ and $\dot{A}$ with Hermitian operator-valued distributions on the space of 1-forms and constrains the canonical field operators by the *equal-time commutation relations*:

(2.15) $$\begin{cases} [\dot{A}_\mu(t, \mathbf{x}), A_\nu(t, \mathbf{y})] = i\eta_{\mu\nu} \delta^3(\mathbf{x} - \mathbf{y}) \\ [A_\mu(t, \mathbf{x}), A_\nu(t, \mathbf{y})] = [\dot{A}_\mu(t, \mathbf{x}), \dot{A}_\nu(t, \mathbf{y})] = 0. \end{cases}$$

The space on which the field operators act is a *bosonic Fock space* that consists of one-photon, two-photon, etc., states of the field and is obtained from the symmetric tensor algebra over the one-photon Hilbert space $*_0$ ($\oplus$ = direct sum):

(2.16) $$*_{bos.} = S^*(*_0) = \mathbb{C} \oplus *_0 \oplus S^2(*_0) \oplus S^3(*_0) \oplus \ldots$$

Commutation relations for field operators serve a two-fold purpose:

*i*) In the eyes of theory they define the Lie algebra of the group of canonical transformations that act on the space of field states to facilitate the time evolution of those states.

*ii*) As far as computation of scattering amplitudes is concerned, it helps to know how to replace products in one order with an expression involving the product in the opposite order.

One then expresses the quantized electromagnetic field $A$ as the sum of the individual quantum oscillators – in the form of plane waves – of various amplitudes and wave numbers, i.e., a Fourier transform with operator coefficients:

(2.17) $$A(x) = \int \frac{d^3k}{2k_0(2\pi)^3} \sum_{\lambda=0}^{3} a_\lambda(k)e^{-i\mathbf{k}\cdot\mathbf{x}}\varepsilon^\lambda(k) + a_\lambda^\dagger(k)e^{i\mathbf{k}\cdot\mathbf{x}}\varepsilon^{*\lambda}(k)$$

in which $k_0 = |\mathbf{k}|$, and $\{\varepsilon^\lambda(k), \lambda = 0, 1, 2, 3\}$ is an orthonormal 4-coframe field on the positive light cone ($k^2 = 0$) in momentum space that one refers to as the *polarization frame* for the wave.

One also restricts the operators $a_\lambda(k)$, which annihilate a plane wave of positive energy and wave number $k$, (or create a plane wave of negative energy) and the operators $a_\lambda^\dagger(k)$, which create them, by the *canonical commutation relations:*

(2.18) $$\begin{cases} [a^{(\lambda)}(k), a^{(\lambda')\dagger}(k')] = -g^{\lambda\lambda'} 2k^0(2\pi)^3 \delta^3(\mathbf{k}-\mathbf{k}') \\ [a^{(\lambda)}(k), a^{(\lambda')}(k')] = [a^{(\lambda)\dagger}(k), a^{(\lambda')\dagger}(k')] = 0. \end{cases}$$

One sees that the aforementioned fact that quantum harmonic oscillators have a non-zero ground state means that each term in this infinitude of terms will contribute a non-zero contribution to the total energy of the system. This infinite *zero-point energy* is reminiscent of the infinite self-energy of the electromagnetic field of a point charge, but it can be eliminated from the quantum expression by shifting the definition of zero potential accordingly. We shall return to this aspect of the quantum vacuum later in the context of the Casimir effect, but first we need to define the term "quantum vacuum" more precisely.

One defines the *bosonic vacuum state* to be an element $|0> \in *_0$ that is characterized by the fact that it contains no photons, in the sense that any of the photon annihilation operators will take it to 0:

(2.19)  $\qquad\qquad a_\lambda(k)|0> = 0, \qquad$ for all $k$.

For any second-quantized – i.e., operator-valued – field, such as $A$, the passage to the classical limit is obtained by taking the *vacuum expectation value* $<0|A|0>$ of the field. One then regards the more general quantized field as representing this classical limit along with quantum fluctuations about the classical vacuum state.

Some of the key features of the bosonic vacuum state are:
i)    It is not merely the 0 state.
ii)   For the electromagnetic interaction it is a unique element of $*_0$.
iii)  $|0>$ generates $*_0$, in that every one-photon state is of the form:
$$a_\mu^\dagger(k)|0>$$
for some suitable $k$.

*c. Quantization of the field of a charged fermion.* One of the subtle ambiguities in the way that quantum electrodynamics treats electromagnetic fields is in the way that the field of a charged spinning particle is classically represented by a Dirac 4-spinor:

(2.20) $\qquad\qquad \psi: Spin(M) \to \mathbb{C}^4, \qquad x \infty\ \psi^\mu(x)$

when the field of an electromagnetic wave is represented by a 1-form, viz., $A$. Here the notation *Spin(M)* refers to the spin bundle that covers the bundle $SO_0(3,1)(M)$ of oriented time-oriented Lorentz frames on $M$ ([5]). What makes this potentially somewhat confusing is that the electromagnetic wave was presumably the result of a disturbance in the field of a charge when its acceleration became non-vanishing. Hence, whereas the Maxwellian picture would involve only a change in character of the field of the source charge from an electrostatic field to an electromagnetic one, the quantum picture also seems to change a 4-spinor into a 1-form in the process. What is even more confusing is that the Dirac equation for a mass $m$:

(2.21) $\qquad (i\partial\!\!\!/ - m)\psi = 0, \qquad \partial\!\!\!/ \equiv \gamma^\mu \partial_\mu, \mu = 0, 1, 2, 3,$

can be regarded not only as a Lorentz invariant form of the Schrödinger equation, whose solutions are complex amplitudes whose moduli-squared represent the spatial probability density function for the particle, but one can also relate the Dirac equation to the Maxwell equations for $A$ when $m = 0$.

Since $\psi$ is complex-valued one must consider its conjugate transpose $\psi^\dagger$ to represent a distinct field. The corresponding form of the Dirac equation that it satisfies is:

(2.22) $\qquad \overline{\psi}(i\partial\!\!\!/ - m) = 0, \qquad \overline{\psi} \equiv \psi^\dagger \gamma^0.$

In this expression, the left action of the operator $i\partial\!\!\!/$ means that one takes the product of the row vector and the Dirac matrix before differentiating.

The fields $\psi$ and $\overline{\psi}$ are next assumed to be distributions on 4-spinor fields that take their values in an algebra of Hermitian operators that act on a Hilbert space, which now amounts to a *fermionic Fock space* of one-electron, two-electron, etc., scattering states, and by the Pauli exclusion principle amounts to the exterior algebra over the one-particle Hilbert space $*_f$ ($\oplus$ = direct sum):

(2.23) $\qquad *_{ferm.} = \Lambda^*(*_f) = \mathbf{C} \oplus *_f \oplus \Lambda^2(*_f) \oplus \Lambda^3(*_f) \oplus \ldots$

In order to complete the canonical variables, we need to define the momenta that are conjugate to the fields $\psi$ and $\overline{\psi}$. However, when one defines the Dirac Lagrangian:

(2.24) $\qquad \mathcal{L} = \frac{i}{2}[\overline{\psi}\overset{\mathbf{r}}{\partial\!\!\!/}\psi - \overline{\psi}\overset{\mathbf{s}}{\partial\!\!\!/}\psi] - m\psi\overline{\psi}$

and computes the obvious conjugate momenta $\partial\mathcal{L}/\partial\dot{\psi}$ and $\partial\mathcal{L}/\partial\dot{\overline{\psi}}$ one finds that when the fields $\psi$ and $\overline{\psi}$ satisfy the Dirac equations they also make $\mathcal{L} = 0$. Hence, canonical quantization does not work in configuration space.

One then passes to momentum space by way of a Fourier analysis of the field operators:

(2.25) $\qquad \begin{cases} \psi(x) = \int \dfrac{d^3k}{(2\pi)^3} \dfrac{m}{k_0} \displaystyle\sum_{\alpha=1,2} [b_\alpha(k)u^{(\alpha)}(k)e^{-ik\cdot x} + d_\alpha^\dagger(k)v^{(\alpha)}(k)e^{ik\cdot x}] \\ \overline{\psi}(x) = \int \dfrac{d^3k}{(2\pi)^3} \dfrac{m}{k_0} \displaystyle\sum_{\alpha=1,2} [b_\alpha^\dagger(k)\overline{u}^{(\alpha)}(k)e^{ik\cdot x} + d_\alpha(k)\overline{v}^{(\alpha)}(k)e^{-ik\cdot x}]. \end{cases}$

---

[5] We naively pass over the topological details of this construction, such as the fact that if $M$ is a non-compact four-dimensional Lorentz manifold that admits such a spin structure then it must be parallelizable. [**20**].

The spinor fields $u^{(1)}(k)$, $u^{(2)}(k)$, $v^{(1)}(k)$, $v^{(2)}(k)$ are defined by starting with the canonical basis vectors for $C^4$ and applying a boost transformation with velocity $v = |\mathbf{k}|/k_0$, which one represents in $GL(4; C)$ through the action of the Dirac matrices; more specifically, $SO_0(3,1)$ is represented in the group of units of the Clifford algebra that is generated by the $\gamma_\mu$.

Rather than defining canonical commutation relations for the operators $b_\alpha(k)$ and $d_\alpha(k)$, which annihilate plane waves of positive energy and wave number $k$ (or create ones with negative energy), and the operators $b_\alpha^\dagger(k)$ and $d_\alpha^\dagger(k)$, which create them (or annihilate them, resp.), since the operators act on fermionic Fock space, one defines *canonical anti-commutation relations:*

$$(2.26) \quad \begin{cases} \{b_\alpha(k), b_\beta^\dagger(k')\} = \{d_\alpha(k), d_\beta^\dagger(k')\} = (2\pi)^3 \dfrac{k^0}{m} \delta^3(\mathbf{k}-\mathbf{k}')\delta_{\alpha\beta} \\ \{b_\alpha(k), b_\beta(k')\} = \{b_\alpha^\dagger(k), b_\beta^\dagger(k')\} = \{d_\alpha(k), d_\beta(k')\} = \{d_\alpha^\dagger(k), d_\beta^\dagger(k')\} = 0. \end{cases}$$

As in the case of the bosonic Fock space, this fermionic Fock space is also presumed to have a unique state $|0> \neq 0$ that represents the state of zero fermion number, i.e., the *fermionic vacuum state,* and is therefore annihilated by all of the fermion annihilation and anti-fermion creation operators:

$$(2.27) \quad b_\alpha(k)|0> = d_\alpha(k)|0> = 0 \quad \text{for all } \alpha, k.$$

In a manner that is analogous to the bosonic case, the one-electron states can all be obtained from the action of the creation operators on the fermionic vacuum state. Similarly, the vacuum expectation value of a quantized field, such as the *current* 1-form that one associates with a fermionic state $|\psi>$:

$$(2.28) \quad J_\mu = -\tfrac{1}{2} ie[\overline{\psi}\gamma_\mu \psi - c.c.]$$

represents the classical limit, and the actual field $J_\mu$ includes the usual quantum fluctuations about the vacuum state.

*d. Electromagnetic scattering amplitudes.* So far, we have discussed only *free* quantum electromagnetic fields. For the better part of the Twentieth Century, the actual *interacting* field model was almost always approached by means of a perturbing operator-valued interaction Lagrangian $L_I$ or Hamiltonian $H_I$ that one added to the sum of the free particle Hamiltonians $H_0$:

$$(2.29) \quad H = H_0 + H_I.$$

Rather than examine the resulting system of – generally nonlinear – field equations directly, one then went directly to the construction of the scattering amplitude for a given scattering event. There are two popular ways of obtaining such amplitudes:

The first method is to express the amplitude as a power series in the coupling constant that is obtained from $H_I$ by using the operator-valued fields and evaluating:

$$(2.30) \quad S_{if} = \lim_{t \to +\infty} U(t) = T \exp\left[-i\int_{-\infty}^{+\infty} dt'\, H_I(t'; i, f)\right],$$

where:

(2.31) $$U(t) = \sum_{n=0}^{\infty} \frac{(-i)^n}{n!} \int_{-\infty}^{t} dt_1 \int_{-\infty}^{t} dt_2 \mathbf{L} \int_{-\infty}^{t} dt_n T[H_I(t_1)\mathbf{L} H_I(t_n)].$$

In these expressions, the operator $T$ means that the operators that appear in the expansion are ordered by increasing time from right to left.

In the case of electromagnetic interactions the interaction Hamiltonian is:

(2.32) $$H_I(t; i, f) = \int i_\mathbf{t}(J \wedge *A_i) = \int d^3x\, J_\mu A_i^\mu,$$

where the integration is over a proper time simultaneity hypersurface that the time orientation $\mathbf{t}$ is normal to. Moreover, the current 1-form $J_\mu$ defined above (2.25) is the one that is associated with the 4-spinor field that represents the charged particle that gets coupled to $A$.

The other common approach to computing scattering amplitudes is to performing an infinite-dimensional integration – whose mathematical rigor is, however, dubious – of the field action over the gauge-equivalence classes of the fields:

(2.33) $$S_{if} = \int \$A \exp\{-iS[A,\psi,J]/\hbar\}.$$

The notation $\$A$ refers to the somewhat ill-defined volume element on the infinite-dimensional manifold of gauge equivalence classes of fields $A$ and the notation $S[A, \psi, J]$ refers to the total action functional for the fields $A$, $\psi$, and $J$, which includes the free field terms in its Lagrangian density along with the interaction terms. In this formulation, the fields are not operator-valued. The role of vacuum state is played by the classical extremal solution and the quantum fluctuations about the vacuum state are represented by the non-extremal field configurations in a neighborhood of the classical extremal. This method of approaching quantum field theory is particularly suited to the methods of gauge field theories, since it is more sensitive to the topology and geometry of spacetime and the field space, especially the methods of Morse theory, than the operator-valued field approach.

Although the method of functional − or Feynman – integrals sounds more straightforward than the perturbation expansions of canonical field theory, nevertheless, one finds that, in practice, one must often resort to perturbation expansions of the scattering amplitude that one obtains from a functional integration just the same. (Indeed, experimental particle physicists are often of the opinion that it is the Feynman rules themselves that are what is important, not the choice of theory that produced them.)

In this expansion, the exponent on the coupling constant describes the number of vertices in the Feynman diagram. Once one has at least two vertices ([6]), it is possible for a "loop" to appear in the diagram, and, with it, an unwanted infinity in the scattering amplitude. No-loop diagrams are the "tree level" of the scattering process; this corresponds to the classical approximation. One can also reorganize the aforementioned power series into a series of terms that represent successively more loops in the diagrams:

---

[6] In the case of the strong interaction, one can even find loops at the one-vertex level, in the form of "tadpoles."

(2.34) $$S_{if} = \Gamma_0(i,f) + \Gamma_1(i,f)\,h + \Gamma_2(i,f)\,h^2 + \ldots$$

which amount to "radiative" corrections to the tree-level, and also represent a power series in powers of that one commonly calls a *loop expansion*. In this way of representing the series, one sees how the tree level corresponds to the classical limit of $h \to 0$. In the functional integral formalism, it also shows how the quantum transition from a given initial state to a given final state involves not only the classical variational extremal path, but non-extremal "quantum fluctuations" about the classical extremal.

The use of the symbol $\Gamma$ above is suggestive of the fact that one can also regard the $S$ operator as an integral operator on configuration space with a kernel in the form of a Green function. The Fourier transform of this Green function is the momentum space *propagator*. Because this function will amount to an algebraic expression, it is generally more advantageous in applications to construct this propagator directly, although the Green function plays a more prominent role in the functional integral approach, which is more sensitive to some of the theoretical subtleties that are associated with spacetime geometry and topology.

*e. Anti-matter.* The first hint that quantum physics got that there was a hitherto unobserved symmetry in elementary matter, viz., matter/anti-matter duality, was in the attempt of Klein and Gordon to define the relativistic form of the Schrödinger equation by generalizing the non-relativistic expression for the kinetic energy of a massive particle of rest mass $m_0$ to its relativistic form:

(2.35) $$E = \frac{1}{2m_0}p^2 \;\to\; E^2 = p^2 c^2 + m_0^2 c^4$$

and applying the canonical quantization laws([7]): $E \propto i h \frac{\partial}{\partial t}$, $p_i \propto \frac{h}{i}\frac{\partial}{\partial x^i}$ to obtain the Klein-Gordon equation:

(2.36a) $$\mathbf{u}\psi + \left(\frac{mc}{h}\right)^2 \psi = 0,$$

which can also be written:

(2.36b) $$\mathbf{u}\psi + k_C^2 \psi = 0,$$

if one introduces the Compton wave number $k_C$ of the mass in question.

Since the transition from non-relativistic kinetic energy to relativistic energy also carries with it a transition from a differential operator that is of first order in time to one that is of second order in time, one finds that negative kinetic energy solutions must be just as possible as positive energy ones. At that point in history, quite some time before the positron was first discovered, this was regarded as a physical absurdity. Along with the fact that the conserved current that one associated with the *U*(1) gauge invariance of the field Lagrangian was not positive definite, a fact that clashed with the statistical interpretation of that current as a probability current, Dirac thought maybe the solution was to look for a wave equation that was first order in time, in the hopes that both problems would be eliminated.

---

[7] Note that one can also see this process as simply the association of the energy-momentum 1-form $Edt + p_i dx^i = h(k_C dt + k_i dx^i)$ with the frame field $\{h\frac{\partial}{\partial t}, -h\frac{\partial}{\partial x^i}, i=1,2,3\}$ – or rather, the association of $k = k_C dt + k_i dx^i$ with $\{\frac{\partial}{\partial t}, -\frac{\partial}{\partial x^i}, i=1,2,3\}$ – if one regards the factor of $i$ as related to the fact that one usually considers the Hermitian operators to be physically meaningful, not the anti-Hermitian ones, which define the Lie algebra of the unitary transformations, i.e., the dynamical group.

His approach was to look for a "square root" of the Klein-Gordon operator, i.e.:

$$(2.37) \qquad \left(a^\mu \frac{\partial}{\partial x^\mu}\right)\left(a^\nu \frac{\partial}{\partial x^\nu}\right) = \eta^{\mu\nu} \frac{\partial}{\partial x^\mu}\frac{\partial}{\partial x^\nu}.$$

In this expression, we have set $x^0 = ct$ and $\eta^{\mu\nu} = \text{diag}(+1, -1, -1, -1)$ is the Lorentz scalar product on Minkowski space. Upon expanding the left-hand side of (2.37), we see that the symbols $a^\mu$ that we introduce formally must have the property:

$$(2.38) \qquad a^\mu a^\nu + a^\nu a^\mu = 2\eta^{\mu\nu},$$

which also characterizes the Clifford algebra of Minkowski space if $a^\mu$ is an orthonormal basis. Customarily, one represents these four basis vectors by four 4×4 complex matrices $\gamma^\mu$, which are called the *Dirac matrices*, instead. Note that the four Dirac matrices are *generators* of the 16-dimensional Clifford algebra, not a basis for the underlying vector space.

Furthermore, the wavefunction $\psi$ must take its values in $\mathbb{C}^4$ in order that the Dirac matrices can act upon it naturally; such wavefunctions are called *Dirac 4-spinors*. The Lorentz group can be represented in the group of Clifford units (invertible elements of the Clifford algebra), which also acts linearly on $\mathbb{C}^4$ by way of the 4×4 complex matrices. Wavefunctions that take their values in the particular representation that we have chosen describe particles with a spin of 1/2.; since the action of the Lorentz group on C itself is only by way of the identity transformation, one says that the Klein-Gordon wavefunction describes particles of spin 0.

The resulting relativistic form of the Schrödinger equation for spin 1/2 particles is then the *Dirac equation:*

$$(2.39) \qquad \partial \psi + k_C \psi = 0,$$

where we have defined:

$$(2.40) \qquad \partial = \gamma^\mu \frac{\partial}{\partial x^\mu}.$$

An amusing aspect of Dirac's reformulation of the relativistic Schrödinger equation is that he actually failed to eliminate the negative energy solutions; they reappeared in another form by way of the charge conjugation operator. It was not until the discovery of the positron some years later that Dirac's theory was finally embraced as a definitive advance in the understanding of elementary matter. The gist of the charge conjugation symmetry in the solutions of the Dirac equation is that to every spin-1/2 particle there is an anti-particle with the same rest mass and spin, but the opposite charge. The Feynman interpretation of these negative energy anti-particles is that they represent positive energy particles propagating backwards in time.

A further consequence this duality was the possibility that since a particle-antiparticle pair had the same total charge and spin as a photon, viz., 0 and 1, resp., there was the possibility of a phase transition from the pair to a photon whose energy would be $2mc^2$, or the opposite transition. These two processes were called *pair annihilation* and *pair creation*, respectively. One can also visualize these processes as the association of a two-particle state in the fermionic Fock space with a one-particle state in bosonic Fock space.

*f. The hole theory of the vacuum state.* This scenario of pair creation and annihilation brought with it a radical paradigm shift in the way that theoretical physics regarded the vacuum state of spacetime. In particular, one had to account for the appearance of non-zero mass out of what used to be regarded as "empty space." Of course, this is not precisely correct since it was not truly empty as long as it contained a photon; i.e., electromagnetic waves represent energetically excited states of the spacetime vacuum that one obtains by applying an appropriate creation operator to the bosonic vacuum state.

The picture that Dirac suggested, which is often referred to as the "Dirac Sea," is that:

*i*) An antiparticle is really a negative energy "hole" that was left behind when a positive energy particle was excited out of the vacuum ground state.

*ii*) The vacuum ground state is actually an infinitude of negative energy holes that are filled with positive energy particles in a one-to-one correspondence.

*iii*) A hole is filled by pair annihilation; a hole is created by pair creation.

A drawback to this conception of the vacuum ground state is that the infinitude of filled holes in the vacuum ground state leads to an infinite energy sum that does not converge; this relates to the zero-point energy of the photon field, and also appears in the problem of the self-energy of the photon.

*g. Vacuum fluctuations.* One can envision a state of the spacetime vacuum that is between the ground state and the state of actual pair creation, namely, a state of *vacuum polarization*. Like the polarization of charges that one finds in a dielectric when it is subjected to an electric field, presumably a strong enough electric field should bring about a polarization of matter/anti-matter pairs in the Dirac Sea into electric dipoles short of the point at which they actually "ionize" into distinct particles. This, in turn, could affect the strength of the electric field, which would represent a departure from the linearity of the Maxwell equations. Of course, the electric field strengths that are actually necessary to bring about such phenomena in a noticeable way are so enormous that they could only exist close to elementary charges.

Since quantum field theory, like quantum physics in general, takes a statistical approach to everything, one must distinguish between "real" and "virtual" processes. The basis for the distinction is whether they are experimentally observable or only something that goes on inside of the black box. An example of such a virtual process, which is still included in the electromagnetic vacuum state, is the creation and subsequent annihilation of a pair.

In terms of Feynman diagrams, this is represented as:

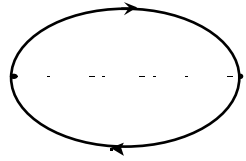

Fig. 2. Virtual pair creation and annihilation in the vacuum state.

There are infinitely many other such vacuum diagrams, with increasing numbers of vertices and loops. Their effect on the scattering amplitude for a given scattering process that one is examining is an overall factor of $e^{i\alpha}$, in which the phase factor $\alpha$ is, however, infinite. Hence, one customarily simply disregards all of these diagrams.

Although much of the foregoing discussion of the vacuum state tends to give the impression that the only differences between macroscopic electrodynamics and quantum electrodynamics take the form of submicroscopic fluctuations about the classical electromagnetic fields, there is, in fact, at least one consequence of the quantum picture that has tangible macroscopic ramifications. This takes the form of the *Casimir effect* [**20**], which can be physically defined by the results of the experiment that confirmed its validity. Basically, when two perfectly conducting planes of unit area are placed parallel to each other in a perfect vacuum at a separation distance of *a,* they attract each other with a force of magnitude:

$$(2.41) \qquad F = -\frac{\pi^2}{240 a^4}.$$

In an actual experiment where the area was 1 cm$^2$ and $a = 0.5$ μ$m$, the measured value was $F = 0.2 \times 10^{-5}$ N. Under the assumption that the force is conservative, i.e., $F = -dU/da$ for some $U = U(a)$, this corresponds to a potential energy between the plates that varies with *a* as:

$$(2.42) \qquad U(a) = -\frac{1}{3}\frac{\pi^2}{240 a^3}.$$

This phenomenon was attributed to the notion that when one restricts the system of quantum oscillators that a quantized electromagnetic field represents to a compact three-dimensional submanifold of space with boundary, then, just as the ground state solution to the one-dimensional harmonic oscillator boundary value problem of quantum mechanics must be a non-zero function (more specifically, a Gaussian function) even though it vanishes on the boundary of the region, in order to be consistent with the fact that the ground state energy is greater than zero, similarly, the vacuum expectation value of any electromagnetic field in such a submanifold must be greater than zero. This also suggests some sort of symmetry-breaking process going on, which leads us to suspect that the phase transition of vacuum polarization has much in common with the phase transition to a magnetized state in a ferromagnet, at least at a qualitative level.

Actually, in order to relate the Casimir effect to vacuum polarization, one must consider the vacuum expectation values of the energy-momentum tensor of the electromagnetic field, i.e.:

(2.42)                              $< 0 | T_{\mu\nu} | 0 >.$

The Hamiltonian density that is associated with this tensor is then:
(2.43)                              $\ast = T_{00},$

and if the plates are parallel to the *yz*-plane then the Casimir force (per unit area) on that plane is:
(2.45)                              $F = < 0 | T_{11} | 0 >.$

At this stage of explanation, the Casimir effect essentially amounts to the difference between taking vacuum expectation values over all of space (which is presumably non-compact and without boundary) and taking vacuum expectation values over a compact region with boundary. This is reminiscent of the fact that there will be difference between the free particle spectrum of a particle on a line, which will be a continuous spectrum, and the spectrum of a free particle on a closed bounded interval with periodic boundary conditions, which will be a discrete, but infinite, spectrum. This would suggest that the symmetry breaking is from R to Z, which produces a vacuum manifold of R/Z = $S^1$. Later we shall discuss the role of the topology of spacetime in the Casimir effect in more detail.

Another aspect of the Casimir effect that is worthy of mention is the fact that it seems to closely parallel the circumstances of the early work on blackbody radiation that led Planck to postulate that the energy spectrum of an electromagnetic field in a compact region of space with boundary must be discrete, not continuous, in order to eliminate the unphysical infinity in the energy flux at absolute zero temperature. As we shall observe, quantum electrodynamics has its share of unphysical infinities, as well.

*h. Regularization and renormalization.* As we just observed, the first one-loop correction to the tree-level scattering amplitudes, namely vacuum polarization, is associated with an unphysical infinity. Since it took the form of a phase, one simply ignored its multiplicative contribution to the final expression. The next three radiative corrections that we shall discuss, viz., self-energy of the photon and electron, and vertex parts, are also associated with infinite values of things, such as masses and charges, that should actually be finite in real physical terms. The manner by which QED attempts to eliminate these contradictions is to first redefine the integral for the scattering amplitude in question to something that converges, which is the process of *regularization*, and then redefine the quantities that are incorrectly infinite in such a manner that they become finite, which is the process of *renormalization.* Furthermore, the way that one renormalizes a spurious infinity depends on the way that one regularized the integral.

For instance, in the method of regularization by means of a cutoff, one first restricts the domain in momentum space over which the integral is performed to a bounded one by introducing a momentum cutoff. One then adjusts the quantity that needs fixing, such as mass, charge, the field itself, or even the coupling constant, by subtracting off the appropriate expression, which is finite when the cutoff is included, and then hopefully one will arrive at an expression that is independent of the cutoff that was introduced. Presumably, that will allow the expression to remain valid when one looks at the limit as the cutoff goes to infinity.

Another method of regularization that one might encounter are *dimensional* regularization, which involves changing the dimension of the space to one in which the integral converges and then eventually looking at the limit as the dimension goes back to four. This method is more popular when considering quantum electrodynamics as a particular type of gauge field theory, since it leads to a gauge-invariant renormalization process.

Despite the "klugey" character of the process of regularization and renormalization, the traditional justification is that although these methods may have dubious mathematical rigor they still ultimately produce expressions for physical quantities that are in unarguable agreement with the experiments. Hence, until the next great revolution in physics reveals the deeper first principle that obviates the kluge, one must humble oneself to more practical realities. Nevertheless, it does seem as if the long-term result of the phenomenological approach to fundamental physical processes is that most of physics has long since abandoned the search for such a deeper first principle out of pure agnosticism.

*i. Self-energy of the electron.* As mentioned above in the Abraham-Lorentz model of the electron, the attempt to attribute the mass of the electron to the total energy contained in its electrical field failed because of the infinity in this energy, which was traceable to the point-like nature of the electron model and the pole in the Coulomb potential. This problem of an unwanted infinite self-energy for the electron persists even in quantum electrodynamics ([8]).

The simplest way of representing the situation is by a second-order (one loop) Feynman diagram:

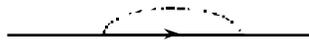

Fig. 3. Electron self-energy.

The divergent additive contribution that such processes make to a given scattering amplitude is corrected by *mass renormalization*, which amounts to subtracting off the infinite self-energy of the field from the infinite mechanical mass of the electron to obtain the finite rest mass $m_0$ of a "bare" electron ([9]).

*j. Self-energy of the photon.* The concept of the photon having a self-energy has no classical counterpart unless one considers the attempts to give the photon a nonzero mass.

---

[8] The moral of this situation is clear: you cannot always solve your problems by quantizing them.
[9] This follows from the well-known result of modern analysis that $\infty - \infty = m_0$ (!!!) However, as we shall see, in conjunction with regularization, it is more like $\lim_{x \to \infty} x - x + m_0 = m_0$.

The quantum origin of such a self-energy is in the aforementioned creation and annihilation of a virtual pair by a photon, which can be represented by the diagram:

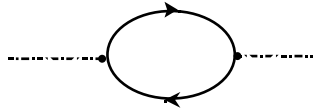

Fig. 4. Photon self-energy.

The presence of the loop implies a divergent contribution to a scattering amplitude, which, like the electron self-energy, is also additive. The manner by which this divergence is corrected is by replacing the massless photon with a massive vector boson, renormalizing the mass, and demanding that the renormalized mass be zero, in order to assure the gauge invariance of the resulting amplitude.

*k. Vertex part corrections.* A vertex part is a third order process that defines a loop. In a sense, it is also a sort of self-interaction in which the incoming electron interacts wit the outgoing one after colliding with a photon. The diagram is shown in Fig. 5.

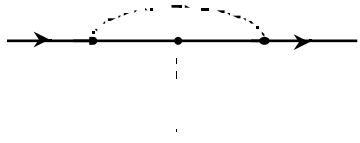

Fig. 5. Vertex part.

The additive divergence that such a process contributes is removed by *charge renormalization.*

*l. Photon-electron scattering.* In classical electrodynamics, the scattering of a photon by collision with a charged particle would simply not be possible, since photons have no charge to couple to the field of the charged particle. However, one of the early contributions to the foundations of quantum mechanics was the experimental observation that a photon that scattered off of a bound atomic electron could lose momentum to the electron, with a resulting shift in wavelength:

(2.46) $$\Delta\lambda = \lambda_C(1 - \cos\theta), \quad \lambda_C = \frac{2\pi}{k_C} = \text{Compton wavelength of } m_e,$$

which is called the *Compton shift*, and is consistent with the de Broglie hypothesis that sometimes gets referred to as "wave/particle duality." It basically says that massive matter and massless matter are both forms of the same sort of wavelike object, and can therefore collide like particles while showing the interference and diffraction effects that would be consistent with wave motion.

In QED, the generalized form of this process is called *Compton scattering*, which also includes the scattering of photons from free charges, as well as bound ones. Its tree level diagram is, as one would expect:

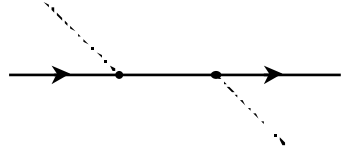

Fig. 6. Compton scattering.

Beyond the very existence of such an interaction, the other non-Maxwellian aspects of Compton scattering show up at the higher numbers of vertices and loops in the Feynman diagrams. In particular, a photon can give rise to an electron-positron pair in the presence of a strong enough electric field, so there are "scattering channels" to other outgoing states than the elementary one, such as more outgoing electrons or a pair. An interesting aspect of the radiative corrections to Compton scattering (cf. Jauch and Rohrlich [**17**]) is that Lorentz and gauge invariance imply that the net effect of all such corrections to the scattering amplitude reduces to a multiplicative factor in the non-relativistic limit (*Thomson scattering*), and the factor is precisely the renormalization factor that is obtained from mass and charge renormalization. Moreover, the observable features of this factor vanish in the non-relativistic limit.

*m. Electron-electron scattering.* Again, we are using the term "electron" loosely to mean "elementary charges." In particular, when the incoming charges are both electrons one refers to *Møller scattering*, and when they consist of an electron and a positron, one calls that *Bhaba scattering*. In simplest form, the diagrams – up to the directions of the external lines – are:

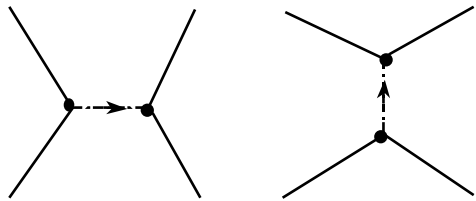

Fig. 7. Electron-electron scattering.

Bhaba scattering can involve outcomes that are distinct from those of Møller scattering, such as the low energy production of electron-positron bound states, or *positronium*, and the internal annihilation and re-creation of the incoming pair. Similarly, the fact that Møller scattering involves indistinguishable incoming particles will make it distinct from Bhaba scattering. Nevertheless the two processes are closely kinematically equivalent, and the scattering cross-section for the latter process can be obtained from that of the former process by changing the appropriate signs of the momenta that are associated with the external lines that described anti-particles, which presumably propagate backward in time. Ultimately, the non-relativistic limit of that cross-section is just the one that Rutherford had computed for classical Coulomb scattering.

A classical phenomenon that one would expect in the collision of two charges is *bremsstrahlung*, or "braking radiation," since one would expect that the charges involved would undergo decelerations and accelerations in the process of colliding. At the tree level, one simply replaces an external line for a particle with a line that had a photon line leaving it:

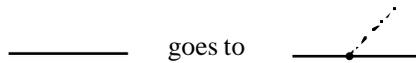

Fig. 9. Brehmsstrahlung.

to obtain a third-order correction to the tree-level scattering amplitude. We shall return to the subject of brehmsstrahlung in the context where one of the incoming charges is regarded as so relatively massive that it serves only to define an external field.

*n. Photon-photon scattering.* The very fact that two photons can collide as if they were massive particles is one of the most compelling arguments for nonlinear electrodynamics at the elementary level. This is because in the linear theory of Maxwell the only effects that one should expect from superposing two electromagnetic waves should be the interference effects of conventional optics. Not surprisingly, the realm in which one expects to find experimental evidence of photon-photon scattering is the realm of large |**E**|. Presumably what one should look for, at that rate, is a distortion of the interference pattern that would be consistent with linear superposition.

In the eyes of quantum electrodynamics, the root of the nonlinear interaction of two photons is the vacuum polarization that might occur if the **E** fields in question had magnitudes in excess of that which would bring about the splitting of a photon into an electron-positron pair (real or virtual). The various one-loop diagrams for this process are:

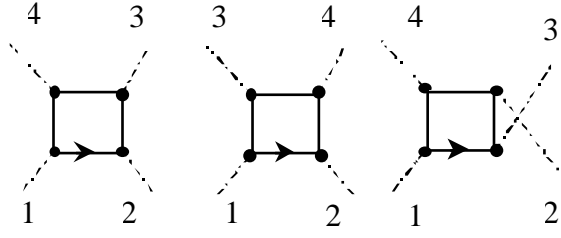

Fig. 10. Photon-photon scattering.

along with the diagrams that one obtains by giving the central loop the opposite orientation, i.e., replacing the particles with their anti-particles.

A direct application of the Feynman rules for the electromagnetic interaction to these diagrams gives a fourth rank tensor $\Pi_{\mu\nu\lambda\sigma}(k_1, k_2, k_3, k_4)$ that one calls the *vacuum polarization tensor,* although it is should not be confused with the fourth rank tensor that one obtains from a linear constitutive law for the Minkowski field strength 2-form $F$ and the electromagnetic induction 2-form $H$, since vacuum polarization would suggest a nonlinear constitutive law. This tensor necessarily has a divergent contribution from the loop, which shows up as an additive contribution. As it turns out, the coefficient of this infinite contribution is zero, anyway. We shall have more to say about this tensor later when we discuss the work of Euler and Heisenberg.

One can define an effective interaction Lagrangian for this process:

(2.47) $$+_I = c_1 \|F\|^2 + c_2 \|F \wedge F\|^2 \mathcal{9}, \qquad \mathcal{9} = \sqrt{-g}\, dx^0 \wedge \ldots \wedge dx^3,$$

which contributes the terms:

(2.48) $$\frac{\delta +_I}{\delta A} = \frac{\partial +_I}{\partial A} - d\frac{\partial +_I}{\partial (dA)} = -d\frac{\partial +_I}{\partial F}$$
$$= -2d(c_1\|F\|{*}F + c_2\|F\wedge F\|F)$$
$$= -2(c_1 d\|F\| \wedge {*}F + c_1\|F\|{*}\delta F + c_2 d\|F\wedge F\|\wedge F)$$

to the Maxwell field equations, which become:

(2.49) $$\begin{cases} 0 = dF \\ 0 = (1 - 2c_1 \|F\|)\,{*}\delta F - 2c_1 d\|F\|{\wedge}{*}F - 2c_2 d\|F\wedge F\|\wedge F. \end{cases}$$

One can show that in the low-energy limit before the polarization gives way to pair creation the empirical constants $c_1$ and $c_2$ must be:

(2.50) $$c_1 = \frac{5}{180}\frac{\alpha^2}{m^2}, \qquad c_2 = -\frac{7}{90}\frac{\alpha^2}{m^4}.$$

Another process that can result from the collision of two photons is actual pair production, which would presumably be the extreme form of simple vacuum polarization. The Feynman diagram for this process is:

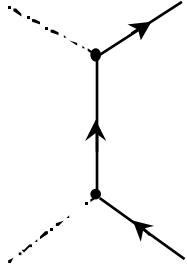

Fig. 11. Pair production during photon-photon scattering.

This diagram is actually kinematically equivalent to a rotation of the diagram for Compton scattering, so one can obtain the scattering amplitude for it by an appropriate substitution of momenta in the other.

   *o. Scattering in an external field.* When there is little momentum transfer between interacting charges, such as in the interaction of photons or electrons with sufficiently large atomic nuclei, it is generally more convenient to regard one of incoming particles as fixed in space and contributing only an external field to the dynamics of the other one. This carries with it an approximation in the perturbative evaluation of scattering amplitudes, called the *external field approximation.* It amounts to simply retaining the classical limit of the external field, i.e., its real-valued potential 1-form $\phi$, in the interaction Hamiltonian, which means the total quantized potential 1-form $A = a + \phi$ for a system of interacting particles will consist of an operator-valued part $a$ that represents the radiation field of the system along with a real-valued part $\phi$ that represents the external field in which they interact. Since the effect of such an interaction would usually produce excitations of the classical field $\phi$, by regarding $\phi$ as real-valued, we are ignoring the change of state that the external field itself could undergo during interaction.

   In keeping with the overall focus of this study, we shall comment only upon those external field problems that offer some direct insight into the nonlinearity of the electromagnetic interaction at the atomic level.

   First, we need to point out that the presence of an external field can change the nature of the fermionic vacuum state, but not in an unambiguous way. The ambiguity arises from the gauge invariance of $\phi$, which makes the predictable shift in the energy spectrum of fermionic plane wave states arbitrary. Furthermore, the free-field fermionic plane wave spectrum usually involves a gap of width $2m$ between the positive and negative energy states. In the event that the external field produces bound states, such as an attractive Coulomb potential would, the bound state energies might penetrate the gap to a significant degree. This situation is related to the so-called *Klein Paradox*, which suggested that in a strong enough field the formation of pairs would tend to neutralize the field that produced them, and potential barriers of height $2m$ might become easily penetrable by electrons. The solution to the paradox is simply to realize that it only defines the limits of validity for the external field approximation. Furthermore, the gauge ambiguity is resolved by choosing the gauge that the gap is still centered at 0, but this time we are requiring the vacuum state to be annihilated by the electron annihilation and positron creation operators that pertain to the *bound interaction picture.* This approach to the interaction Hamiltonian allows the state | ψ > of the interacting system to evolve in time only due to the interaction of the matter field and the radiation field. This subjects the matter field operator ψ and the current operator *J* to a canonical transformation that effectively absorbs the external field into the definition of ψ, and therefore the electron annihilation and positron creation operators.

Next, since it is the validity of Coulomb's law that is being called into question, one must naturally consider the point of departure for Coulomb scattering that leads into the quantum form of that process. In fact, one finds that one is looking at the same considerations that we have been discussing all along: higher-loop radiative corrections such as the self-energy of the charge, creation and annihilation of virtual pairs, vertex part contributions, etc. The fact that these quantum corrections to the Rutherford scattering formula are borne out by experiment can be construed as convincing proof of the existence of limits to the validity of Coulomb's law at the level of |**E**| near the classical nuclear radius.

Note that if one sets out to create actual matter-antimatter pairs in experimental practice, one should know that pair production and annihilation requires the presence of a strong external field – such as the nuclear electrostatic field − not only to bring about the polarization of the photon to begin with, but to contribute the momentum that is necessary in order to give overall conservation of momentum in the interaction. The diagram for pair creation in an external field is actually kinematically equivalent to the one for brehmsstrahlung by an electron in the same field so the scattering amplitudes are related by a substitution of momenta.

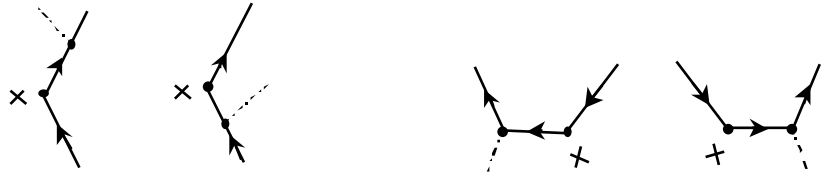

Fig. 12. Brehmsstrahlung vs. pair creation in an external field.

When a photon passes close enough to a spatially fixed charge to feel an electrical field strength that is sufficiently strong for vacuum polarization to take place, the resulting virtual charged particle pair can interact with the external field. This purely quantum effect can lead to *Delbrück scattering*, in which subsequent annihilation of the virtual pair gives the overall effect of the scattering of a photon by an external potential.

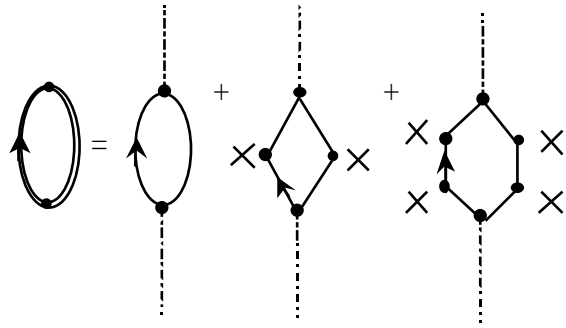

Fig. 13. The first few diagrams for Delbrück scattering.

One can also see that Delbrück scattering can be regarded as a radiative correction to Compton scattering of a photon by a nucleus, or the scattering of a photon from a set of virtual ones.

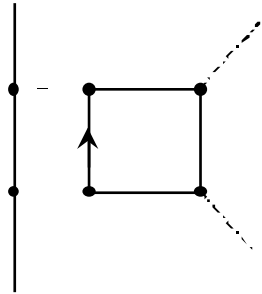

Fig. 14. One of the radiative corrections to Compton scattering from a nucleus.

Clearly, the very fact that Delbrück exists as an experimentally established fact of quantum electrodynamics is one more vote of confidence for concentrating on the large field strength nonlinearities in electrostatics as a definitive extension of the scope of Maxwellian electrodynamics.

*p. Bound states.* Although it seems as though the majority of the results that one obtains from the methods of quantum field theories pertain to high-energy scattering events, nevertheless, some of the important consequences of quantum electrodynamics are found in the low-energy realm of bound states ([10]).

Perhaps the simplest scenario in which one can show the departure of quantum electrodynamics from classical dynamics is in the simple case where one starts with an electron and a proton at rest and close enough together that when they are released they would not gain very much total kinetic energy as a result of their mutual attraction. In Maxwellian – i.e., Coulomb – electrostatics the final state of this system would be the coupling of the electron with the proton, which is essentially a potential energy of negative infinity, unless you give at least one of the two charges a non-zero radius. However, one of the simplest quantum aspects of the hydrogen atom is that the lowest energy level to the electron-proton bound states that define that structure is actually –13.6 eV, not $-\infty$. This corresponds to a radial separation of $5.3 \times 10^{-9}$ cm, which is called the *Bohr radius*; note that it is equal to $18,900 R_e$.

---

[10] By contrast, the phenomenon of *asymptotic freedom* in the coupling constant of the strong interaction is what makes the low-energy realm the most problematic; the perturbation expansions for the scattering amplitudes converge only for a small enough coupling constant, which one finds only at *high* c.o.m. collision energies for the strong interaction.

Hence, one should suspect that the limits of Coulomb's law are not necessarily all found within the $R_e$ limit that was previously mentioned, even though the application of Coulomb's law to the Schrödinger or Dirac equation produces very precise results as far as experimental agreement is concerned. One should remember that to some extent the success of the Schrödinger equation was originally more theoretically fortuitous than rigorous.

As we mentioned previously, a bound state of an electron and a positron is called *positronium*. Because the two particles have the same mass, the spectrum of bound states differs somewhat from the hydrogen atom, in which the proton can be treated as an infinite mass, in the one-body approximation, and the electron mass replaced with the reduced mass of the system. Furthermore, the possibility of pair annihilation makes positronium an unstable bound state. Note that charge conjugation symmetry says that a bound state of a positron and an anti-proton, which one might call *anti-hydrogen*, is just as stable as its hydrogenic analog. Indeed, the spectrum of radiation that is emitted and absorbed by an anti-hydrogen atom is indistinguishable from a sufficient distance, which begs the question of how one would distinguish a star from an "anti-star" on the basis of astronomical measurements alone, since both would emit the same sort of photons with the same spectral type.

Atomic electron energy levels above the ground state can have degeneracies that arise from the symmetry of the nuclear electrostatic potential field, which results in non-zero values of orbital angular momentum, which one described by the eigenvalues $l$ and $m$, for the electron shells, along with the intrinsic angular momentum of $\pm h/2$. These angular momenta result in associated magnetic moments for the shell. When coupled to an external magnetic field, the resulting torque breaks the symmetry and resolves the degeneracy in the eigenvalue $m$, which refers to the $z$-component of the orbital angular momentum; this effect is called the *Zeeman effect*. There are also contributions to the energy levels in the absence of external fields in the form of shifts that arise from the coupling of the electron spin to its orbital angular momentum (the *spin-orbit coupling*) and the coupling of the electron spin to the nuclear spin (the *hyperfine structure.*)

One of the most significant of the early experimental successes of quantum electrodynamics concerned a further shift in the bound state spectrum of atomic electrons due to the *anomalous* magnetic moment of an atomic electron, which one calls the *Lamb shift*. The origin of this correction was in the radiative corrections to the Coulomb scattering (although one only concerns oneself with bound states) of the atomic electron. In particular, one first considered the effects of vacuum polarization and fluctuations due to the creation and annihilation of virtual pairs. Compared to the classical Bohr magneton of $\mu_0 = e/2m_e$, by accounting for the first few radiative corrections, one can obtain a value of:

(2.51) $$\mu = \left(1 + \frac{\alpha}{2\pi} - 0.328 \frac{\alpha^2}{\pi^2}\right)\mu_0 = 1.0011596\mu_0,$$

which compares favorably to the experimental value of $(1.0011612 \pm 0.0000024)\mu_0$.

The way that Lamb and Retherford verified this effect was to compare the measured frequencies for the $2\ ^2s_{1/2}$ to $2\ ^2p_{1/2}$ transition for hydrogen and deuterium atoms. In the eyes of the Dirac equation – i.e., to tree level – they should be equal, but, in fact, the measured values were $1057.77 \pm 0.01$ MHz for hydrogen and $1059.00 \pm 0.01$ MHz for deuterium, with a shift of $1.23 \pm 0.15$ MHz. The best theoretical values (as of Jauch and Rohrlich) are $1057.99 \pm 0.2$ MHz for hydrogen and $1059.33 \pm 0.2$ MHz for deuterium, with a shift of $1.35 \pm 0.04$ MHz.

*q. Spectral bounds.* One of the drawbacks to the phenomenological approach of quantum field theory is that often it obscures genuine physical issues behind a smoke screen of "virtual reality." More precisely, genuine physical issues, such as the existence of bounds to the spectrum of physically possible photons are treated as mere computational conveniences, such as ways of treating infrared divergences in scattering amplitudes and the imposition of ultraviolet cutoffs in the process of regularization. The usual excuse for not taking such things seriously in physically fundamental terms is that one will allow the limits imposed to go to zero or infinity in the final expression, and thus cease to matter.

Actually, one could say that the ultimate bounds on the spectrum of electromagnetic radiation are cosmological in origin. For instance, the low frequency end of the spectrum is limited by the longest wavelength that will fit into a spacelike simultaneity hypersurface $\Sigma$, which relates to the issue of whether the geometry of such a spacelike submanifold is bounded, i.e., is there a maximum value $\lambda_{max}$ attained by the distance $d(p_1, p_2)$ when $p_1, p_2 \in \Sigma$? Assuming the simplest dispersion relation $\lambda v = c$, this places a lower bound of $v_{min} = c/\lambda_{max}$ on the spectrum. However, this dispersion relation is based on the assumption that the electric susceptibility of the vacuum is truly constant, which neglects the effects of vacuum polarization.

The upper bound on the spectrum is defined either by the energy available in $\Sigma$ for creating photons, since the energy in a photon is $hv$, or by the smallest possible wavelength, if there is one.

As far as the first option is concerned, at the very greatest, $hv$ cannot exceed the total energy in $\Sigma$. However, this is probably not a "sharp" upper bound since photons are created only by physical processes; such processes do not, presumably, include the conversion of all energy in the universe into one photon. Of the known processes, the one that is most efficient in converting mass into photons is pair annihilation, which produces photons with a frequency of $v = 2mc^2/h$. For instance, if the particles both have a mass of 1 GeV/$c^2$ then the resulting photon has a frequency of $5 \times 10^{20}$ Hz, which is in the range of gamma rays. The expression for $v$ would seem to be limited only by the mass of the antiparticles that are annihilated, but this also means that one might consider not only very massive particles, such as a plutonium nucleus colliding with an anti-plutonium nucleus or even a black hole colliding with an anti-black hole, but also high kinetic energies, since *m* is the relative mass of each colliding particle.

The second possibility – a lower bound on $\lambda$ – is more geometrical in character. Indeed, it seems to be the primary issue in the difference between classical mechanics and quantum mechanics at its most fundamental level. In effect, the classical limit of quantum wave mechanics is the limit in which $\lambda$ is allowed to go to zero, at which point the wavelike character of the motion of a massive particle, as evidenced by interference and diffraction effects, vanishes, and one can treat the bicharacteristics of the wave motion as the classical trajectories. This is the also the limit in which wave optics gives way to geometrical optics: the limit in which light waves give way to light rays. The conclusion must then be that the classical case of $\lambda = 0$ for massive matter must only be an approximation for some non-zero bound that characterizes wave motion in the eyes of quantum physics. At the very least, one wonders if the lower bound on $\lambda$ imposed by the Planck length ($\sim 10^{-33}$ cm.) is sharp.

One might also consider that if the energy of a photon is still related to the usual expression in $|\mathbf{E}|^2$ and $|\mathbf{B}|^2$, which would imply that a given frequency can be associated with only one field strength, then the existence of an upper bound on field strength will imply an upper bound on frequency. As we shall see, the existence of such an upper bound on field strength is a common issue in nonlinear electrodynamics.

An unavoidable issue that must be addressed whenever one imposes bounds on the electromagnetic spectrum is that of Lorentz invariance, since this is presumably a property of the frequency-wave-number 1-form $k = \omega/c\theta^0 + k_i\theta^i$. This is because the requirement that $\omega_{min} \leq \omega \leq \omega_{max}$ is inconsistent with the fact that one can find a Lorentz frame $\theta^\mu$ in which $\omega$ is arbitrarily large. However, one should keep in mind that this only means that $\omega_{min}$ and $\omega_{max}$ are not frame-invariant scalars, but should probably be regarded as associated with each choice of Lorentz frame. Hence, one might postulate a weaker condition on the spectral bounds, such as the postulate that for every Lorentz frame $\theta^\mu$ such a pair of numbers, $\omega_{min}(\theta^\mu)$ and $\omega_{max}(\theta^\mu)$, exists, but that these numbers vary with the choice of frame in accordance with the postulate of Lorentz invariance.

In quantum electrodynamics, the so-called infrared divergences arise from either the vanishing photon mass in the integration of radiative corrections or from the fact that the probability that "soft" photons, i.e., low frequency ones, will be produced increases with the wavelength. Indeed, it turns out that the infrared divergence is a remnant of a corresponding divergence in the Rutherford scattering cross section for Coulomb scattering, which, in turn, relates to the infinite range of the Coulomb interaction, or, equivalently, the vanishing mass of the photon.

The ultraviolet divergences appear when one goes beyond the tree level in the perturbation expansion since a loop is associated with too high of a power of *k* in the denominator to be compensated for by external lines in the limit as *k* goes to zero. One addresses the appearance of such divergences by the introduction of an ultraviolet cutoff on *k* during the regularization process in such a way that after renormalization the resulting expression will be cutoff independent.

**3. Euler and Heisenberg's theory of vacuum polarization.** Since vacuum polarization is a consequence of the Dirac theory of charged particles, Euler and Heisenberg [**22**] took up the task of specifically computing the polarization tensor field that results from the quantized form of Dirac's theory, i.e., quantum electrodynamics. Actually, they stopped short of the specific computation of the polarization tensor field, but pointed out that it sufficed to compute the potential energy density *U* that is associated with *F*:

(3.1) $$U(F) = \frac{1}{4\pi}[*\mathbf{D} \wedge E - +] = \frac{1}{4\pi}[\frac{\partial +}{\partial E} \wedge E - +],$$

where $+ = +(F)$ is the electromagnetic field Lagrangian.

Furthermore, they considered only the case of very slowly varying *F*. Like Mie and Born, whose theories we shall also discuss, they also assumed that $U$ − hence $+$ − was Lorentz invariant, and chose the same invariants as Born, viz., $\|F\|^2$ and $F \wedge F$.

*a. The density matrix of a Dirac wavefunction.* Their starting point was to look at the density matrix *R* that one associates with a Dirac wavefunction $\psi(t, x, k)$:

(3.2) $$< t,x,k \mid R \mid t',x',k' > = \sum_n \psi_n^*(t',x',k') \psi(t,x,k),$$

in which the sum is over all definite states, as well as the related matrix $R_S$:

(3.3) $$< t,x,k \mid R_S \mid t',x',k' > = \tfrac{1}{2}[< t,x,k \mid R \mid t',x',k' >_{\text{def.}} - < t,x,k \mid R \mid t',x',k' >_{\text{indef.}}],$$

in which the second expression is the sum over all indefinite states. One lets $S_0$ denote the value that $R_S$ takes when the wavefunctions propagate in field-free and matter-free space, *A* denotes the electromagnetic potential 1-form, $\alpha_0 = -\alpha^0 = 1$, $\alpha_i = \alpha^i$ are the Dirac matrices, and if *P* and *P'* denote the points $(t, x)$ and $(t', x')$ and $\tau(P, P')$ refers to the proper time interval that separates *P* from *P'*:

(3.4) $$\tau(P,P')^2 = \|d\|^2 = (ct - ct')^2 - (x^1 - x'^1)^2 - (x^2 - x'^2)^2 - (x^3 - x'^3)^2,$$

and one further defines:

(3.5) $$u[P, P'] = -\frac{i}{2\pi^2}\exp\left(i\frac{e}{\hbar c}\int_P^{P'} A\right),$$

in which the integral is taken along the straight line through *P* and *P'*. This puts the density matrix definition into the form:

(3.6)
$$< t,x,k \mid R_S \mid t',x',k' > = \frac{1}{\tau(P,P')^4} x_\rho \alpha^\rho u[P,P'] - \frac{v}{\tau(P,P')^2} + w\log|\tau(P,P')^2|.$$

Actually, the density matrix that they computed was:

(3.7) $$r = R_S - S,$$

where:

(3.8) $$S[P, P'] = \exp\left(i\frac{e}{\hbar c}\int_P^{P'} A\right) S_0 + \frac{a}{\tau(P,P')^2} + b\log\left|\frac{\tau(P,P')^2}{C}\right|,$$

in which the matrices *a* and *b* are elaborate expressions involving *F, d,* and the Dirac matrices (the morbidly curious can confer the literature). Moreover:

(3.9) $$C = 4\left(\frac{h}{\gamma mc}\right)^2, \qquad \gamma = \text{Euler constant} = 1.781\ldots$$

*b. The Euler-Heisenberg effective Lagrangian.* Ultimately, Euler and Heisenberg (see also Schwinger [**23**]) obtained an effective Lagrangian that had the general form that Mie had used, viz., $\tfrac{1}{2} F^{\wedge *} F + \dagger_I$, where:

(3.10) $$\dagger_I = 16\pi^2 mc^2 \left(\frac{mc}{h}\right)^3 \lim_{n' \to \infty} \{\ldots\}$$

with:

(3.11) $\{\ldots\} =$

$$\frac{1}{16} - \frac{b^2}{8} - \frac{[1+(2n'+1)b]^2}{16}(1 - 2\log[1+(2n'+1)b]) - \frac{b^2}{24}\log[1+(2n'+1)b]$$

$$- \frac{b}{2}\sum_{n=1}^{n'}(1+2nb)\log(1+2nb) + \frac{ba^2}{12}\left(1 + 2\sum_{n=1}^{n'}\frac{1}{1+2nb} - \frac{1}{b}\log[1+(2n'+1)b]\right)$$

$$+ \frac{ba^4}{90}\left(1 + 2\sum_{n=1}^{n'}\frac{1}{(1+2nb)^3}\right) + \sum_{m=3}^{\infty} c_m \frac{a^{2m} b}{2m-1}\left(1 + 2\sum_{n=1}^{n'}\frac{1}{(1+2nb)^{2m-1}}\right),$$

in which we have left aside, for the moment, the matter of the definition of $c_m$ and the actual convergence of the power series in the last term.

If one regards $\dagger_I = \dagger_I(a, b)$ as a function of the two invariants $a^2 - b^2$ and $a^2 b^2$ then, since $\dagger_I(a, 0) = \dagger_I(0, ia)$, one can indirectly compute the $c_m$ by what Heisenberg and Euler assure us is a straightforward, but tedious, calculation.

For the weak field case, viz., $a \ll 1, b \ll 1$, one has the approximation:

(3.12) $$\lim_{n' \to \infty} \{\ldots\} = \frac{(a^2-b^2)^2 + 7(ab)^2}{180} + \frac{13(ab)^2(a^2-b^2) + 2(a^2-b^2)^3}{630} + \mathbf{L}$$

For the strong field case, viz., $a \gg 1, b \gg 1$, one has the approximation:

(3.13) $\lim_{n' \to \infty} \{\ldots\} =$

$$b^2\left[\frac{1}{12}\log b - 0.191\right] + \frac{b}{4}[\log b - 0.145] + \frac{\log b}{8} + 0.202 - \frac{a^2}{12}[\log b + 0.116] + b\left[\frac{a^2}{12} + \frac{a^4}{90} + \mathbf{L}\right].$$

Ultimately, Euler and Heisenberg arrive at an integral representation for the vacuum polarization energy density:

(3.14) $\dagger_I(E^2 - B^2, E \cdot B) =$

$$\frac{e}{hc}\int_0^\infty \frac{d\eta}{\eta}\left\{i\eta^2(E\cdot B)\frac{\cos\left(\frac{\eta}{E_c}\sqrt{E^2 - B^2 + 2i(E\cdot B)}\right) + c.c.}{\cos\left(\frac{\eta}{E_c}\sqrt{E^2 - B^2 + 2i(E\cdot B)}\right) - c.c.} + |E_c|^2 + \frac{\eta^2}{8}(B^2 - E^2)\right\}.$$

Note that there is a critical electric field strength $E_c$ in this theory that corresponds to the field strength at the "surface" of the electron.

In order to make closer contact with the Born-Infeld Lagrangian, we substitute $\wedge = \frac{1}{2}(B^2 - E^2)$ and $\& = E \cdot B$. This gives:

$$(3.15) \quad \pounds_I(\wedge, \&) = \frac{e}{hc}\int_0^\infty \frac{d\eta}{\eta}\left\{i\eta^2 \& \frac{\cos\left(\frac{\eta}{E_c}\sqrt{-2(\wedge - i\&)}\right) + c.c.}{\cos\left(\frac{\eta}{E_c}\sqrt{-2(\wedge - i\&)}\right) - c.c.} + |E_c|^2 - \frac{\eta^2}{4}\&\right\}.$$

Schwinger provides the following expansion:

$$(3.16) \quad \pounds_I(\wedge, \&) = \frac{2\alpha}{45}\frac{(h/mc)^3}{mc^2}[4\wedge^2 + 7\&^2] + \ldots$$

with:

$$(3.17) \quad \alpha = \frac{e^2}{4\pi hc} = \frac{1}{137},$$

whereas a corresponding expansion of the Born-Infeld Lagrangian would give:

$$(3.18) \quad \pounds_I(\wedge, \&) = -\frac{1}{2E_c}[\wedge - \frac{1}{2E_c^2}(\wedge - \&)^2 + \ldots$$

which has a passing similarity to the Euler-Heisenberg expansion.

The difference between the expression that was derived by Euler-Heisenberg and that of Born-Infeld is chiefly in the fact that Born and Infeld started with the phenomenon of vacuum polarization as an assumption for the definition of a nonlinear theory of electromagnetism, whereas Euler and Heisenberg started with the assumption of a linear theory and showed how quantum corrections produced the vacuum polarization as a result, in the form of an effective Lagrangian for a nonlinear electrodynamical theory.

**4. Discussion.** One must expect that a considerable degree of resolution in the details of the elementary processes of matter in interaction will be absent from the formulas that describe the scattering process since quantum field theory introduces two primary sources of blurring in that resolution:

*i*) The statistical interpretation of wave mechanics and its association of Hermitian operators with measurements performed on an essentially stochastic space of system states; i.e., a measurement is associated with a spectrum of outcomes, not a unique outcome.

*ii*) The passage to the limit $t \to \pm\infty$ that is characteristic of the scattering process will necessarily sacrifice a great deal of fine detail concerning a scattering event by demoting it to the status of "virtual" processes that may be indicative of actual physical processes, such as pair creation and annihilation, or merely spurious consequences of the mathematical methodology.

A powerful tool in the modern theory of nonlinear differential equations is the method of the scattering transform and its inverse ([11]). Schematically:

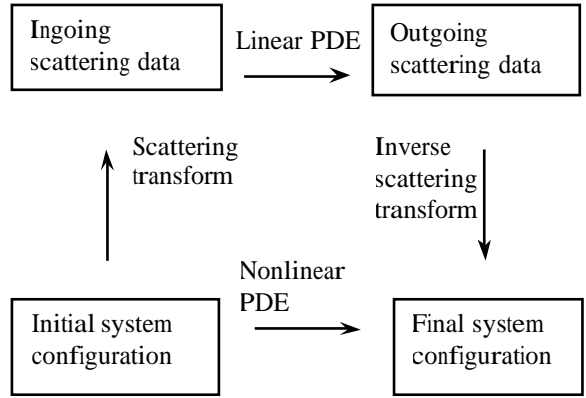

Fig. 15. Scattering and inverse scattering transforms.

In terms of Fig. 15, the main successes of quantum field theory have been in the top level of the diagram. The objective of the present study if to see if it is possible to use those successes as a means of building a more sophisticated model for the elementary structure of charged particles and their electromagnetic interactions by examining the avenues of nonlinearity in electromagnetism and their consequences. One expects that the first attempts will have to be effective nonlinear field theories, in advance of deeper intuitions that might suggest unifying first principles of a broader scope in Nature. Hence, one might consider the effective Lagrangians and effective potentials that one obtains from quantum electrodynamics to be strongly worded hints at more fundamental nonlinear field Lagrangians that should reproduce the scattering data in that limit ([12]). For instance, we shall see that the effective Lagrangian for vacuum polarization due to a constant electric field that was obtained by Euler and Heisenberg starting from quantum electrodynamics closely resembles the Lagrangian that Born and Infeld used as the basis for a nonlinear theory of electrodynamics by starting with assumptions about how vacuum polarization will affect the maximum possible electric field strength.

A consistent theme of the foregoing synopsis of quantum electrodynamics was that Coulomb's law must undoubtedly break down in the realm of large field strengths, at which point vacuum polarization takes over, and any further increase in field strength will result in the phase transition of pair creation. Indeed, as we shall see, these facts also played a key role in the earliest attempts at nonlinear electrodynamics. A possible consequence of modifying Coulomb's law might be the elimination of the spurious infinity in the self-energy of a charged particle that seems to originate in the pole at zero in the Coulomb field.

---

[11] In the popular literature one simply deals with the "inverse scattering transform," since one is converting a problem in nonlinear PDE's into a problem that takes the form of reconstructing a scattering potential from scattering data. This seems potentially confusing, since one certainly does not refer to both the Fourier transform and its inverse by the same term; however, what we are calling the "scattering transform" is not as rigorously defined as the Fourier transform.

[12] Along these lines, cf. Dittrick [**13**].

Furthermore, a key role must be played by the empirical nature of photon-photon scattering at high field intensities, since one would find therein the most definitive statement of nonlinearity in the superposition of such fields, as well as in the scattering of photons and electrons by external fields of large magnitudes.

Part II: Nonlinear electrodynamics

**5. Electrodynamics in continuous media.** When one is looking at the electrodynamics of continuous media [**24**], the issues that distinguish such media from the classical electromagnetic vacuum manifold arise from the fact that the atoms that comprise massive macroscopic material media are decomposable into smaller charged particles that might interact with the applied electric and magnetic fields, and in so doing, change the applied fields themselves. The key changes that come about in material media are the formation of electric dipoles in dielectric media under the action of an electric field and the formation of magnetic dipoles in magnetic materials under the action of a magnetic field.

In the electric case, one often introduces an *electric displacement* field **D** that measures the degree of polarization of the material due to the formation of electric dipoles, or rather, the *polarization* vector field does:

(5.1) $$\mathbf{P} = \frac{1}{4\pi}(\mathbf{D} - \mathbf{E}).$$

Hence, **P** takes the form of a density of electric dipole moments.

**E** and **D** are related in a manner that is analogous to strain and stress in continuous media, respectively, i.e., by an electrical constitutive relation that is analogous to the mechanical ones:

(5.2) $$\mathbf{D} = \mathbf{D}(\mathbf{E}).$$

Not surprisingly, for small enough values of |**E**|, one often makes the predictable empirical approximation that the relationship is not only linear but isotropic, and introduces a *dielectric constant* ε that is associated with the material in question:

(5.3) $$\mathbf{D} = \varepsilon \mathbf{E}.$$

Just as in the elastodynamics, one can also consider more general linear relationships, such as ones with non-diagonal matrices, etc., depending on the properties of the material.

Similarly, one can introduce a magnetic induction **H** that is associated with the magnetic field strength **B** by another constitutive equation:

(5.4) $$\mathbf{H} = \mathbf{H}(\mathbf{B}),$$

which is often give by a *magnetic permeability* constant μ for the case of a linear isotropic magnetic material:

(5.5) $$\mathbf{H} = \mu \mathbf{B}.$$

One also defines a *magnetization* vector field:

(5.6) $$\mathbf{M} = \mathbf{H} - \mathbf{B},$$

which is analogous to the electric polarization vector field. **M** is treated as an *order parameter* in the condensed matter approach to superconductivity, and represents a density of magnetic dipole moments.

An important point to note, which we will return to in the context of nonlinear wave motion, is the fact that the speed of propagation $c$ of an electromagnetic wave in such a medium is given by:

(5.7) $$c^2 = \frac{1}{\varepsilon\mu}$$

or, if one lets $c_0$ represent that propagation speed in vacuo, one could also relate the *index of refraction* $n = c_0/c$ of the medium to $\varepsilon$ and $\mu$, as well, viz., $n = c_0\sqrt{\varepsilon\mu}$

As is well known, the electric and magnetic fields can be assembled into an antisymmetric second rank covariant tensor field – i.e., a 2-form – in the form of the Minkowski electromagnetic field strength 2-form $F$:

(5.8) $$F = \theta \wedge E - *_S B,$$

in which $\theta = i_t g$ is the timelike unit 1-form that is metric dual to a chosen time orientation, in the form of a timelike unit vector field **t**. We have also introduced the notation:

(5.9) $$*_S B = *(\theta \wedge B) = i_B \mathcal{9}_S, \qquad (\mathcal{9}_S = i_t \mathcal{9})$$

for the 2-form that is spatially dual to $B$ by way of the spatial volume element $\mathcal{9}_S$. Moreover, $E = i_E g$, $B = i_B g$ are the 1-forms that are metric-dual to the vector fields **E**, **B**.

Similarly, one can construct the electromagnetic induction 2-form $H$ from **D**, **H** as:

(5.10) $$H = \theta \wedge D - *_S H, \qquad D = i_D g, \; H = i_H g.$$

Conversely, one if one is given $F$ and $H$, one decomposes them into spacelike 1-forms (modulo **t**) by way of:

(5.11) $$\begin{cases} E = i_t F & B = i_t * F \\ D = i_t H & H = i_t * H. \end{cases}$$

In this relativistic form, the constitutive law takes the form:

(5.12) $$H = H(F, *F).$$

Note that we are treating $F$ and $*F$ as distinct fields, much as one would consider a complex number $z$ and its conjugate $z^*$ to be distinct fields. Indeed, for a four-dimensional Lorentz manifold $(M, g)$ the $*$-isomorphism that one gets from $g$ defines an almost complex structure on the vector bundle $\Lambda^2(M)$ since $*^2 = -1$ when that map is applied to 2-forms. Furthermore, we are leaving open the possibility that $D = D(E, B)$ and $H = H(E, B)$ at this point.

$F$ and $H$ are second rank tensors, so even in the case of a linear constitutive law there can be considerable complexity to the relationship, since a linear map from $\Lambda^2(M)$ to itself is essentially a fourth rank tensor field on $M$. In the linear theory, one generally reduces such complexity by symmetry assumptions, such as spatial isotropy, as in the case of continuum mechanics, but we must keep in mind that it is the nonlinear case that is of immediate interest.

We may also generalize the definition of polarization accordingly to the *electromagnetic polarizability 2-form*:

(5.13) $$P = \frac{1}{4\pi}(H - F).$$

In continuous media, one expects the linearity to break down for large enough |**E**|, such as when one is dealing with a high-intensity laser pulse propagating through the material, as in nonlinear optics, or large enough |**H**|, such as with the Meissner effect in superconductivity. General the onset of nonlinearity is also associated with a phase transition, as well. Hence, at least in the case of continuous media, we would seem to be dealing with a situation in which electrodynamics has some of the same basic ingredients for a type of nonlinearity that is closely analogous to the continuum-mechanical case of elastic materials.

Maxwell's equations, including the electromagnetic inductions, are:

(5.14) $$dF = 0, \qquad \delta H = \frac{4\pi}{c} J, \qquad H = H(F, *F).$$

The *current* 1-form $J$ plays the role of the source of the $H$ field, and generally takes the form $J = \rho_0 u$, where $\rho_0$ is the charge density for a distribution of compact spatial support, as seen in the rest frame for the 1-form $u = i_v g$, which is the covelocity that is metric dual to the velocity vector field **v** that describes the motion of the charge distribution. In general the timelike part of $J$ will be the source of the $D$ field and the spacelike part of will be the source of the $H$ field (the 1-form, of course).

If we write $J = \rho\theta + \hat{j}$, where $r$ is the charge density in an arbitrary frame then the 3+1 form of Maxwell's laws are:

(5.15) $$dB + \frac{\partial}{\partial\tau} *_s E = 0 \qquad \delta_s B = 0$$
$$dD - \frac{\partial}{\partial\tau} *_s H = \frac{4\pi}{c}\hat{j} \qquad \delta_s D = \rho.$$

We have introduced the notations $\partial/\partial\tau$ and $d_s$ for exterior derivative of a differential form in the direction **t** and the spatial of that exterior derivative by the general rule:

(5.16) $$d\alpha = \theta \wedge \frac{\partial}{\partial\tau} + d_s\alpha.$$

Furthermore, we have defined the spatial divergence of a differential form by:

(5.17) $$\delta_s\alpha = *_s d_s *_s \alpha.$$

Rather than include the constitutive law explicitly, we may include it implicitly by way of the polarization 2-form $P$. Maxwell's equations then become:

(5.18) $$dF = 0, \qquad \delta F + 4\pi\delta P = \frac{4\pi}{c} J.$$

This can be put into the form of a forced nonlinear wave equation:

(5.19) $$\mathbf{u}F + 4\pi d\delta P = \frac{4\pi}{c} dJ.$$

From Maxwell's equations in the form (5.13) one derives the conservation of $J$ as a consequence:

(5.20) $$\delta J = 0.$$

One might also start with this latter equation as a defining equation for the system of fields that is composed of $F$ and $J$:

(5.21) $$dF = 0, \qquad \delta J = 0.$$

One then has to introduce some way of coupling the source $J$ to the field $F$.

Note that the latter pair of equations express that $F$ is *closed* and $J$ is *co-closed*, respectively. If $M$ has vanishing second de Rham cohomology $H^2(M, \mathbf{R})$ then there is a non-unique potential 1-form $A$ such that:

(5.22) $$F = dA \qquad \text{(modulo a closed 1-form)},$$

and if $H^1(M, \mathbf{R})$ vanishes there is a non-unique 2-form $H$ such that:

(5.23) $$J = \delta H \qquad \text{(modulo a co-closed 2-form)}.$$

This suggests that the electromagnetic induction 2-form $H$ might play the same role in relation to the source current 1-form $J$ that the potential 1-form $A$ does in relation to $F$.

If we use the definitions (5.22) and (5.23), include the constitutive relation by way of $P$, and choose the Lorentz gauge $\delta A = 0$, then Maxwell equations become the nonlinear wave equation in $A$:

(5.24) $$\mathbf{u}A + 4\pi\delta P = \frac{4\pi}{c} J.$$

In this form, the possibility for introducing nonlinearity rests solely upon the nature of $\delta P$.

One can also obtain the constitutive relation between $F$ and $H$ by starting with a field Lagrangian $\mathcal{L} = \mathcal{L}(F)$ and define $H$ by way of:

(5.25) $$H = \frac{\delta \mathcal{L}}{\delta F} = \frac{\partial \mathcal{L}}{\partial F}.$$

Since:

(5.26) $$\delta\mathcal{L} = \frac{\partial \mathcal{L}}{\partial F} \wedge \delta F = H \wedge \delta F,$$

this gives us an antisymmetric form of the corresponding statement in continuum mechanics that associates a virtual work $\delta\mathcal{O}$ with a virtual strain $\delta e$ by way of the stress-strain constitutive relation $\sigma = \sigma(e)$:

(5.27) $$\delta\mathcal{O} = \frac{\partial \mathcal{L}}{\partial e}(\delta e) = \langle\sigma, \delta e\rangle.$$

Of course, in this case, the tensors involved, $e$ and $\sigma$, are generally symmetric. For the moment, we shall not elaborate on the nature of the scalar product of two such tensors. In any event, one may also express the Lagrangian in the form:

$$(5.28) \qquad += \tfrac{1}{2} H \wedge *F = \tfrac{1}{4} H_{\mu\nu} F^{\mu\nu},$$

which includes the contributions from **D•E** and **H•B** that one finds in the conventional formulation.

The modern approach to the properties of materials is by way of condensed matter physics [**25-27**], in which one derives the bulk properties of continuous matter, such as $k$, $\varepsilon$, $\mu$, etc., from more detailed statistical models at the quantum – i.e., atomic – level. A good direction to pursue, at least with the hindsight of quantum electrodynamics, is whether analogous condensed matter models apply to the spacetime vacuum state in its own right; in particular, one expects that vacuum polarization should play the central role in nonlinear electrodynamics.

If we are willing to restrict ourselves to Minkowski space, just to avoid deep mathematical complexities that arise when one tries to do Fourier analysis on manifolds that do not have any radius vector field or translation invariance, then we can examine the form that Maxwell's equations take when we assume that $F$ and $H$ take the form of a plane wave with a frequency-wavenumber 1-form given by $k = \omega/c\, dt + k_i dx^i$. Hence:

$$(5.29) \qquad F = F_0 \exp(ik_\mu x^\mu) = F_0 \exp[i(\omega t - k_\mu x^\mu)] = \qquad H = H_0 \exp(ik_\mu x^\mu).$$

Substitution into (5.13) gives:

$$(5.30) \qquad k \wedge F = 0, \qquad i_k H = \frac{4\pi}{c} J,$$

where $k = i_k g$. These are essentially the Fourier transforms of the Maxwell's equations, and are characterized notably by the fat that for each $k$ one obtains a system of algebraic equations for $F$ and $H$, rather than differential equations.

The 3+1 decomposition of these equations is:

$$(5.31) \qquad \begin{aligned} \hat{k} \wedge E - \omega *_s B &= 0, & \omega D - *_s(\hat{k} \wedge H) &= \frac{4\pi}{c} \hat{j} \\ \hat{k} \cdot B &= 0, & \hat{k} \cdot D &= \frac{4\pi}{c} \rho_0. \end{aligned}$$

In these equations, we have introduced the notations $\hat{k}$ and $\hat{j}$ for the spacelike parts of $k$ and $J$, respectively.

The first equation imples that $E \cdot B = 0$, and the third one says that $B$ is orthogonal for $\hat{k}$. Outside the support of $\rho_0$ – i.e., in vacuo – we find that $D \cdot H = 0$ and $\hat{k} \cdot D = 0$. Hence, the 1-forms $B$ and $D$ must lie in the spacelike plane that is orthogonal to $\hat{k}$. One is tempted to make orthogonal frames out of ($\theta$, $E$, $B$, $\hat{k}$) and ($\theta$, $D$, $H$, $\hat{k}$), but there are two problems with this: In the first place the electric and magnetic fields are not always non-zero, and in the second place, unless one knows that $E$ is collinear with $D$ and $B$ is collinear with $H$, which depends upon the constitutive relation, one has no guarantee that $E$ or $H$ are not parallel with $\hat{k}$. Of course, in a linear isotropic medium, one will have the aforementioned collinearity, which will imply that for any point where all four fields are non-zero, either (θ, E, B, $\hat{k}$) or (θ, **D**, H, $\hat{k}$) will define orthogonal frames at that point.

The 2-forms that describe wave fields have a deep property that is best described in terms of the concept of the *rank* of an exterior form, which is defined as the codimension of the annihilating subspace of the form. In the case of 2-forms, this must be even dimensional, so on a four-dimensional manifold a 2-form, say $F$, can have rank 0, 2, or 4; rank 0 would mean that $F = 0$. A consequence of the definition of rank is that in the case where $F$ has rank four, the minimum number of linearly independent 1-forms that it takes to express $F$ is four, so $F$ takes the form:

(5.32) $\qquad F = a \wedge b + c \wedge d$

for some linearly independent 1-forms $a, b, c, d$.

In the case where $F$ has rank two, there are linearly independent 1-form $a$ and $b$ such that:

(5.33) $\qquad F = a \wedge b.$

Another way of characterizing this difference in rank is by forming the product $F \wedge F$:

(5.34) $\qquad F \wedge F \begin{cases} = 0 & \text{rank}(F) = 2 \\ \neq 0 & \text{rank}(F) = 4. \end{cases}$

When we apply the 3+1 decomposition, we see that:

(5.35) $\qquad F \wedge F = 2\theta \wedge E \wedge *B = 2g(E, B)\mathbf{9}.$

Hence, the rank two case is characterized by the orthogonality of $E$ and $B$, which also characterizes the wavelike solutions of the Maxwell equations. Note that the orthogonality of $E$ and $B$ is not sufficient to establish that a solution wavelike; for instance, a pair of orthogonal static $E$ and $B$ fields will satisfy Maxwell's equations.

In the next two sections, we shall examine some of the early attempts at formulating nonlinear theories of electrodynamics. The first one that we shall discuss, which is due to Gustav Mie, was a largely formal mathematical structure that only reduced the ambiguity in the choice of field Lagrangian without reaching a unique conclusion. The second theory, which is due to Max Born and Leopold Infeld, attempted to incorporate some of the thinking of quantum electrodynamics concerning the structure of the vacuum, especially as it pertained to vacuum polarization. In the final section, we shall examine some of the phenomena of linear optics with the intent of building some intuition of what to expect of the propagation of electromagnetic waves in polarized media in the nonlinear regime of the constitutive laws.

**6. Mie's theory of electrodynamics.** Gustav Mie [**29**] made one of the first attempts to extend beyond the linearity of Maxwell's equations by assuming that the Lagrangian of the electromagnetic field could be decomposed into the sum of the conventional electromagnetic field Lagrangian and a second term that was only significant in the realm of atomic to subatomic distances:

(6.1) $\qquad +(g, A, F) = \frac{1}{2} F \wedge *F + +_I(g, A, F).$

*a. Mie's equations.* In order to start from as general a place as possible, Mie made the nonlinear part of the electromagnetic field Lagrangian depend upon not only $g$ and $F$, as does the conventional Lagrangian, but also upon a choice of potential 1-form $A$. Mie assumed, moreover, that $+_I$ would have to be Lorentz invariant. From this, he (incorrectly) concluded that $+_I$ must be a function of the four (actually five) Lorentz invariant scalar or 4-form expressions that one could form from $g, A, F$:

(6.2)
1. $I_1 = \|F\|^2 = F \wedge *F$,
2. $I_2 = F \wedge F$,
3. $I_3 = \|A\|^2$,
4. $I_4 = \|i_\mathbf{A} F\|^2$,
5. $I_5 = \|i_\mathbf{A} *F\|^2$,

in which $\mathbf{A}$ is the vector field that is metric-dual to $A$, i.e., $A = i_\mathbf{A} g$. Note that if $\mathbf{A}$ is collinear with $\mathbf{t}$ then the last two invariants are proportional to $\mathbf{E}$ and $\mathbf{B}$.

One of the leading criticisms of Mie's theory centered around his decision to include the gauge form $A$ in the Lagrangian explicitly, which would subsequently break the gauge invariance of the Lagrangian. Nowadays, one recognizes that the first invariant is the essence of a mass term for a gauge field for a spontaneously broken symmetry of the vacuum state when one applies the Higgs mechanism. In the present case, the gauge field is the electromagnetic field, so any theory that involves the first invariant must deal with the previously mentioned question of the photon mass.

An immediate variation of the Lagrangian gives:

(6.3) $$\delta +_I = \frac{\delta +_I}{\delta g_{\mu\nu}} \delta g_{\mu\nu} + \frac{\partial +_I}{\partial A} \wedge \delta A + \frac{\partial +_I}{\partial F} \wedge \delta F,$$

which can be put into the form:

(6.4) $$\delta +_I = T^{\mu\nu} \delta g_{\mu\nu} + *J \wedge \delta A + *H \wedge \delta F,$$

with the obvious definitions:

(6.5) $$T^{\mu\nu} = \frac{\delta +_I}{\delta g_{\mu\nu}}, \quad *J = \frac{\partial +_I}{\partial A}, \quad *H = \frac{\partial +_I}{\partial F}.$$

The Euler-Lagrange equation, namely:

(6.6) $$0 = \frac{\delta +_I}{\delta A} = \frac{\partial +_I}{\partial A} - d\left(\frac{\partial +_I}{\partial (dA)}\right) = \frac{\partial +_I}{\partial A} - d\left(\frac{\partial +_I}{\partial F}\right)$$

that results from varying $A$ and $F$ in this Lagrangian is:

(6.7) $$\delta H = J.$$

Note that this equation automatically implies the conservation of $J$:

(6.8) $$\delta J = 0.$$

We can augment the former equation with another one that gives the pair the overall form of the Maxwell equations when one includes inductions if we introduce the integrability condition $dF = 0$, which is, of course, consistent with the constraint that $F = dA$. Similarly, we should include the definition $*J = \partial +_I / \partial A$ in the full set of field equations, as well as the constitutive law $*H = \partial L_I /\partial F$: Hence, we can summarize the full set of Mie equations as:

(6.9)  $\qquad F = dA, \qquad \delta H = J, \qquad *H = \dfrac{\partial L_I}{\partial F}, \qquad *J = \dfrac{\partial L_I}{\partial A}.$

This makes it clear that it would be incorrect to say that the field theory is complete at this stage, since we still have said nothing about the specific nature of $L_I$ beyond some general axiomatic statements. Hence, one of the major criticisms of Mie's theory, beyond the lack of gauge invariance, was the fact that one still had to find a more specific form for the supplementary Lagrangian $L_I$.

*b. Electrostatic spherically symmetric solutions.* One of Mie's attempts at deducing a suitable form for $L_I$ that would lead to a field equation whose spherically symmetric electrostatic solutions would suggest some extension of Coulomb's law into the "interior of the electron" attracted little attention at the time. Perhaps that is because when he was discussing the physical nature of that solution many of the things that are regularly discussed in theoretical physics – antiparticles, mass terms in Lagrangians, singularities in fields, and the like – had no basis in experimental practice, or at least no common usage in theory, and were easily overlooked. Hence, just for the sake of completeness, we summarize one of Mie's attempts at such a solution.

The electrostatic form of Mie's equations are obtained by letting:

(6.10)  $\qquad F = \theta \wedge E, \qquad A = \varphi\theta, \qquad H = \theta \wedge D, \qquad J = \rho\theta,$

namely:

(6.11)  $\qquad E = -d\varphi, \qquad \delta D = \rho, \qquad \mathbf{D} = *\dfrac{\partial L_I}{\partial E}, \qquad \rho = *\dfrac{\partial L_I}{\partial \varphi}.$

Hence, they have the Maxwell form for continuous media, with appropriate definitions for the constitutive law $\mathbf{D} = \mathbf{D}(E)$ and the definition of the charge density $\rho$.

What Mie envisioned was an extension of the empirical $1/r$ Coulomb potential into a neighborhood of $r = 0$ that would involve what he called "knot-like" singularities ("*Knottenstellen*") in the field lines of $\mathbf{D}$ with a very thin "atmosphere" surrounding the source. By the modern understanding of the adjective "knot-like" (or even its usage by Tait back then), he was not actually defining knots in the modern sense of embeddings of circles in a three-dimensional manifold, so perhaps it would be more accurate to say that he was simply imagining a singular behavior in the field lines of $\mathbf{D}$ about the source charge that was more elaborate than the character of the Coulomb field would suggest.

He further restricted the Lagrangian by the assumptions that it would be an even function of $\rho$ and that it would be at least continuous. It is interesting to note that if he had made the former assumption on $\varphi$ instead of $\rho$, then he would have been close to a Landau-Ginzburg sort of theory, with its associated spontaneous breaking of the symmetry in the vacuum state, and the Higgs mechanism that might give mass to the photon.

Mie made the intriguing observation that the role of $\rho$ in regard to the equilibrium state in electrostatics might be analogous to the perturbations of the steady flow state in hydrodynamics. Unfortunately, he then went off into a blind alley by applying this to an earlier, long since discredited, model of atoms as balls of positive and negative charges intermingled, which also included the assumption that positive and negative charges could not cancel each other our or be created from nothing (remember, this was 1912!).

He then got back to more meaningful considerations – at least, by modern standards – namely, the stability of the equilibrium state.  In the case of spherical symmetry, he arrived at the conclusion that one necessary condition for that stability is that $\partial \varphi / \partial \rho$ must have the same sign as $\varphi/\rho$ everywhere. This is indeed the case under the prior assumption that $+_I$ is an even function of $\rho$. He also showed that another necessary condition was that $\partial \|E\|/\partial \|\mathbf{D}\|$ have the same sign everywhere as $\|E\|/\|\mathbf{D}\|$. Apparently, it was not also known if the necessary conditions were sufficient, but it was certain that unstable equilibrium field configurations were possible. Furthermore, the assumption that the potential energy at the surface of a ball of radius $r$ about the source − namely, $\varphi(r)$ − should equal $Q/4\pi r$, as with Coulomb's law, when $r$ is sufficiently large implies the further restriction on $+_I$ that $\varphi(r)$ must go to zero at infinity faster than $r^{-3}$.

Ultimately, he arrives at the following differential equation for the equilibrium state:

$$(6.12) \qquad r \frac{\partial^2 +_I}{\partial \varphi_r^2} \varphi_{rr} + 2 \frac{\partial +_I}{\partial \varphi_r} + r \frac{\partial}{\partial \varphi}\left( \frac{\partial +_I}{\partial \varphi_r} \varphi_r - +_I \right) = 0,$$

in which $\varphi_r$ and $\varphi_{rr}$ are the first and second derivatives of $\varphi$ with respect to $r$, respectively. He then showed that there was an infinitude of $+_I$ that solved this equation and did not contradict the properties of conventional electrostatics outside of an appropriate neighborhood of the singularity.

*c. A particular case.* A particular case for $+_I$ that Mie chose to examine was that of:

$$(6.13) \qquad +_I = -\tfrac{1}{2}\eta^2 + \tfrac{1}{6}a\chi^6$$

with:

$$(6.14) \qquad \eta = \sqrt{\varphi^2 - j^2}, \qquad \chi = \sqrt{E^2 - \mathbf{D}^2}, \qquad J = \varphi\theta + j.$$

In the static case, one has:

$$(6.15) \qquad \mathbf{D} = E, \qquad \rho = a\varphi^2,$$

which makes the stability equation take the form:

$$(6.16) \qquad r\varphi_{rr} + 2\varphi_r + ar\varphi^5 = 0.$$

By various transformations, this gives the equation:

$$(6.17) \qquad \left(\frac{dv}{d\xi}\right)^2 = Cv + v^2 - \frac{4}{3}av^4$$

with:

$$(6.18) \qquad v = r\varphi^2, \qquad \xi = \ln\frac{r}{r_0},$$

and $C$, $r_0$ are integration constants. This equation, in turn, is solvable by an elliptic function of $\xi$ that has an essential singularity at $x = \infty$, which corresponds to essential singularities at $r = 0$ and $r = \infty$. The nature of $r_0$ is that it defines the null surface of $\varphi(r)$.

Mie then classified the real solutions for $\varphi$ in terms of the sign of $C$: $-$, 0, or $+$.

For the case of $C > 0$, the qualititative nature of the electrostatic solution was described by saying that the behavior of $E$ near $r_0$ is very close to that of the Coulomb field strength, but if one splits one charged spherical surface into two charged spheres and separates them then for increasingly weak $E$ the distance at which the "atmosphere" that surrounds each shell produces a noticeable departure from Coulomb's law goes infinite, but for increasingly strong $E$, the distance gets infinitesimal. A striking difference between this solution for $\varphi(r)$ and Coulomb's law was the fact that the center of the field would be surrounded by alternating positive and negative shells of charge, which would also grow closer together as one approached $r = 0$, and this characterizes the essential singularity at 0. As for the essential singularity at infinity, one characterizes it by the fact that there are infinitely many maxima for $\varphi(r)$ between any finite $r$ and $\infty$.

In the case of $C < 0$, the main difference is that the oscillation becomes more stepwise, and in the case of $C = 0$, one gets the closed-form solution:

$$(6.19) \qquad \varphi(r) = \sqrt{\frac{3r_0^2}{a}} \frac{1}{\sqrt{r^2 + r_0^2}},$$

which has no more singularities.

*d. The case of the electron.* If we start with eq. (6.19) and try to adapt it to the known properties of the electron, first we note that $r_0$ must be quite small to remain consistent with known data on the electron. The corresponding charge density that we obtain from $\varphi$ is:

$$(6.20) \qquad \rho(r) = \sqrt{\frac{3r_0^2}{a}} \frac{3r_0^2}{\sqrt{(r^2 + r_0^2)^5}}.$$

By integrating from $r = 0$ to infinity, one obtains a total charge of:

$$(6.21) \qquad Q[0 \leq r] = 4\pi \sqrt{\frac{3r_0^2}{a}}.$$

We point out that, actually, either the positive or negative root can be used in this expression.

In order to get some idea of the orders of magnitude in the electrical atmosphere, one computes the total charge between some $r_1$ and infinity:

$$(6.22) \qquad Q[r_1 \leq r] = \left[1 - \left(\frac{r_1}{\sqrt{r_1^2 + r_0^2}}\right)^3\right] Q[0 \leq r],$$

which shows that the charge is quite localized about $r = 0$.

In order to be objective about the concluding remarks of Mie in the second part of his treatise, we suspend our insistence that he is only talking about electrons in particular and consider that he was unwittingly talking about the nature of matter and anti-matter. One then sees that there is something compelling about the fact that in a universe where electrostatics is governed by our present choice of $+_I$, one finds that charges cannot come close to each other in the equilibrium state. Contrary to the usual expectations, when they have the same sign they will merge into a larger charge, and when they have opposite signs they will move apart indefinitely.

Of course, if one is looking for a nonlinear contribution to the Coulomb potential that would serve as a cohesive force to bind a spatially extended elementary charge distribution into a stable configuration despite the mutual Coulomb repulsion of its constituents then this is the right sort of character for such a binding force. Similarly, if one is looking for the manner by which protons can coexist in stable nuclei, a similar model might apply. At the risk of unbounded speculation, one wonders if the strong interaction − at least at the nuclear level – might really be the nonlinear contribution to the Coulomb force that becomes dominant at the subatomic level in order to bind like charges into stable extended configurations.

*e. Mie's elementary electric dipole.* One of the more fascinating passages in Mie's work, which gets little modern discussion moreover, is the passage in the third installment of the treatise that deals with the behavior of elementary dipoles in the aforementioned nonlinear model. One can axiomatically characterize them as regions in which one finds *E* and **D** have large magnitudes and opposite directions, with a large magnitude to *d***D**, but a small magnitude to δ**D**. Notice that such a situation could not happen in linear electrodynamics unless the electric susceptibility were of consistently negative sign; furthermore, *d***D** would have to be zero, along with *dE*.

Since δ**D** is related to the total charge in a region, and *d***D** plays the role of the "vorticity" of the field lines for the vector field **D**, the picture that Mie described is one of an elementary electric dipole surrounded by closed **E** field lines like a permanent magnet, and such that in the interior the **D** field is in opposition to the **E** field. Although Mie commented that finding the conditions for the existence of such entities would be difficult, one must realize how little was known in his time about the role of topology in vortex theory.

Mie points out that in Maxwell's theory, which is still presumed to be valid outside of a certain neighborhood of an elementary charge, spherical electromagnetic waves represent electrical dipoles, but they do not represent the equilibrium, or static, field case. Indeed, the tendency of the spherical wave is to expand outward with lightlike velocity. To Mie, the next question to address is that of the nature of the phase transition that transforms an elementary electric dipole into an expanding spherical wave, and, conversely, a contracting spherical wave into an elementary electric dipole. Once again, he points to the stability of the equilibrium state for the elementary electric dipole to be the key issue to examine.

When Mie addresses the matter of discreteness in the spectrum of the radiation that is produced by exploding electric dipoles, he only goes so far as to show that if the elementary dipoles are discretely "quantized" then so are the electromagnetic pulses that are produced by the explosion.  Under the additional assumption that the wave produced always has the same form, he then proves that one can characterize Planck's constant *h* as proportional to the square of the number of closed electrical field lines in the spherical wave, which he envisions as something like an expanding hollow spherical vortex shell.

Mie also suggests that there could be such entities as elementary magnetic dipoles, as well.  He characterizes these solutions by the presence of a region in which $d\mathrm{H} = 0$, but $\delta \mathrm{H} \neq 0$, even though $\delta B = 0$, as usual.  Moreover, H will be oppositely directed to *B* in the interior of such a region.

**7. Born and Infeld's theory.**  Originally, the stated intent of Born and Infeld's theory [**30**] of electromagnetism was to ameliorate the infinitude of the self-energy of the electron by considering an electromagnetic field Lagrangian that implicitly contained an upper bound $E_c$ on the electric field strength.

*a. Born-Infeld equations.*  Specifically, the Lagrangian that Born and Infeld suggested was:

(7.1) $$ \mathcal{L} = (\sqrt{E_c^2 + I_1 - I_2} - E_c)\mathcal{V} $$

with:

(7.2) $$ \wedge \equiv I_1 = \|F\|^2, \qquad \& \equiv I_2 = F \wedge F, \qquad \mathcal{V} = \sqrt{-g}\, dx^0 \wedge \ldots \wedge dx^3. $$

The form of this Lagrangian was suggested by one of the various attempts at unifying the theories of gravitation and electromagnetism, namely, the Einstein-Schrödinger theory [**31**] in which the symmetric spacetime metric *g* was replaced by an asymmetric second-rank covariant tensor field *a*, whose symmetric part equaled *g* and whose antisymmetric part equaled *F*.  The presence of the constant $E_c$ will later be seen to imply the boundedness of the electric field strength, even for a pointlike charge.

To relate the Born-Infeld theory to Mie's theory, note that the general form of the Born-Infeld Lagrangian is:

(7.3) $$ \mathcal{L} = \mathcal{L}(g, I_1, I_2) = L(g, I_1, I_2)\mathcal{V}. $$

Hence, the Born-Infeld Lagrangian does not contain the gauge-fixing photon mass term, in particular.

Some years later, Born [**32**] decided that a better starting point for the theory was that of vacuum polarization, which accounts for a maximum electric field strength in a more specific way as a maximum electric field strength beyond which vacuum polarization yields to pair creation.  Regardless of interpretation, the choice of Lagrangian gives some immediate consequences.  We shall first state the field equations collectively and then discuss them individually:

(7.4) $$ \begin{cases} dF = 0 \\ \delta H = 0 \\ *H \equiv \dfrac{\partial \mathcal{L}}{\partial F} = \dfrac{\partial \mathcal{L}}{\partial \wedge} F + \dfrac{\partial \mathcal{L}}{\partial \&} *F. \end{cases} $$

The first equation, which is common to the Maxwell theory, is actually an *assumption*, since Born and Infeld felt that the electromagnetic field strength tensor should still admit a potential. The second equation is also common to the Maxwell theory if one regards $H$ as the 2-form of electromagnetic induction that is associated with $F$ by way of the constitutive equation that the third equation represents. However, the second equation is obtained from the Lagrangian as the associated Euler-Lagrange equation, so its appearance is not quite so axiomatic.

For the particular choice of Lagrangian, the electromagnetic induction 2-form becomes:

(7.5) $$H = (E_c^2 + \wedge - \&)^{-1/2}(F - \&*F),$$

so the associated polarization 2-form is:

(7.6) $$P = H - F = -\frac{1}{L + E_c}(LF + \&*F).$$

The canonical energy-momentum tensor that one obtains from this Lagrangian is:

(7.7) $$T_\nu^\mu = L\delta_\nu^\mu - H^{\lambda\mu}F_{\lambda\nu}$$

One can also express the Lagrangian in terms of $g$ and $H$, or use these variables as conjugate variables in the Hamiltonian:

(7.8) $$* = + - H \wedge *F = (-g)^{-1/2}[\sqrt{-|g + *H|} - \sqrt{-|g|}]$$

*b.* 3+1 *form of the Born-Infeld equations.* If we have defined a time orientation on spacetime by way of a unit timelike vector field **t** then one can define the usual spacelike field strength and induction 1-forms of electrodynamics (or their metric-dual vector fields) by:

(7.9) $$\begin{cases} E = i_t F & B = i_t *F \\ \mathbf{D} = i_t H & H = i_t *H. \end{cases}$$

If we rescale the Lagrangian density to:

(7.10) $$L = \sqrt{1 + \wedge - *} - 1$$

then we get:

(7.11) $$\wedge = \frac{1}{E_c^2}(B^2 - E^2), \qquad \& = \frac{1}{E_c^2}g(B, E).$$

Specifically, we get the constitutive equations:

(7.12) $$\mathbf{H} = E_c^2 *\left(\frac{\partial +}{\partial B}\right) = (1 + \wedge - \&)^{-1/2}(B - \&E)$$

$$\mathbf{D} = E_c^2 *\left(\frac{\partial +}{\partial E}\right) = (1 + \wedge - \&)^{-1/2}(E - \&B).$$

The Born-Infeld field equations become:

(7.13)
$$\begin{cases} dE + \dfrac{1}{c}\dfrac{\partial B}{\partial t} = 0 & \delta B = 0 \\ dH - \dfrac{1}{c}\dfrac{\partial D}{\partial t} = 0 & \delta D = 0. \end{cases}$$

These are formally identical with Maxwell's sourceless equations for continuous media. The nonlinearity is, of course, hidden in the second set of equations, and only appears when one explicitly substitutes the electric displacement **D** and magnetic induction H with their expressions in terms of $E$ and $B$.

When one converts the energy-momentum tensor into 3+1 form, one gets:

(7.14)
$$T^{\mu\nu} = \begin{cases} \text{Energy:} & E_c L + \dfrac{1}{4}\mathbf{D}\cdot\mathbf{E} \\ \text{Left momentum:} & c\mathbf{G} = \dfrac{1}{4}\mathbf{D}\times\mathbf{B} \\ \text{Right momentum:} & c\mathbf{P} = \dfrac{1}{4}\mathbf{E}\times\mathbf{H} \end{cases} \quad \sigma^{ij} = E_c L \delta^{ij} - \tfrac{1}{4}D^i E^j,$$

in which the spacelike stress tensor $\sigma$ is defined by $\sigma^{ij} = T^{ij}$. Note, in particular, the asymmetry of the tensors $T$ and $\sigma$.

*c. Electrostatic spherically symmetric solutions.* In order to see how Born and Infeld changed the character of Coulomb's law, set $B = H = 0$ and assume that $E$ and **D** are spherically symmetric and independent of $t$. The remaining equations are:

(7.15) $\qquad\qquad dE = 0, \qquad \delta\mathbf{D} = 0.$

If one solves the second equation in spherical coordinates, one gets:

(7.16) $\qquad\qquad \mathbf{D}_r = \dfrac{e}{r^2},$

which is, of course, the Coulomb law, under the assumption that $e$ is the source of the **D** field, i.e.:

(7.17) $\qquad\qquad e = \dfrac{1}{4\pi}\int_{S^2} *_s \mathbf{D};$

The constitutive law gives:

(7.18) $\qquad\qquad E_r = \sqrt{1 - \left(\dfrac{E_r}{E_c}\right)^2}\, \mathbf{D}_r,$

and if we assume that:

(7.19) $\qquad\qquad E_r = -\dfrac{d\phi}{dr}$

then the potential that one obtains for $E$ is of the form:

(7.20) $\qquad\qquad \phi(r) = f\left(\dfrac{r}{r_0}\right)\dfrac{e}{r_0}$

where $r_0$ is implicitly defined by:

$$（7.21） \qquad E_c = \frac{e}{r_0^2}$$

and plays the role of a critical distance from the point source at which the Coulomb field strength has magnitude $E_c$, and the function $f$ is more complicated, viz.:

$$(7.22) \qquad f(x) = f(0) - \tfrac{1}{2} F\left(\frac{1}{\sqrt{2}}, \bar{\beta}\right)$$

in which $F(k, b)$ is the elliptic function of the first kind with $k = 1/\sqrt{2}$ and $\bar{\beta} = 2\tan^{-1}x$; this also makes:

$$(7.23) \qquad f(0) = F\left(\frac{1}{\sqrt{2}}, \frac{\pi}{2}\right) = 1.8541.$$

Hence, the maximum value of the potential is at $r = 0$, which is still a finite value, namely:

$$(7.24) \qquad \phi(0) = 1.8541 E_c.$$

The qualitative form of $\phi(x)$ is surprisingly similar to the Coulomb potential, except at $x = 0$:

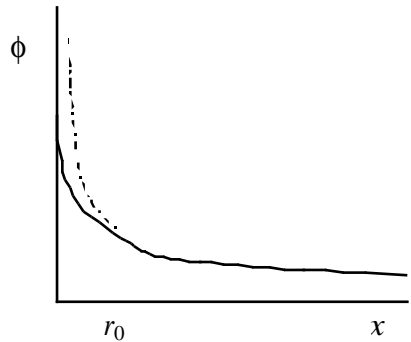

Fig. 16. Born-Infeld potential vs. Coulomb potential.

In particular, past $r_0$ it is in close agreement with the Coulomb potential.

The corresponding radial field strength is easier to express in closed form:

$$(7.25) \qquad E_r = \frac{E_c}{\sqrt{1 - \left(\dfrac{r}{r_0}\right)^2}},$$

which presents an aspect that is analogous to the relativistic theory of kinetic energy, except that the $r/r_0$ seems to play the role of $v/c$, this time.

In order to get some numerical figures for a specific charged particle, Born and Infeld chose the electron, and deduced that:

(7.26) $$\begin{cases} r_0 = 1.2361 \dfrac{e^2}{m_e c^2} = 2.28 \times 10^{-13} \, \text{cm.} \\ E_c = 9.18 \times 10^{15} \, \text{e.s.u.} \end{cases}$$

This also gives and idea of the regime in which Maxwell's vacuum equations are expected to depart from linearity.

**8. Electromagnetism in curved spaces.** So far, the only spacetime manifold that we have been implicitly concerned with is Minkowski space. Indeed, many of the most powerful techniques of quantum electrodynamics and optics center around the applicability of the Fourier transform. However, the Fourier transform is defined in terms of an object, in the form of the radius (or position) vector field:

(8.1) $$\mathbf{r} = x^\mu \frac{\partial}{\partial x^\mu}$$

that can only be defined locally on a more general manifold.

Furthermore, the replacement of a Green function of the form $G(x,y)$ with one of the form $G(x-y)$ depends upon the possibility of making the expression $x - y$ well-defined. At best, this suggests an affine space, since more general actions of the translation group of $R^n$ on a manifold would have non-trivial isotropy subgroups. Hence, there would sometimes be more than one displacement vector associated with a given pair of points.

Nevertheless, since the stated intent of this monograph is to explore the limits of classical linear electrodynamics, as they relate to the successes of quantum electrodynamics, it is only appropriate that we discuss some of the "classical" ideas – in the sense of the word "classical" that means "non-quantum" – that were not of strictly classical origin in the historical sense of that word.

*a. Spacetime topology and electrodynamics.* Electrodynamics is a rich source of applications for differential topology because some of its most elementary notions have immediate interpretations in terms of the topology of spacetime itself.

The first issue that we shall address is the fact that one generally expects any solution of the electrodynamical field equations to vanish at "spatial infinity." This is closely related to the expectation that the field solution have finite energy, and since energy is going to be an integral over "space," we can see that the asymptotic condition on solutions is equivalent to assuming that if we start with space in the form of $R^3$ (the $t = 0$ subspace of Minkowski space) then we are assuming that any field solution $F$ can be extended from $R^3$ to the one-point compactification of $R^3$ by the addition of a "point at infinity," which is diffeomorphic to $S^3$. However, note that the most elementary solution, viz., the Coulomb field of a point charge, is not defined over all of $R^3$ to begin with, but only over $R^3 - \{0\}$, which is diffeomorphic to $S^2 \mathfrak{Z} R$. When one adds the point at infinity to this, one produces a space that is once again diffeomorphic to $R^3$! The non-compactness of this resulting space is the reason for the divergence of the self-energy integral for the field. Hence, one generally prefers that either the space part of spacetime be compact or that one consider only fields of compact spatial support, such as source distributions.

We need to be more precise about the term that we just used, namely "space." In Minkowski space, it is elementary to identify a three-dimensional spacelike submanifold that will play that role, namely the $t = 0$ subspace. However, that is because – once again – we are implicitly using the action of the translation group to associate that subspace with any other $t$, as well. In general, a four-dimensional Lorentz manifold is only locally expressible in the form $R \exists \Sigma$, where $\Sigma$ is a three-dimensional Riemannian manifold. Indeed, such a form suggests some sort of canonical definition of simultaneity, which violates the spirit of special relativity.

We shall refer to such a product manifold representation of spacetime $M = R \exists \Sigma$ as defining a *cylindrical foliation* of codimension one whose leaves, which are therefore all diffeomorphic to $\Sigma$, represent simultaneity hypersurfaces for a given choice of rest frame. In general, the question of spacetime simultaneity foliations is very involved and more details would take us beyond the scope of the present investigation, so we refer the reader to [**33**] for some of those details. For now, we simply point out that the cylindrical foliation seems to be inevitable if one wishes to express spacetime as the Cauchy development of some initial submanifold (cf. [**34**]). This may be simply due to the fact that when the time evolution of something is governed by a flow, one must accept that flows consist of diffeomorphisms, which are not going to change the topology of the initial submanifold. However, one way to get around this is to drop the requirement that the initial Cauchy submanifold be maximal, which could produce a more interesting foliation.

Another topic in elementary electrodynamics that introduces topological considerations is that of the very existence of potentials for the electric and magnetic fields individually, as well as the existence of a potential for the electromagnetic field $F$. These problems take on different shades of topological nuance depending upon whether one is considering statics or dynamics.

In electrostatics, we first regard the electric field as a 1-form $E$ on our three-dimensional Riemannian space manifold $\Sigma$. Its defining properties are:

(8.2) $\qquad dE = 0, \qquad \delta E = 4\pi/c\rho,$

where $\rho$ is the source charge distribution. To say that there exists a potential function for $E$ is to say that there exists a 0-form $\phi$ such that:

(8.3) $\qquad E = d\phi.$

This is certainly sufficient to insure that $dE = 0$, but the question is whether it is also necessary. The question of necessity then comes down to $H^1(\Sigma, R)$, the first de Rham cohomology space for $\Sigma$, which consists of equivalence classes of closed 1-forms that differ by an exact 1-form. If $H^1(\Sigma, R)$ vanishes then all closed forms admit global potential functions, but if this is not the case then some closed forms will admit only locally defined potentials.

An example of a $\Sigma$ for which $H^1(\Sigma, \mathbb{R}) \neq 0$ is a $\Sigma$ that is not simply connected. For instance, of $\Sigma = S^1 \mathfrak{Z} \mathbb{R}^2$, such as $\mathbb{R}^3 - \{z\text{-axis}\}$, then $\pi_1(\Sigma) = \mathbb{Z}$, and correspondingly $H^1(\Sigma, \mathbb{R}) = \mathbb{R}$. This example of fundamental importance in vortex theory or if we assume that our source charge distribution is a linear distribution along the $z$-axis.

In magnetostatics, the potential that one looks for is a potential 1-form. For this reason, it is better to think of the fundamental object for a static magnetic field is the 2-form $*B$, instead of the 1-form $B$. The defining equations for $B$ are:

(8.4) $$\delta B = 0, \qquad dB = 4\pi/cj,$$

where $j$ is the current distribution that constitutes the source of the magnetic field. We rewrite the equations in terms of $*B$ as:

(8.5) $$d*B = 0, \qquad *\delta*B = 4\pi/cj.$$

The first of these suggests that there might be a potential 1-form $A$ for $*B$, i.e., a 1-form $A$ such that:

(8.5) $$*B = dA.$$

Whether this necessary or merely sufficient is a question of the *second* de Rham cohomology space $H^2(\Sigma, \mathbb{R})$, this time. An example of a $\Sigma$ for which $H^2(\Sigma, \mathbb{R})$ is non-vanishing is $S^2 \mathfrak{Z} \mathbb{R}$, which is essentially $\mathbb{R}^3 - \{0\}$; in this case, $H^2(\Sigma, \mathbb{R}) = \mathbb{R}$.

One such magnetic field on $S^2 \mathfrak{Z} \mathbb{R}$ takes the form of a Coulomb field with a pointlike magnetic charge $q_m$ at the origin:

(8.6) $$*B = *\left(\mu \frac{q_m}{r^2} dr\right) = \mu q_m \cos\phi \, d\theta \wedge d\phi$$

since:

(8.7) $$\mathfrak{9}_\Sigma = r^2 \cos\phi \, dr \wedge d\theta \wedge d\phi.$$

We have $d*B = 0$ and $dB = 0$, and since the magnetic flux through any 2-sphere about the origin is:

(8.8) $$\Phi_m[S^2] = \int_{S^2} *B = 4\pi q_m,$$

even though Gauss's theorem is not applicable to this example, since the region bounded by the sphere is not compact, one thinks of such a case as representing a *magnetic monopole* of magnetic charge $q_m$. Ordinarily, the requirement in Maxwell's equations that $dB = 0$ is interpreted as the assumption that no such thing as a magnetic monopole exists (indeed, they have yet to be experimentally exhibited), but now we see that this is also based upon the assumption that $H^2(\Sigma, \mathbb{R}) = 0$.

It is more customary to discuss the issues of de Rham cohomology at the level of the fully unified theory of electromagnetism that starts with the Minkowski field strength 2-form $F$ as its fundamental object and Maxwell's equations:

(8.9) $$dF = 0, \qquad \delta F = \frac{4\pi}{c} J,$$

as the defining ones. The first one brings up the existence of a potential 1-form *A* and, once again, one sees that this comes down to the nature of the second de Rham cohomology space, but this time we need to look at $H^2(M, \mathbb{R})$. Of course, if we are assuming that $M = \Sigma \Im \mathbb{R}$ then $H^2(M, \mathbb{R}) \cong H^2(\Sigma, \mathbb{R})$, and we are back to the previous considerations. However, we ought to check to see what sort of result we should expect when we integrate *F* over a closed spacelike surface *S*. By means of the *E*–*B* decomposition:

(8.10) $$F = \theta \wedge E - *B,$$

we see that the spacelike part of *F* is $*B$ again, and:

(8.11) $$\int_S F = \int_S *B$$

takes on the character of the total magnetic flux through *S*. Similarly, since:

(8.12) $$*F = \theta \wedge B + *E,$$

we also have that:

(8.13) $$\int_S *F = \int_S *E,$$

which represents the total electric flux through *S*.

Since we have already seen that topology can be the source of a magnetic field for which no source current exists, Misner and Wheeler [**35**] pointed out that the same sort of topological mechanism can be responsible for "charge without charge," i.e., non-vanishing electric flux through a closed spacelike surface when there is no electric charge distribution to serve as its source. The suggested construction by which spacetime might have non-trivial second de Rham cohomology is by the attachment of what they called *wormholes* ([13]). This amounts to forming the connected sum $\Sigma \# (S^1 \Im S^2)$, where the object being attached is somewhat reminiscent a 3-handle, in the sense of the Smale handlebody construction, which would be $S^1 \Im D^2$, where $D^2$ is the open 2-disk. We illustrate the attaching construction in Fig. 17:

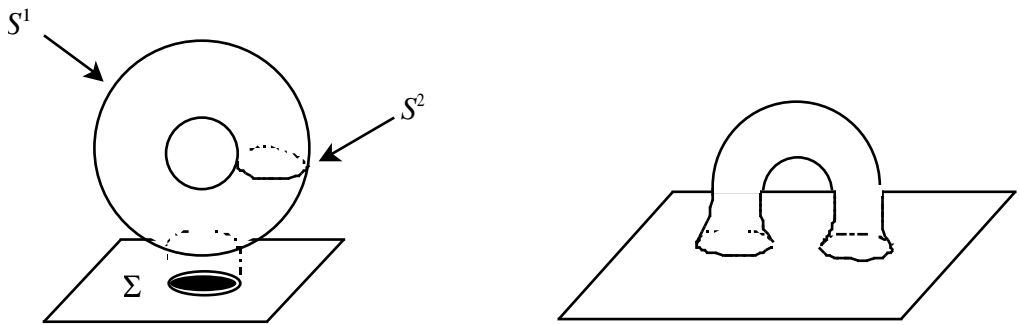

Fig. 17. The attachment of a wormhole to a spacelike section Σ.

---

[13] See also Visser [**36**].

Note that we need to assume that our four-dimensional spacetime *M* admits a cylindrical foliation or this attaching construction might pertain to only one simultaneity leaf at a particular moment of proper time. In a more general *M* one would have t find a different way to introduce 2-holes in the topology, i.e., non-zero generators of $H^2(M, \mathbb{R})$. since the only thing you can attach to a four-dimensional manifold by connected sum and produce another four-dimensional manifold must be four-dimensional.

Although one school of thought regards the field strength 2-form *F* as the physically fundamental object and the potential 1-form *A* as an auxiliary construction, due to the gauge ambiguity in its definition, the approach of gauge field theory is to start with the potential *A*, or rather, the gauge equivalence class that it defines, as the fundamental object and then introduce $F = dA$ as a derived variable. The advantage of this approach is to make possible a far-reaching analogy between the gauge field theories of physics and the mathematical methods of differential geometry and topology.

One can start with the definition that makes two 1-forms on *M*, $A_1$ and $A_2$, *gauge equivalent* iff they have the same exterior derivative:

(8.14) $\qquad A_1 \sim A_2$ iff $dA_1 = dA_2$.

One can see that this is equivalent to saying that they differ by a closed 1-form:

(8.15) $\qquad A_1 \sim A_2$ iff $A_1 - A_2 \in Z^1(M, \mathbb{R})$

where $Z^1(M, \mathbb{R})$ is the space of closed 1-forms. With the usual apologies to de Rham cohomology and the Poincaré lemma, physicists usually strengthen this condition to saying that they differ by an *exact* 1-form, i.e.:

(8.16) $\qquad A_1 \sim A_2$ iff $A_1 - A_2 = d\lambda$ for some $\lambda \in \Lambda^0(M)$.

Without loss of generality, one can assume that $\lambda = \ln g$ for some $g \in \Lambda^0(M)$ and this says that when $A_1$ and $A_2$ are gauge equivalent one has:

(8.17) $\qquad A_2 = g^{-1}A_1 g + g^{-1}dg$.

Admittedly, the inclusion of the factors $g^{-1}$ and $g$ in the first term is superfluous in the present context, but that is only because all of the functions concerned take their values in an Abelian Lie group or Lie algebra, viz., $(\mathbb{R}^*, \cdot)$ or $(\mathbb{R}, +)$. In a more general – i.e., non-Abelian – context (8.17) describes the way that the local representative of a connection 1-form *A* on a *G*-principal bundle $P \to M$ over *M* transforms when one changes from one local section to another by means of a transition function *g* that is defined on the intersection of the domains of the two local sections. In this general context *g* takes its values in the Lie group *G*, which represents the *gauge group* of internal symmetries for the physical field that one is dealing with, and *A* takes its values in the Lie algebra of *G*, which represent the infinitesimal generators of the internal symmetries. The *G*-principal bundle *P* is then referred to as the *gauge structure* for the field theory. The field strength *F* that we have defined as an exterior derivative is more generally defined as the curvature 2-form that is associated with *A*:

(8.18) $\qquad F = dA + \tfrac{1}{2}[A, A]$,

where:

(8.19) $\qquad [A, A](X, Y) = [A(X), A(Y)]$.

In the present case, since $A$ takes the values in R, there are two possible Lie groups that one could use for $G$, namely, (R, 3) and $S^1 = SO(2) = U(1)$. By analogy with quantum wave mechanics, in which wave functions are defined up to a $U(1)$ phase factor, it is generally assumed that this is the proper choice for $G$. Hence, we have promoted our original 2-form $F$ that was defined on $M$ to the status of local representative for a curvature 2-form on a $U(1)$-principal bundle $P$ over $M$, that one obtains from a choice of local gauge $\phi: U \, , \, M \to P$, and the potential 1-form $A$ then becomes the local representative of a connection 1-form on $P$ that takes its values in the Lie algebra of $U(1)$, which is the imaginary line.

Since we have introduced the intermediary of a local section as a local choice of gauge the immediate question to address is whether the local section can be extended to a global one. However, a global section of a principal bundle exists iff it is trivial, i.e., isomorphic as a $G$-principal bundle to the projection $G3M \to M$. Whether or not this is always the case is determined by the topology of $M$. For instance, if $M$ is contractible, like Minkowski space, any $G$-principal bundle will be trivial. More generally, the obstructions to triviality are defined by the *Chern classes* of $P$, which are even-dimensional elements of the integer cohomology $c_i[P] \in H^{2i}(M, Z)$. By the Chern-Weil homomorphism they can be represented in the de Rham cohomology by closed 2-forms on $M$ that one obtains from forming certain symmetric polynomials in $F$. For instance, in four dimensions, one need only consider:

(8.20) $$c_1[P] = \frac{1}{2\pi} F, \qquad c_2[P] = \frac{1}{2\pi} F \wedge F,$$

Of equal importance to topology are the integrals of these forms over two-dimensional and four-dimensional compact orientable submanifolds:

(8.21) $$\int_S c_1[P] = \frac{1}{2\pi} \int_S F, \quad \int_M c_2[P] = \frac{1}{2\pi} \int_M F \wedge F.$$

When $S$ is spacelike and $F$ has its electromagnetic meaning, the first expression gives twice the magnetic flux through $S$. The second expression is more involved. Since one can relate $c_2[P]$ with Euler class of $T(M)$. Chern's generalization of the Gauss-Bonnet theorem says that the left-hand side should equal the Euler-Poincaré characteristic $\chi[M]$ of $M$. If $M$ is compact then this must vanish if $M$ is to admit a Lorentz structure. Similarly, in order for a 2-form $F$ on a four-dimensional Lorentz manifold to represent an electromagnetic wave solution one must have, more specifically:

(8.22) $$F \wedge F = (E \cdot B) 9 = 0.$$

Note that we have already encountered $c_2[P]$ in the form of one of Mie's list of Lorentz invariants from which one can construct a general theory of electrodynamics, as well as in the form of one of the two such invariants that Born and Infeld chose for their own Lagrangian. Clearly, the invariant will only be significant for the non-wavelike solutions of the electromagnetic equations, such as, presumably, the static fields of elementary charges. One also encounters the same 4-form $c_2[P]$ in other field theories where the fields are real in the form of the first Pontrjagin class of the $SO(2n)$-principal bundle that defines the gauge structure. The integral of that form over a compact orientable four-dimensional submanifold *N* of *M* is then referred to as the "topological charge" in that submanifold.

When one is concerned with the topology of one-dimensional and three-dimensional submanifolds of *M*, one can also note that if $F = dA$ (locally) then $F \wedge F = dA \wedge dA = d(A \wedge dA)$, so both $c_1[P]$ and $c_2[P]$ can be locally expressed in non-unique ways as exact forms. Hence, by Stokes's theorem, whenever *S* is bounded by a loop *C*, or *N* is bounded by a closed orientable three-manifold Σ, the integrals:

(8.23) $$\int_c A, \quad \int_\Sigma A \wedge dA$$

are also topological invariants. The forms *A* an *A^dA* are then *Chern-Simons secondary characteristic classes* that are associated with *P*. In hydrodynamics, if *A* represents the covelocity 1-form of a flow then the first integral is the ciculation of the flow around *C* and the 3-form *A^dA* relates to the integrability of the flow. In differential geometry, the first integral will give the angle through which a vector that is acted on by *A* will be rotated when it is parallel-translated around *C*; i.e., it gives the *holonomy transformation* of the loop.

One might naturally wonder whether topology plays a significant role in the definition of the constitutive relation between the field strength 2-form *F* and the induction 2-form *H*. This seems to be essentially reducible to the question of whether topology plays an essential role in the process of vacuum polarization. Since one is dealing with physical processes that are most significant in the realm of small neighborhoods of elementary matter, one might conjecture that vacuum polarization could involve some sort of "topology-changing" processes, such as the attachment of a wormhole to account for the formation of a bound charge-anticharge pair. If this were the case then the "Dirac Sea" would start to look more like "spacetime foam."

However, since the total charge remains zero, one could also imagine less topologically dramatic transitions, such as simply a change in the number of singularities of the vacuum electromagnetic field itself. One must keep in mind that topological invariants such as the Euler-Poincaré characteristic are usually fairly coarse in their resolution of details. For instance, by the Poincaré-Hopf theorem, a manifold *M* can have $\chi[M] = 0$ and still admit vector fields with an arbitrary number of zeroes, as long as their indices still balance out under summation. Hence, a vector field with no zeroes can turn into a vector field with a source and a sink without requiring a change in the underlying topology of the space.

*b. Geometric sources of nonlinearity.* The primary interfaces between geometry and nonlinear electrodynamics, in the sense of field theories whose Lagrangians are constructed from Mie's Lorentz invariants, seems to be in the fact that most of the invariants, with the exception of the purely topological *F^F*, depend on the spacetime metric *g*, if only by way of the spacetime volume element ϑ. Strictly speaking, one could define a volume element on a manifold that did not admit a Lorentz metric, such as a compact orientable four-manifold with non-zero Euler-Poincaré characteristic – e.g., $S^4$ – but since the Lorentz metric itself seems to originate in the characteristic equation of the propagation of electromagnetic radiation, one tends to assume that spacetime has both a volume element and a Lorentz metric. In fact, one usually defines the volume element locally in a coordinate chart $(U, x^\mu)$ by means of $\mathcal{V} = \sqrt{-g}\, dx^0 \wedge dx^1 \wedge dx^2 \wedge dx^3$. Hence, one can make associations such as $A \wedge *A = g(A, A)\mathcal{V}$ and $F \wedge *F = (E^2 - B^2)\mathcal{V}$.

By way of these invariants, $g$ can show up in either the field equations or the constitutive equation. Note that in either the Maxwell or Born-Infeld equations the only place in which the metri figures is in the codifferential operator, by way of the $*$ operator, which involves $\mathcal{V}$. In local form, one has, for a 2-form $\alpha = \frac{1}{2}\alpha_{\mu\nu}\, dx^\mu \wedge dx^\nu$:

(8.24) $$\delta\alpha = *d*\alpha = *d[\tfrac{1}{2}\alpha_{\mu\nu} *(dx^\mu \wedge dx^\nu)]$$
$$= *\left[\frac{\partial(\sqrt{-g}\,\alpha_\nu^\mu)}{\partial x^\mu}\, dx^0 \wedge \ldots \wedge \overline{dx^\nu} \wedge \ldots \wedge dx^3\right]$$
$$= \frac{1}{\sqrt{-g}} \frac{\partial(\sqrt{-g}\,\alpha_\nu^\mu)}{\partial x^\mu}\, dx^\nu,$$

in which the overbar indicates that the differential does not appear in the term.

One can give this last expression a different form in terms of the Levi-Civita connection $\nabla$ that is defined by $g$:

(8.25) $$\delta\alpha = \left[\frac{\partial\alpha_\nu^\mu}{\partial x^\mu} + \frac{\partial(\ln\sqrt{-g})}{\partial x^\mu}\alpha_\nu^\mu\right] dx^\nu = (\nabla^\mu\alpha_{\mu\nu})dx^\nu,$$

since:

(8.26) $$\frac{\partial(\ln\sqrt{-g})}{\partial x^\mu}\alpha_\nu^\mu = -\frac{1}{2}\frac{1}{\sqrt{-g}}\frac{\partial(\det g)}{\partial x^\mu}\alpha_\nu^\mu = -\Gamma^\rho_{\rho\mu}\alpha_\nu^\mu.$$

One can similarly show that if $\alpha = \alpha_\mu\, dx^\mu$ is a 1-form then its codifferential is:

(8.27) $$\delta\alpha = \partial^\mu\alpha_\mu - \Gamma^\rho_{\rho\mu}\alpha^\mu = \nabla^\mu\alpha_\mu.$$

Hence, even the equations $\delta F = 0$, $\delta H = 0$, $\delta J = 0$ will involve a contribution from the metric, or at least its determinant. Of course, in order for this to introduce a nonlinearity in an equation for $F$, one would have to have that $g$ itself depended on $F$. In the case of $H$, the nonlinearity could still originate in the constitutive equation without $g$ being explicitly dependent upon $F$.

There are solutions of the Einstein field equations for gravitation that use the Faraday energy-momentum tensor $T_{em}$ for an electromagnetic field as the source of the gravitationl field, such as in Rainich's theory, or add it to the energy of the mass distribution, as in the Reissner-Nordstrøm solution. Of course, one should keep in mind that these gravitational solutions were derived from starting with the assumption that the correct field equations for the electromagnetism were those of Maxwell, so unless one also assumes that the nonlinearity in one's nonlinear theory of electrodynamics is of purely geometric origin (i.e., by way of the $*$ operator), these solutions are of limited significance in nonlinear electrodynamics, beyond the fact of their existence.

It is not surprising that Einstein eventually suspected that the unification of the theories of gravitation and electromagnetism would probably depend upon first finding a nonlinear extension of electromagnetism that would be more consistent with the nonlinear theory of gravitation that general relativity represents.

**9. Nonlinear optics.** One rich source of experimentally defined insight into the way that nonlinearity affects the propagation of waves in Nature is the burgeoning field of nonlinear optics [**37**, **38**]. If one wishes to be practical in the name of nonlinear optics, one recognizes that the macroscopic material media through which electromagnetic waves might propagate generally tend to be either dielectric materials, which would affect the electric field strength, or ferroelectric materials, which would affect the magnetic field strength. Hence, the literature of nonlinear optics generally tends to treat the two cases separately; indeed, it is the propagation of electromagnetic waves in dielectric media that occupies the center of attention, whereas their propagation in ferroelectric materials is treated as a more specialized topic. However, since the medium that we are most concerned with is the spacetime vacuum manifold when it is in its wave phase (i.e., an *SO*(2) reduction of the bundle of linear frames that facilitates the propagation of waves, cf. [**39**]), we shall try to maintain a certain objectivity that one would necessarily surrender if one were solely concerned with fiber optics or more conventional applications.

*a. Nonlinear susceptibility tensors.* For dielectric materials, nonlinearity generally takes the form of the breakdown of linearity in the susceptibility tensor for the optical medium at high electrical field strengths. Although the specific mechanism by which linearity in a dielectric medium breaks down probably has idiosyncrasies that are not shared by the breakdown of the spacetime vacuum manifold during vacuum polarization, nevertheless, one suspects that the qualitative nature of the phenomena that one observes during the propagation of light waves through nonlinear optical media must be of value in building an intuition of what might be happening in analogous quantum electrodynamical situations, such as the scattering of light by light.

In the realm of "weak nonlinearity," one regards the polarization 1-form as the sum of a linear part and a nonlinear part:

(9.1) $$P = P^{(L)} + P^{(NL)},$$

where

(9.2a) $$P^{(L)} = P^{(0)} + \chi^{(1)}(\mathbf{E}),$$

(9.2b) $$P^{(NL)} = \chi^{(2)}(\mathbf{E}, \mathbf{E}) + \chi^{(3)}(\mathbf{E}, \mathbf{E}, \mathbf{E}) + \dots$$

It is worthwhile to point out that the zeroth order term $P^{(0)}$ might play a significant role in vacuum polarization, although it represents a polarization of the medium that exists in the absence of an applied field. Certainly this is well established in the electrodynamics of macroscopic media, since it accounts for the existence of polarizing filters, but before one dismisses it as absurd in the context of the spacetime vacuum manifold, one must first recall the Casimir effect.

The linear susceptibility tensor field $\chi^{(1)}$ is a symmetric second rank doubly covariant tensor field – i.e., $\chi^{(1)} \in T^{2,0}(M)$ – and for isotropic dielectrics it simply takes the form $\chi g_\Sigma$, where $g_\Sigma$ is the restriction of the spacetime metric tensor field $g$ to a proper time simultaneity leaf. The nonlinear susceptibilities $\chi^{(2)}$, $\chi^{(3)}$, …, are all completely symmetric covariant tensor fields of successively higher rank, so one has defined, in effect, a Taylor series expansion of the nonlinear map:

(9.3) $$P: T(M) \to T^*(M), \quad \mathbf{E} \infty\ P(\mathbf{E})$$

with:

(9.4) $$\chi^{(n)}|_x = \frac{1}{n!} D^n P\big|_x$$

about a given $x \in M$.

If one loosens the latter local definition of $P$ somewhat by regarding $P$ as a map that takes vector fields to covector fields, i.e.:

(9.5) $$P: X(M) \to \Lambda^1(M),$$

then the various susceptibility tensor fields can also be regarded as linear operators that take completely symmetric contravariant tensor fields to covector fields:

(9.6) $$\chi^{(n)}: S^n(M) \to \Lambda^1(M).$$

At the risk of ambiguity, but to remain consistent with established notations, we represent these linear operators as integral operators whose kernel − or Green function − is also denoted by $\chi^{(n)}$:

(9.7) $$P^{(n)}(x) = \int_{x' \in M} \chi^{(n)}(x',x)(\mathbf{E}, \mathbf{L}, \mathbf{E}), \quad n > 1.$$

From this point onward, since most of the established literature of nonlinear optics makes the predictable transition to the frequency-wave number domain, we shall abandon the full generality of differentiable manifolds, which would mean surrendering our right to define the translation invariance of the $\chi^{(n)}$ that would make:

(9.8) $$\chi^{(n)}(x', x) = \chi^{(n)}(x' - x).$$

Hence, we shall retreat to Minkowski space, so $M = \mathrm{M}$, in order to make Fourier analysis more natural. A word of caution is appropriate: ordinarily, the true power of Fourier analysis is felt most strongly in linear systems theory, since the Fourier transform has the power to turn linear differential equations into algebraic equations, which can be solved for the Fourier transform of the solution to the differential equation. However, when the differential equation is nonlinear, so the operator that takes forcing terms, such as $\mathbf{E}$, to solutions, such as $P$, cannot be represented by a convolution, this is no longer the case. Nevertheless, one can still get considerable insight into the effects of nonlinearity by observing what each of the successive linear operators $\chi^{(n)}$ do in the frequency domain.

First, one Fourier transforms $\mathbf{E}$ and $P$:

(9.9a) $$\mathbf{E}(\mathbf{r}, t) = \int \frac{d^3k\, d\omega}{(2\pi)^4} \hat{\mathbf{E}}(\mathbf{k}, \omega) e^{i\mathbf{k}\cdot\mathbf{r}} e^{-i\omega t}$$

(9.9b) $$P(\mathbf{r}, t) = \int \frac{d^3k\, d\omega}{(2\pi)^4} \hat{P}(\mathbf{k}, \omega) e^{i\mathbf{k}\cdot\mathbf{r}} e^{-i\omega t}.$$

This allows us to resurrect our original multilinear relationship between $\mathbf{E}$ and $P$ in the frequency domain:

(9.10) $$\hat{P}^{(n)}(\mathbf{k}, \omega) = \hat{\chi}^{(n)}(\hat{\mathbf{E}}, \mathbf{L}, \hat{\mathbf{E}}),$$

where:

(9.11a) $\quad \hat{\chi}^{(1)}(\mathbf{k},\omega) = \int d^3r\, dt\, \chi^{(1)}(\mathbf{r},t) e^{-i\mathbf{k}\cdot\mathbf{r}} e^{i\omega t}$

(9.11b) $\quad \hat{\chi}^{(2)}(\mathbf{k}_1\omega_1,\mathbf{k}_2\omega_2) =$
$\int d^3r_1\, dt_1\, d^3r_2\, dt_2\, \chi^{(2)}(\mathbf{r}-\mathbf{r}_1, t-t_1; \mathbf{r}-\mathbf{r}_2, t-t_2) e^{-i\mathbf{k}_1\cdot(\mathbf{r}-\mathbf{r}_1)} e^{i\omega_2 t_2} e^{-i\mathbf{k}_2\cdot(\mathbf{r}-\mathbf{r}_2)} e^{i\omega_2 t_2}$

(9.11c) $\hat{\chi}^{(3)}(\mathbf{k}_1\omega_1,\mathbf{k}_2\omega_2,\mathbf{k}_3\omega_3) =$
$\int d^3r_1\, dt_1\, d^3r_2\, dt_2\, d^3r_3\, dt_3\, \chi^{(2)}(\mathbf{r}-\mathbf{r}_1, t-t_1; \mathbf{r}-\mathbf{r}_2, t-t_2, \mathbf{r}-\mathbf{r}_2, t-t_2)$
$\times e^{-i\mathbf{k}_1\cdot(\mathbf{r}-\mathbf{r}_1)} e^{i\omega_1 t_1} e^{-i\mathbf{k}_2\cdot(\mathbf{r}-\mathbf{r}_2)} e^{i\omega_2 t_2} e^{-i\mathbf{k}_3\cdot(\mathbf{r}-\mathbf{r}_3)} e^{i\omega_3 t_3},$

etc.

Note that there is a close analogy between the way that physics approaches the analysis of nonlinear optical operators and the analysis of the nonlinear interactions of elementary particle physics, i.e., by decomposing the nonlinearity into a power series of multilinear terms. This would suggest that the frequency domain susceptibility tensors play a role in nonlinear optics that is analogous to the role of momentum-space propagators in quantum electrodynamics. Of course, our main concern in the present context is investigating the extent to which the polarization of the spacetime vacuum manifold might be analogous to the polarization of nonlinear dielectric media under high electric field strengths.

*b. Nonlinear superposition effects.* Naturally, one of the primary concerns of nonlinear optics is to understand the ways in which nonlinear superposition differs from linear superposition in natural processes. Generally, the way one approaches this is to consider the linear superposition of two plane waves:

(9.12) $\quad\quad\quad \mathbf{E} = \mathbf{E}_a \cos(\mathbf{k}_a\cdot\mathbf{r} - \omega_a t + \phi_a) + \mathbf{E}_b \cos(\mathbf{k}_b\cdot\mathbf{r} - \omega_b t + \phi_b)$

to be an "input" to the nonlinear susceptibility operator and observe the way that the successive multilinear operators in the frequency domain treat such a wave.

When one first examines the quadratic susceptibility, one finds that the effect is to produce new terms that correspond to all possible sums and differences of the frequency-wave number 1-forms:

(9.13) $\quad\quad k_a = \omega_a/c\, dx^0 - (k_a)_i dx^i$ and $k_b = \omega_b/c\, dx^0 - (k_b)_i dx^i$.

Each incoming wave will produce two *self-interaction terms*, corresponding to $\omega + \omega = 2\omega$ and $\omega - \omega = 0$. The first process is referred to as *second-harmonic generation* and the second one represents a DC term in the output, which amounts to a spatially uniform time-invariant electric field; one calls this effect *optical rectification.* The interaction of the two waves produces waves with frequency-wavenumber 1-forms equal to $k_a + k_b$ and $k_a - k_b$.

It is also entirely permissible for the frequency of one of the fields − say $\mathbf{E}_b$ − to be 0, in which case, one will be considering the interaction of a photon with an external field, by way of the medium. This results in a term in $P$ that is linear in $\mathbf{E}_a$, which is equivalent to changing the value of $\chi^{(1)}$ by a term that is proportional $\mathbf{E}_b$. This, in turn, implies that the refractive index of the medium is a function of the DC field, which is referred to as the *Pockels effect.*

If we replace the DC field with an electromagnetic wave whose field strength $E_b$ and frequency $\omega_b$ are both higher than those of $\mathbf{E}_a$ then the quadratic mixing of the frequencies first gives a beat frequency $\omega_I = \omega_b - \omega_a$, which then beats with $\omega_b$ to produce $\omega_a$ again. This has the effect of producing another term in $P$ that is proportional to $E_a$, which effectively changes $\chi^{(1)}$ by a contribution that is proportional to $E_b$. This process is called *parametric amplification*; i.e., the parameter $\chi^{(1)}$ is being modulated by the action of $E_b$.

In order to examine the cubic term, one uses an input wave that consists of a linear superposition of plane waves with frequency-wavenumber 1-forms, $k_a + k_b + k_c$. Once again, the effect of the term at this order is to produce all possible sums and differences of these three 1-forms. This will necessarily subsume the cases described for the quadratic term, when one includes the pairs $(k_a, k_b)$ and $(k_a, k_c)$. In addition, one will have the generation of third harmonics for each term in the sum, and one must further look at all of the ways that we can combine these 1-forms three at a time, a process that one calls *four-wave mixing*, the fourth wave being the output wave. Note that these triples of 1-forms do not have to be distinct. For instance, $k_a + k_a + k_b = 2k_a + k_b$ is a possibility. One can represent the various mixing processes diagrammatically; for instance, the processes that can produce an output that corresponds to $2k_a - k_b$ are given by the following diagrams, in which the rhombus or triangle simply refers to a summing junction:

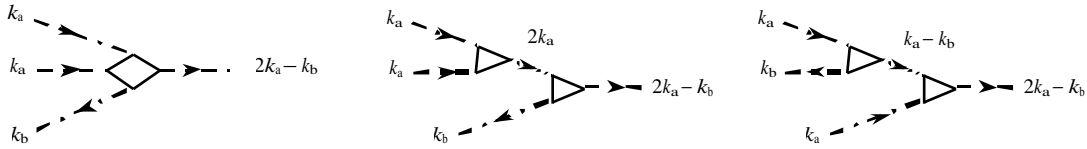

Fig. 18. Four-wave mixing processes that produce $2k_a - k_b$.

In the case of a single field propagating in the medium, so all three incoming frequencies are the same, one obtains a polarization term that is proportional to $E^3$, and one can equivalently alter $\chi^{(1)}$ to have the form $\chi^{(1)} + \chi^{(3)}E^2$. Hence, one can think of the refractive index as dependent upon the intensity of the incident wave.

When a wave propagates through a medium that is subject to a DC field, the cubic term makes a contribution that is equivalent to changing the refractive index by a term that is proportional to the square of the DC field; This is referred to as the *Kerr effect*.

*c. Resonance nonlinearities.* Whenever a medium has both a distributed spin density and a distributed charge density, such as the atomic ions of crystal lattices, there will be resonant frequencies that are associated with not only the coupling of the resulting electric dipole moment density – i.e., polarization – to time-varying external electric fields, but also due to the collective modes of the distribution due to spin-spin coupling. The former case is what one finds in spin resonance phenomena, and essentially represents linear interaction. In the latter case, one can get the production of spin waves in the medium that interact with the propagation of electromagnetic waves in a nonlinear way.

There is a second way that a distribution of electric dipoles can respond to an applied electric field **E**: there can be a displacement of the dipoles themselves. Suppose, for simplicity, that the dipoles are distributed on a spatial lattice and let us represent the displacement of the $i^{th}$ dipole by the vector $\mathbf{u}^{(i)}$ then we can expand the polarization 1-form in such a case into:

(9.14) $\qquad p_\alpha[\mathbf{u}^{(i)}, \mathbf{E}] = p_\alpha^0[\mathbf{u}^{(i)}] + \chi_{\alpha\beta}^{(1)}[\mathbf{u}^{(i)}]E_\beta + \mathbf{L}$

In this expansion, the first term represents the polarization in the absence of an applied field. One can further expand the individual terms as a power series in $\mathbf{u}^{(i)}$:

(9.15a) $\qquad p_\alpha^0[\mathbf{u}^{(i)}] = p_\alpha^0 + \sum_i \frac{\partial p_\alpha^0}{\partial u_\gamma} u_\gamma^{(i)} + \mathbf{L}$

(9.15b) $\qquad p_\alpha^1[\mathbf{u}^{(i)}] = \chi_{\alpha\beta}^{(0)} E_\beta + \sum_i \chi_{\alpha\beta\gamma}^{(R)} E_\beta u_\gamma^{(i)} + \mathbf{L}$

$p_\alpha^0$ represents the static dipole moment. One should note that the expression $\partial p_\alpha^0 / \partial u_\gamma$ has the dimension of a charge density.

The term $p_\alpha^{(R)} = \sum_i \chi_{\alpha\beta\gamma}^{(R)} E_\beta u_\gamma^{(i)}$ is called the *Raman nonlinearity*. By a unitary transformation, one can express the displacement $u_\gamma^{(i)}$ in terms of normal modes $q(\nu)$, $\nu = 1, 2, …, 3n$ of the lattice. One can also expression the Raman nonlinearity for each mode in the form:

(9.16) $\qquad p_\alpha^{(R)} = \sum_i \sum_\nu \chi_{\alpha\beta}^{(R)}(\nu) E_\beta q(\nu)$

Suppose that the displacement represents a normal mode of the dipole distribution with a frequency of $\omega_\nu$, so we can represent it as:

(9.17) $\qquad q(\nu) = Q(\nu)e^{-i\omega_\nu t} + Q^*(\nu)e^{i\omega_\nu t}$

Now, assume that the electric field is also sinusoidal:

(9.18) $\qquad \mathbf{E}(t, x) = [E(x)e^{-i\omega t} + E^*(x)e^{i\omega t}]\mathbf{e}(x)$

where $\mathbf{e}(x)$ is a unit vector at $x$. The Raman nonlinearity becomes:

(9.19) $\qquad p_\alpha^{(R)} = \chi_{\alpha\beta}^{(R)} e_\beta [EQ(\nu)e^{-i(\omega+\omega_\nu)t} + EQ^*(\nu)e^{-i(\omega-\omega_\nu)t}] + c.c.$

In this expression, one immediately sees that the frequency $w$ of the incoming wave has split into the shifted frequencies $\omega \pm \omega_\nu$; this phenomenon is referred to as *Raman scattering*. In the prior discussion of four-wave mixing, symmetry considerations dictated that that $\chi_{\alpha\beta\gamma}^{(2)}$ should vanish, but, in the present case, one would expect that Raman scattering should contribute to the four-wave mixing process whenever the difference of the incoming frequencies is tuned to a Raman mode.

The deformation of a medium by way of a displacement, such as **u**, can also be associated with a change in the density ρ (mass, number, charge) of the medium, which can then result in a change in the dielectric constant ε:

(9.20) $$d\varepsilon(t, \mathbf{r}) = \frac{\partial \varepsilon}{\partial \rho} \, d\rho(t, \mathbf{r}) = \rho \frac{\partial \varepsilon}{\partial \rho} \delta u.$$

When the change in density is driven by thermodynamics fluctuations or external disturbances, the result can be the propagation of a density wave, i.e., a phonon. Since this means a corresponding change in the dielectric constant, hence, in the index of refraction, it is possible for an electromagnetic wave to be scattered by such a phonon; this phenomenon is referred to as *Brillouin scattering*. It is closely analogous to Raman scattering if one regards the normal modes of one case as analogous to the phonons of the other.

There are some optical materials that are also capable of taking on a non-zero magnetization **M**, either by means of spontaneous magnetization or when they are subjected to an external magnetic field. There are two primary ways that this can affect the propagation of electromagnetic waves through those media:

*i)* The magnetization can modulate the electric susceptibilities.
*ii)* The magnetization might be the source of spin waves in the medium, which can result in the diffraction of electromagnetic waves.

In the first case, it can make a difference whether the direction of wave propagation is parallel to or perpendicular to the direction of **M**. In the parallel case, it is possible for a precession of the plane of polarization to result, which is called the *Faraday effect*. When the direction of propagation is perpendicular, one finds the appearance of two distinct propagating modes, whose indices of refraction depend of **M**; this is called the *Cotton-Mouton effect*. One mode is characterized by **E**∥**M** and **B** perpendicular, while the roles of **E** and **B** are reversed in the other mode.

Spin waves can be the result of thermal fluctuations in magnetic materials or subjecting the material to a time-varying external magnetic field. Since the magnetic dipole moment density – i.e., the magnetization – of the medium is proportional to the spin density, this means that as a spin wave passes through a region of the medium, its optical properties can change. In particular, the effect is similar to the a moving diffraction grating as far as an electromagnetic wave moving through the medium is concerned. Since this interaction of electromagnetic waves with spin waves is clearly analogous to the aforementioned Brillouin scattering, one can see that the interaction of an electromagnetic wave with such a spin wave is of purely nonlinear origin.

*d. Self-induced transparency.* When an electric field is coupled to an electric dipole the result is a torque that is applied to the dipole. In the quantum mechanical formulation of this scenario, when the electric dipole is due to an atom or nucleus the resulting spin spectrum is discrete. When the imposed electric field is a monochromatic wave of frequency ω resonant coupling can occur when there are spin transitions whose energy difference is close to $\hbar\omega$, in a manner that is analogous to nuclear magnetic resonance due to the coupling of the nuclear magnetic moment to an imposed magnetic field. When one assembles a macroscopic ensemble of such electric dipoles, as a result of the purely linear analysis of quantum mechanics, one expects that such resonances will result in the near-total absorption of the incident radiation. Furthermore, one also expects that near resonance the spreading of a wave packet due to dispersion will be quite pronounced.

However, these last two aspects of the situation originate in the linear approximation that was used in the conventional quantum mechanical treatment. Let us look at the situation for propagation in one space dimension in a medium whose polarization is given by:

(9.21) $\qquad P(t, z) = 2n[\gamma_1 S_1(t, z) + \gamma_2 S_2(t, z)],$

in which $n$ is the index of refraction, $S_1$ and $S_2$ are the $x$ and $y$ components of the dipoles spin vector, and $\gamma_1, \gamma_2$ are empirical constants.

We assume that the electric field takes the form:

(9.22) $\qquad E(t, z) = e(t, z)\cos[\omega t - kz + \psi(t, z)]$

in which $e$ and $\psi$ are slowly varying functions of both $z$ and $t$.

By a transformation to a rotating frame, followed by a rotation to a frame that has one member $\mathbf{e}_\parallel$ parallel to the electric field and two others, $\mathbf{e}_\perp$ and $\mathbf{e}_3$, that are in the plane that is orthogonal to it, we can make:

(9.23) $\qquad \gamma_1 S_1(t, z) + \gamma_2 S_2(t, z) = \gamma[S_\parallel \cos(\omega t - kz + \psi) - S_\perp \sin(\omega t - kz + \psi)]$

for an appropriate constant $\gamma$.

The spin vector is assumed to obey the equations of precession:

(9.24) $\qquad \begin{aligned} \frac{\partial S_\parallel}{\partial t} &= -(\omega_{m0} - \omega)S_\perp \\ \frac{\partial S_\perp}{\partial t} &= -(\omega_{m0} - \omega)S_\parallel - \frac{\gamma}{h}e(t, z)S_3 \\ \frac{\partial S_3}{\partial t} &= \frac{\gamma}{h}e(t, z)S_\perp. \end{aligned}$

in which $\omega_{m0}$ is the resonant frequency of the medium.

If the incoming photon has frequency that is equal to the resonant frequency and the medium is initially in its ground state (i.e., $S_3 = +\tfrac{1}{2}$, $S_\parallel = S_\perp = 0$) then we shall have $S_\parallel = 0$ for all $t$. If we set:

(9.25) $\qquad S_3 = \tfrac{1}{2}\cos\theta, \quad S_\perp = \tfrac{1}{2}\sin\theta$

then the equations of precession reduce to simply:

(9.26) $\qquad \dfrac{\partial \theta}{\partial t} = -\dfrac{\gamma}{h}e(t, z).$

The wave equation for this situation is:

(9.27) $$\left(\frac{\partial^2}{\partial z^2} - \frac{\varepsilon}{c^2}\frac{\partial^2}{\partial t^2}\right)E(z,t) = \frac{4\pi}{c^2}\frac{\partial^2 P}{\partial t^2},$$

in which $\varepsilon$ is the dielectric constant of the medium, and, after he substitutions for $E$ and $P$, setting $k = \sqrt{\varepsilon}\,\omega/c$, and taking (9.26) into account, we obtain:

(9.28) $$\frac{\partial^2 \theta}{\partial z^2} + \frac{\sqrt{\varepsilon}}{c}\frac{\partial^2 \theta}{\partial t \partial z} = -\frac{2\pi n \omega \gamma^2}{\hbar \varepsilon}\sin\theta.$$

By a further transformation of variables:

(9.29) $$\tau = t - \frac{\sqrt{\varepsilon}}{c}(1-\alpha)z, \quad x = (1+\alpha)z = -\frac{c}{\sqrt{\varepsilon}}t, \quad \mu^2 = \frac{2\pi n \omega \gamma^2}{\alpha \hbar \varepsilon}$$

in which $\alpha$ is any non-zero number, we obtain an equation:

(9.30) $$\frac{\partial^2 \theta}{\partial \tau^2} - \frac{c^2}{\varepsilon}\frac{\partial^2 \theta}{\partial x^2} = -\mu^2 \sin\theta$$

that has the same form as the *sine-Gordon equation.*

This equation admits solutions of a type that one refers to as *solitons*, because, in particular, they propagate and interact without any change in shape. The theory of solitons is a field that has seen considerable development in the last few decades (cf. [**40-42**]), so we cannot hope to cover all of the relevant ideas, but for now, we simply point out that the effect of the nonlinearity that we have introduced into the medium as a result of the coupling of the incoming electric field with the electric dipole moment of the medium is to cancel the absorption of the incoming photon at the resonance frequency; this nonlinear effect is generally referred to as *self-induced transparency*.

*e. Self-interaction*. One of the most elementary aspects of nonlinear field theories in general is the possibility that the nonlinear contribution to the field equations might contain a self-interaction term. In nonlinear optics, this can come about quite simply in the form of a medium whose refractive index, which depends upon $\chi^{(3)}$, varies with the intensity of the field. In such a case, a cylindrical light beam whose intensity varies from the center axis to the outer boundary will experience refraction through an angle that varies with the distance from the center. Depending on the nature of the medium, this may result in either *self-focusing* or *self-defocusing*, in which the outermost rays converge or diverge more than the central ray, respectively.

Consider the simple case of the propagation of an electromagnetic plane wave of amplitude $\hat{E}$, frequency $\omega$, and wave number **k** along the $z$-axis of a medium in which the refractive index exhibits a "Kerr law" dependency on the electric field strength:

(9.31) $$n(E) = n_0 + n_2 |\hat{E}|^2.$$

With these simplifications, the two-dimensional wave equation for **E** with nonlinear polarization terms up to third order takes the form (cf. [**37**, **38**]):

(9.32)
$$i\left(\frac{\partial}{\partial z} + \frac{\partial k}{\partial \omega}\frac{\partial}{\partial t}\right)\hat{E} - \tfrac{1}{2}\frac{\partial^2 k}{\partial \omega^2}\frac{\partial^2}{\partial t^2}\hat{E} = -\frac{\omega n_2}{c}|\hat{E}|^2\,\hat{E}.$$

In this expression, one can identify the group velocity $v_g = \partial k/\partial \omega$, as well as a second-order dispersion term $\partial^2 k/\partial \omega^2$. We perform the transformation of variables:

(9.33)
$$u = \tau\sqrt{n_2\frac{\omega}{c}\left|\frac{d^2k}{d\omega^2}\right|^2}\,\hat{E}$$
$$s = \frac{t - z/v_g}{\tau}$$
$$\zeta = \left|\frac{d^2k}{d\omega^2}\right|\frac{z}{\tau^2},$$

in which $s$ effectively defines a transformation to a comoving frame with velocity $v_g$, and $t$ represents a choice of time scale (in practice, this will often be the duration of the initial pulse). This transformation puts (9.32) into the form of the *nonlinear Schrödinger equation*:

(9.34)
$$i\frac{\partial u}{\partial \zeta} + |u|^2\,u = \pm\tfrac{1}{2}\frac{\partial^2 u}{\partial s^2}.$$

The choice of sign on the right-hand side is dictated by the sign of $\partial^2 k/\partial \omega^2$.

Note that although this equation has a formal similarity to the two-dimensional Schrödinger equation with an effective potential energy of:
(9.35)
$$U(\zeta, s) = \pm|u|^2,$$

actually the roles of time and space have been reversed for this application. When the potential is negative, the effective force of self-interaction is attractive and bound state solutions of the self-interaction can exist, which are also solitonic in character. The simplest such solitonic solution of (9.35) can be obtained in closed form:
(9.36)
$$u(\zeta, s) = \sqrt{2\kappa}\,e^{i\kappa\zeta}\,\text{sech}(\sqrt{2\kappa}\,s).$$

The parameter $\kappa$ is essentially related to the pulse width in time, and any positive value is acceptable. When one reverts to the original variables, the soliton takes the form:

(9.37)
$$E_\omega(z, t) = c\left(\frac{\kappa}{3\pi\omega v_p \chi^{(3)}}\right)^{1/2}e^{i\kappa z}\,\text{sech}\left[\frac{1}{v_g}\left(\frac{2\kappa}{\mu}\right)^{1/2}(z - v_g t)\right].$$

In this equation, $v_p = \omega/c$ is the phase velocity, and what we are describing is a pulse of constant shape that propagates at the group velocity $v_g$. However, that is because we are dealing with a "one-soliton" solution; the solutions that involve more than one soliton can change shape in time. At this point, we shall simply point out that the effect of the nonlinear self-interaction term has been to counteract the dispersive nature of the medium with an attractive force that results in a stable shape for the solution of the equation.

**10. Analogies between nonlinear optics and QED.** In the early days of wave mechanics, the driving analogy that physicists such as de Broglie and Schrödinger were pursuing was that perhaps the relationship of wave mechanics to particle mechanics was analogous to the relationship of wave optics to geometrical optics. Particular attention was devoted to the complementary relationship between the motion of the wave fronts (characteristics) and their orthogonal trajectories (bicharacteristics), and the relationship of the Hamilton-Jacobi equation to the Hamilton equations. This analogy was more heuristic than rigorous, and ultimately Schrödinger arrived at his wave equation by a certain amount of trial and error. Subsequently, the acceptance of that equation as the fundamental statement of the non-relativistic motion of wave-particles has followed more from the practical realities of experimental physics than any progress in finding deeper first principles of Nature that would imply the Schrödinger equation as a natural consequence.

A central theme to the present monograph has been that, even in the face of the successes of quantum electrodynamics, the most fundamental problems of quantum physics are still the original problems that necessitated wave mechanics. Contemplating the very nature of charged particles – in particular, the issue of pointlike or extended charge and mass distributions – and how mass, spin, an charge relate to the field of the particles still remains a worthy venture. One can still regard the very structure of atomic spectra as a definitive *reductio ad absurdum* to Maxwell's theory, or, more precisely, the classical theory of radiation.

As we noted, the most suspicious aspect of Maxwell's equations in the context of atomic to sub-atomic processes is their linearity, which one might expect to be an empirical approximation that breaks down in the realm of high field strengths. In particular, the fact of vacuum polarization is fairly well established by now. Indeed, in supercritical QED, one can find phases of the electromagnetic vacuum manifold beyond polarization.

The applicability of an analogy between nonlinear optics and quantum electrodynamics is clearly based upon a deeper analysis of the spacetime vacuum manifold as an optical medium with non-trivial properties, i.e., non-constant susceptibilities. There is already a considerable amount of work that has been done on just the subject of how nonlinear electrodynamics might affect the value of $c_0$ (cf. [**43-48**]). When one regards this "universal constant" as actually a *derived* quantity whose nature hinges upon the nature of the spacetime vacuum state, at least as it relates to the propagation of electromagnetic waves, it is entirely reasonable to consider the constancy of $c_0$ is equivalent to the linearity of electromagnetic wave equations; more generally, one might have to consider $c_0 = c_0(|\mathbf{E}|, |\mathbf{B}|)$. Even in the context of Minkowski space, if one considers the Lorentz scalar product $d\tau^2 = c_0^2 dt^2 - dr^2$ with such a form for $c_0$, one has a nonlinear electrodynamical basis for the introduction of "quantum fluctuations" about the spacetime vacuum metric.

One might want to distinguish the regimes of optical nonlinearity that are defined by subcritical and supercritical field strengths, which would correspond to vacuum polarization and vacuum ionization, or perturbative and non-perturbative QED, respectively. Clearly, one should expect the scattering of photons by external potentials and other photons to be a rich source of nonlinear optical analogies.

Another possible source of nonlinear optical analogies is in the most elementary problem that we just mentioned, namely, the structure of elementary charge distributions. Recall that one of the key issues that seemed to be discredit the extended models of the electrons was the issue of the stability of the distribution when its natural tendency is to expend. We have just seen that the nonlinear self-interaction terms can stabilize wave packets in the face of dispersion to produce solitons. Perhaps there is a corresponding nonlinear self-interaction term for the electrostatic field that would stabilize the source distribution without having to make it pointlike. Since it is the pointlike nature of the distribution that causes the infinity in the self-energy, this suggests that an appropriate stabilizing term would probably eliminate that spurious consequence, as well.

Trying to imagine how nonlinearity might solve the problem of why the orbiting atomic electrons do not radiate under their centripetal acceleration, at least for a discrete set of energy states, seems to be a deeper problem, by comparison. One might consider the possibility that there is something going on that is reminiscent of the self-transparency that makes a nonlinear optical medium pass a wave whose frequency is in resonance with that of the medium when it should have absorbed it; in the present case, we seem to have the opposite effect, a sort of "self-opacity." This suggest that the vacuum might have a discrete set of resonant frequencies near nuclei, which resolves to a question of whether a polarized vacuum surrounds the atomic nucleus (or any charge) with an extended electric dipole moment density that possibly admits spin waves as normal modes. It interesting that although this optical issue namely, resonances in the medium, seemed less relevant to elementary particle processes on first glance, it actually does become significant in supercritical QED (see Greiner, et al. [**14**]).

A "non-radiation" condition for bound electron states might also take the form of $c = 0$ for some set of physical parameters ($|\mathbf{E}|$, or whatever) although since this suggests an infinite dielectric constant or magnetic permeability, this would suggest that one must make a deeper analysis of the speed of propagation of electromagnetic waves in nonlinear media.

# References

1. Thirring, W., *Classical Field Theory,* Springer, Berlin, 1978.
2. Jackson, J.D., *Classical Electrodynamics,* 2$^{nd}$ ed., Wiley, New York, 1976.
3. Abraham, M., Becker, R., *Theorie der Elektrizität,* v. II, Teubner, Leipzig, 1933.
4. Lorentz, H.A., *Theory of Electrons,* Dover, NY, 1952.
5. Bohm, D., Schiller, R., Tiomno, J., "A Causal Interpretation of the Pauli Equation (A)", Supp. Nuovo Cim., **1** (1935), 48-66; Bohm, D., Schiller, J., (B), ibid., 66-91.
6. Wessenhoff, J., Raabe, A., "Relativistic Dynamics of Spin-Fluids and Spin-Particles," Acta Pol., **9** (1947), 8-53.
7. Madelung, E., "Quantumtheorie in hydrodynamischer Form," Zeit. F. Phys. **40** (1927), 322-326.
8. Delphenich, D.H., "The Geometric Origin of the Madelung Potential," LANL Archives gr-qc/0211065.
9. Landau, L.D., Lifschitz, E.M., *Classical Theory of Fields,* Pergamon, Oxford, 1975.
10. Rohrlich, F., *Classical Charged Particles,* Addison-Wesley, Reading, MA, 1965.
11. Chernoff, P., Marsden, J., *Infinite-dimensional Hamiltonian Systems,* Springer LMS 425, 1974.
12. Kijowski, J., Tulczyjew, W., *A Symplectic Framework for Field Theories,* Springer, 1979.
13. Dittrich, W., *Probing the Quantum Vacuum,* Springer, 2000.
14. Greiner, W., Müller, B., Rafelski, J., *Quantum Electrodynamics of Strong Fields,* Springer, 1985.
15. Itzykson, C., Zuber, J., *Quantum field Theory,* McGraw-Hill, NY, 1980.
16. Bjørken, J.D., Drell, S.D., *Relativistic Quantum Mechanics*, McGraw-Hill, NY, 1964.
17. Jauch, J.M., Rohrlich, F., *The Theory of Photons and Electrons,* 2$^{nd}$ ed., Springer, Berlin, 1976.
18. Heitler, W., *The Quantum Theory of Radiation,* Dover, NY, 1954.
19. Streater, R.F., Wightman, A.S., *PCT, Spin and Statistics, and All That,* Benjamin-Cummings, Reading, MA, 1978.
20. Geroch, R., "Spinor Structure of Space-times in General Relativity I," J. Math. Phys., **9** (1968), 1739-1744.
21. Mostepanenko, V.M., Trunov, N.N., *The Casimir Effect and its Applications,* Clarendon Press, Oxford, 1997.
22. Euler, H., Heisenberg, W., "Folgerungen aus der Diracschen Theorie des Positrons," Zeit. f. Phys., **98** (1936), 714-732.
23. Schwinger, J., "On Gauge Invariance and Vacuum Polarization," Phys. Rev., **82** (1951), 664-679.
24. Landau, L.D., Lifschitz, E.M., *Electrodynamics of Continuous Media,* Pergamon, Oxford, 1960.
25. Chaikin, P.M., Lubensky, T.C., *Principles of Condensed Matter Physics,* Cambridge University Press, Cambridge, 1997.
26. Monastrysky, M., *Topology of Gauge Fields and Condensed Matter,* Plenum, NY, 1993.
27. Kleinert, H., *Gauge Fields in Condensed Matter,* World Scientific, Singapore, 1989.
28. Vilenkin, G., Shellard, *Cosmic Strings and Other Topological Defects,* Cambridge Univ. Press, Cambridge, 1994.
29. Mie, G., "Grundlagen einer Theorie der Materie," Ann. d. Phys. **37** (1912), 511-534; *ibid.*, **39** (1912), 1-40; **40** (1913), 1-66.
30. Born, M., Infeld, L., "Foundations of a New Field Theory," Proc. Roy. Soc. A, **144** (1934), 425-451.
31. Lichnerowicz, A., *Théorie relativiste de la gravitational et de l'electromagnetisme,* Masson and Co., Paris, 1955.
32. Born, M., "Théorie non-linéaire du champs Électromagnetique," Ann. Inst. H.P., **7** (1937), 155-265.
33. Delphenich, D.H., "Proper Time Foliations of Spacetimes," LANL Archives gr-qc/0211066.
34. Hawking, S.W., Ellis, C.F.R., *Large-scale Structure of Spacetime,* Cambridge Univ. Press, 1973.
35. Misner, CW., Wheeler, J.A., "Classical Physics as Geometry," Ann. Phys., **2** (1957), 525-603.
36. Visser, M., *Lorentzian Wormholes,* Springer, Berlin, 1996.
37. Mills, D.L., *Nonlinear Optics,* 2$^{nd}$ ed., Springer, Berlin, 1998.
38. Butcher, P.N., Cotter, D., *The Elements of Nonlinear Optics,* Cambridge Univ. Press, 1990.
39. Rajaraman, R., *Solitons and Instantons,* North Holland, Amsterdam, 1989.

40. Delphenich, D.H., "Spacetime G-Structures I: Topological Defects" LANL Archives gr-qc/0304016.
41. Scott, A.C., Chu, F.Y.F., McLaughlin, D.W., "The Soliton: A New Concept in Applied Science," Proc. IEEE, **61**, (1973), 1443-1483.
42. Barone, A., Esposito, F., Magee, C.J., Scott, A.C., "Theory and Applications of the Sine-Gordon Equation," Riv. Nuovo Cim., **1** (1971), 227-267.
43. Dittrich, W., Gies, H., "Light Propoagation in Non-trivial QED Vacua," Phys. Rev. D, **58** (1998), 025004.
44. Dittrich, W., Gies, H., "Applications of the Light Cone Condition for Various Perturbed Vacua," LANL hep-ph/9903469.
45. Dittrich, W., Gies, H., "Vacuum Birerfringence in Strong Magnetic Fields," LANL hep-ph/9806417.
46. Visser, M., Barcelo, C., Liberati, S., "Bi-refringence versus Bi-metricity," LANL gr-qc/02041017.
47. De Lorenci, V.A., Klippert, R., Novello, M., Salim, J.M., "Light Propagation in Non-linear Electrodynamics," LANL gr-qc/0005049.
48. Novello, M., De Lorenci, V.A., Salim, J.M., Klippert, R., "Geometrical Aspects of Light Propagation in Nonlinear Electrodynamics," Phys. Rev. D, **61** (2000) 045001.